\documentclass[author,usenatbib,usecmfonts,reqno]{nrc2}
\pdfoutput=1 

\usepackage[french,english]{babel}
\usepackage{cite}
\usepackage{bm}
\usepackage{amssymb}
\usepackage{amsmath}
\usepackage[figuresright]{rotating}
\usepackage{graphicx}
\usepackage[usenames,dvipsnames]{color}
\usepackage{lipsum}
\usepackage{ulem}
\usepackage[usenames,dvipsnames]{color}

\newcommand{\beq}{\begin{equation}}
\newcommand{\eeq}{\end{equation}}
\newcommand{\bey}{\begin{eqnarray}}
\newcommand{\eey}{\end{eqnarray}}

\newcommand{\oversim}[2]{\protect{\mbox{\lower0.5ex\vbox{%
   \baselineskip=0pt\lineskip=0.2ex
   \ialign{$\mathsurround=0pt #1\hfil##\hfil$\crcr#2\crcr\sim\crcr}}}}} 
\newcommand{\simgreat}{\mbox{$\,\mathrel{\mathpalette\oversim>}\,$}} 
\newcommand{\simless} {\mbox{$\,\mathrel{\mathpalette\oversim<}\,$}} 
\author{Pavel Kroupa} 
\address[label1]{Helmholtz-Institut f\"ur Strahlen- und
  Kernphysik, Nussallee 14--16, D-53115 Bonn, Germany, Institut
  f\"ur Astrophysik, University of Vienna, Austria.\\ {\bf E-mail for
  correspondence:} pavel@astro.uni-bonn.de}

\shortauthor{Pavel Kroupa}

\begin{document}

\title{Galaxies as simple dynamical systems: observational data
disfavor dark matter and stochastic star formation}

\begin{abstract}
  According to modern theory galactic evolution is driven by the
  dynamics of dark matter and stochastic star formation, but galaxies
  are observed to be simple systems. The existence of dark matter
  particles is a key hypothesis in present-day cosmology and galactic
  dynamics.  Given the large body of high-quality work within the
  standard model of cosmology (SMoC), the validity of this hypothesis
  is challenged significantly by two independent arguments: (1) The
  {\it dual dwarf galaxy theorem} must be true in any realistic
  cosmological model. But the now available data appear to falsify it
  when the dark-matter-based model is compared with the observational
  data. A {\it consistency test} for this conclusion comes from the
  significantly anisotropic distributions of satellite galaxies
  (baryonic mass $< 10^8\,M_\odot$) which orbit in the same direction
  around their hosting galaxies in disk-like structures which cannot
  be derived from dark-matter models. (2) The action of dynamical
  friction due to expansive and massive dark matter halos must be
  evident in the galaxy population. The evidence for dynamical
  friction is poor or even absent.  Independently of this, the long
  history of failures of the SMoC have the likelihood that it
  describes the observed Universe to less than $10^{-4}$~\%.  The
  implication for fundamental physics is that exotic dark matter
  particles do not exist and that consequently effective gravitational
  physics on the scales of galaxies and beyond ought to be
  non-Newtonian/non-Einsteinian. An analysis of the kinematical data
  in galaxies shows them to be described in the weak-gravitational
  regime elegantly by scale-invariant dynamics, as discovered by
  Milgrom. The full classical regime of gravitation is effectively
  described by Milgromian dynamics.  This leads to a natural emergence
  of the simple laws that galaxies are indeed observed to obey. Such
  success has not been forthcoming in the dark-matter-based models.
  Observations of stellar populations in galaxies suggest the
  galaxy-wide IMF, the IGIMF, to vary with star formation rate and
  that stochastic descriptions of star formation are inconsistent with
  the data.  This requires a re-interpretation of the stellar-mass
  assembly in galaxies and thus of the accretion rates onto galaxies.
  A consequence of this understanding of galactic astrophysics is that
  most dwarf satellite galaxies are formed as tidal dwarf galaxies in
  galaxy--galaxy encounters, that they follow the mass--metallicity
  relation, that galactic mergers are rare, that galaxies immersed in
  external potentials are physically larger than isolated galaxies and
  that star-forming galaxies form a main sequence.  Eight~predictions
  are offered which will allow the concepts raised here to be
  tested. A very conservative, cold- and warm-dark-matter-free
  cosmological model may be emerging from these considerations.
\end{abstract}

\maketitle

\section{Introduction} 
\label{sec:introd}

The\footnote{This manuscript is part of a special issue whose topic is
  MOND: modified Newtonian dynamics.} standard model of particle
physics (SMoPP) allows the calculation and prediction of particles,
their excited states and their interactions. Although it is based on
many parameters the origin of which are unknown, the mathematical
framework has been shown by many experiments to be an excellent model
of the real elementary particles and their interactions. Experiments
have not yet shown evidence for physics beyond the standard model
(\cite{Yao06, Riotto12}\footnote{But see \cite{Shaham13} for a
  possible indication of new physics at an energy scale $>1\,$Tev and
  \cite{Pohl13} for indications of physics beyond the standard model
  based on the proton radius puzzle.}).

The standard model of cosmology (SMoC) likewise allows the computation
and prediction of galaxies, their perturbations and
interactions. These computations are compared to observed galaxies and
their distribution, but require the star-formation process to be
understood at the level of time-dependent distribution functions that
describe newly formed stellar populations.

The aim of this contribution is to perform two tests which challenge
the SMoC. The two tests are based (i) on the dual dwarf galaxy theorem
and (ii) on dynamical friction. Furthermore, this text attempts to
collate what the empirical evidence implies for the astrophysics of
galaxies, in terms of their dynamics and star-formation processes. The
overall conclusion is that galaxies appear to be simple and computable
systems.

Before introducing both, an outline of the SMoC is given in
Sect.~\ref{sec:SMoC} to allow an appreciation of its structure and
weaknesses. A description of the two different types of dwarf galaxies
that ought to exist, primordial dwarfs and tidal dwarfs (e.g.,
\cite{Kroupa12a}), is given in Sect.~\ref{sec:types}, before
presenting the two tests falsifying the SMoC in
Sect.~\ref{sec:test}. The robustness of the falsification is assessed
in Sect.~\ref{sec:robust}, based on the anisotropic distributions of
galaxy satellites (e.g., \cite{Ibata13, Conn13, Pawlowski12a,
  Pawlowski14}), on the mass-metallicity relation expected for tidal
dwarfs, and on the various internal inconsistencies of the SMoC. The
necessary replacement of Newtonian dynamics by Milgromian dynamics
\cite{Milgrom83,FM12} is presented in Sect.~\ref{sec:sid}. The
build-up of stellar populations in a dark-matter-free universe is
discussed in Sect.~\ref{sec:buildingblocks}.  Here it is argued that
observational evidence disfavors stochastic star formation, and with a
proper understanding of the systematic variation of the galaxy-wide
stellar initial mass function, galaxies also become much simpler than
thought.  Putting the evidence outlined here together, it may be
possible that a highly conservative, Einsteinian cosmological model
without cold or warm dark matter may have already been discovered
(Sec.~\ref{sec:conserv}).  Eight predictions inspired by this whole
framework are presented in Sect.~\ref{sec:predictions}, and
conclusions are drawn in Sect.~\ref{sec:concs}. Appendix~\ref{sec:app}
lists the used acronyms, their meaning and first-time occurrence in
this text.

\section{The Standard Model of Cosmology (SMoC)}
\label{sec:SMoC}

Fundamentally the SMoC is based on two assumptions: {\it
  Hypothesis~0i}: Einstein's general relativistic field equation is
valid and accounts for the space-time--matter coupling.  {\it
  Hypothesis~0ii}: All matter condensed just after the hot Big Bang
(e.g. \cite{Bari13}). Both assumptions are supported by
observational evidence (for example, general relativity has passed all
tests in the Solar System and near neutron stars; hot Big Bang
nucleosynthesis broadly explains the observed abundance patterns,
\cite{ABC48, Burles01, Coc14}).

\begin{figure}
\centering{}\includegraphics[width=9cm, angle=0]{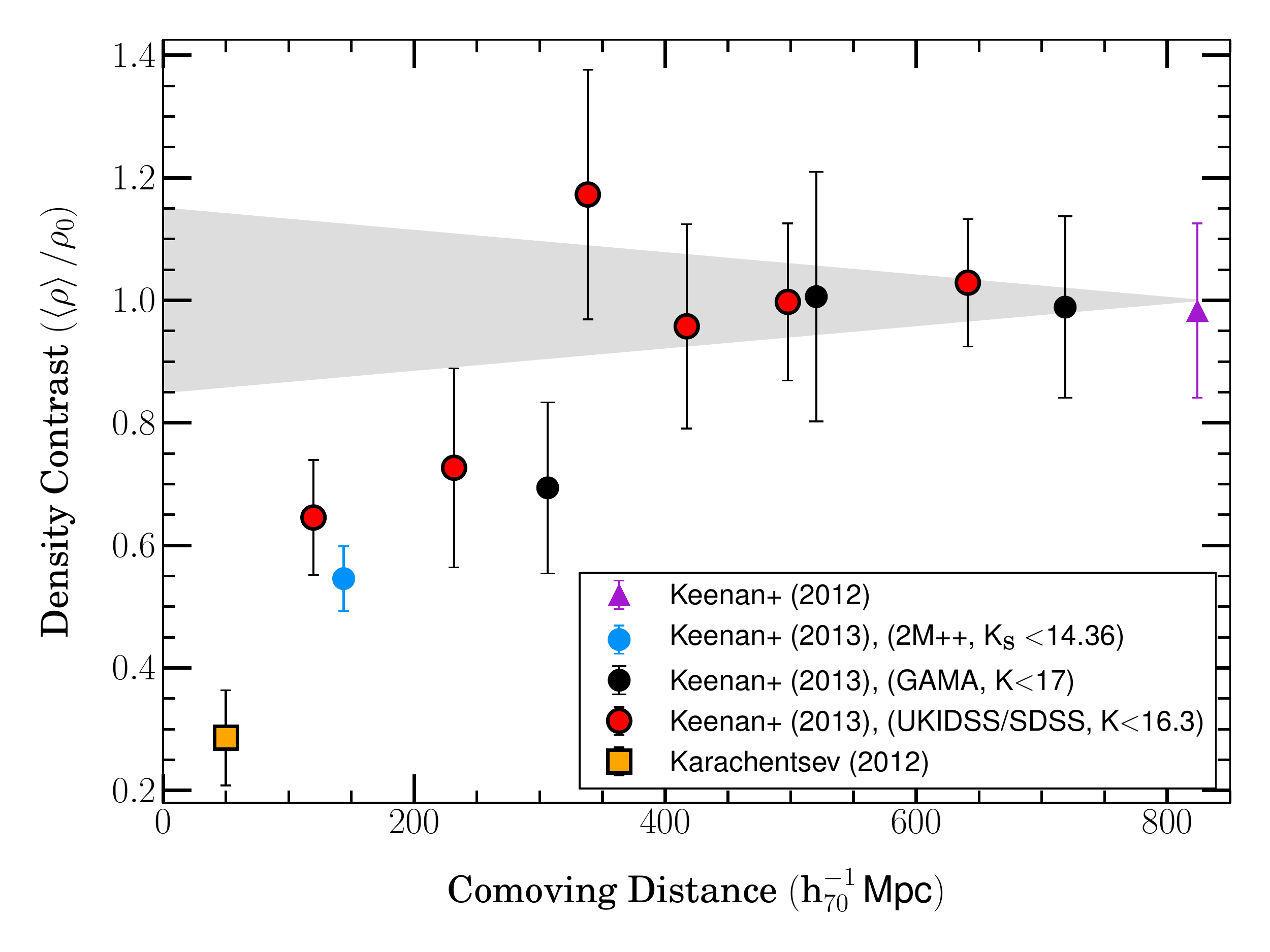}
\caption{\small The Karachentsev--Keenan under density of matter
  around the Sun (see Sec.~\ref{sec:confidencegraph}).  The baryonic
  matter density is plotted as a ratio to the average background
  density in the SMoC vs distance from the Sun to distances of about
  $800 h_{\rm 70}^{-1}\,$Mpc (redshift $z\approx 0.2$). Here the
  Hubble constant is assumed to be $H_0 =
  70\,$km$\,$s$^{−1}\,$Mpc$^{−1}$ (i.e., $h_{70}=1$) . The SMoC, if it
  were a correct representation of the Universe, ought to have
  fluctuations smaller than about 10~\% on scales of 50--400~Mpc
  (\cite{Karachentsev12}, see also fig.~12 in \cite{Keenan13}, here
  indicated by the grey area). The observed under density appears to
  provide the right amount of local inhomogeneity to eliminate the
  need for dark energy within the SMoC. This figure was kindly
  prepared by F. L\"ughausen. }
\label{fig:underdens}
\end{figure}

The computation of an expanding universe under {\it hypotheses 0i and
  0ii} leads to a model that is far more structured and curved than
the observed Universe. The Universe is observed to have a nearly
perfectly flat geometry rather than being curved.  The large-scale
distribution of observed matter (galaxies) and the very high
uniformity of the cosmic microwave background (CMB) radiation imply
the universe to be nearly homogeneous and isotropic\footnote{But see
  Fig.~\ref{fig:underdens}.}.  Since there is no evidence for physics
beyond the SMoPP in the observable Universe, it has also been deduced
that all parts of the known Universe had to have been in causal
contact. Introducing an {\it augmentary hypothesis~1}, inflation,
solves all of these problems at the prize that the physics underlying
inflation is not understood. According to this hypothesis, the volume
of the Universe inflated by a factor of about $10^{78}$ in about
$10^{-33}\,$sec. Evidence for the predicted inflationary gravitational
waves may have been discovered (BICEP2, \cite{BICEP2}), although the
signal may also be due to the Galactic dust foreground
\cite{Flauger14,MS14}.

This resulting model based on {\it hypothesis~0i, 0ii and~1} can
neither account for the observed rotation curves of galaxies nor does
it form structures rapidly enough in comparison with the observed
Universe. This sets limits on the amount of gravitating matter which
can be present in the Universe, and normal (i.e. baryonic) matter
(that is, matter, which is accounted for by the SMoPP) is observed to
contribute significantly too little. Given historical precedents of
successfully introducing new matter components (notably the prediction
of the existence of the planet Neptune because of discrepancies
between calculations and data for the planet Uranus; the prediction of
the neutrino based on missing energy and momentum in the radioactive
beta decay), a further {\it augmentary hypothesis~2} was introduced,
namely that unseen (dark) matter must exist to boost the matter
density, to aid in structure formation and in an attempt to explain
the constant-rotation-velocity curves of galaxies (for a review of
dark matter see e.g. \cite{Einasto13, Trippe14, Popolo14}).  That
galactic rotation curves indeed remain super-Keplerian and nearly flat
at large radii was discovered by Rubin \& Ford (1970, \cite{RF70}) and
established by Bosma (1981, \cite{Bosma81}).  An early argument for
the existence of dark halos was brought up by Ostriker \& Peebels
(1973, \cite{OP73}) who showed that for disk galaxies to be stable
against bar formation their potential needs to be deepened. There are
thus well motivated empirical astronomical arguments suggesting the
presence of a significant dark matter component in galaxies. Bahcall
\& Casertano (1985, \cite{Bahcall85}) argued that the dark halo
profile and the baryonic profile need to be fine tuned in order to
allow for the featureless transition observed between the inner,
visible matter dominated rotation curve and the outer, dark halo
dominated curve. This suggested that the dark halo may be made up of
normal matter.

Possible dark matter candidates such as faint dwarf stars
(e.g. \cite{noWhiteDwarfs13}) and black holes were excluded
\cite{noBHdarkmatter13} by the end of the 1980's. The fine-tuning
pointed out by Bahcall \& Casertano \cite{Bahcall85} was used as an
argument that cold molecular gas may be providing the dark component
\cite{Combes97}. Observational limits on structure formation and
nucleosynthesis however pose a challenge for this proposition, leaving
only the possibility that dark matter (DM) be comprised of new
elementary particles which do not interact significantly with normal
matter except through gravitation \cite{Bergstrom12}. This is
consistent with the constraints on the density of baryonic matter
provided by hot big bang nucleosynthesis.  Broadly two categories are
being considered: cold dark matter (CDM) for massive particles
($>10\,$keV) and warm dark matter (WDM) for light dark matter
particles (of the order of 1--10~keV; axions being an exception as
they are light but cold). Collectively these are referred to here as
{\it exotic dark matter particles}. These are dissipationless
ballistic particles that map out the phase space of the gravitational
Newtonian potential they create. They cannot interact
electromagnetically nor via the strong force as they need to decouple
from normal matter before it recombines in order to form the
gravitational seeds for structure formation.  An excellent review of
the present state-of-the art of the dark matter problem is provided by
\cite{Popolo14}.

The dark-matter cosmological model \cite{Blumenthal84, Davis85}, which
is based on {\it hypotheses~0i, 0ii, 1 and~2}, was found to have a
slower current expansion rate than the real Universe as measured using
supernova of type~Ia (SNIa) as standard candles
\cite{Riess98,Schmidt98,Perlmutter99}. This important discovery forced
the introduction of {\it hypothesis~3} which is a new phase of
inflation driven by dark energy (for reviews see \cite{FTH08,
  Mortonson14}). An undenyingly beautiful aspect of this finding is
that dark energy can be identified readily with the constant $\Lambda$
in GR.

The resulting SMoC (e.g. \cite{Peebles13}, \cite{PLANCK13}),
meaning hereon collectively the $\Lambda$CDM or $\Lambda$WDM models,
has an energy density made up of about 75~\% dark energy, 20~\% exotic
dark matter, and only about 5~\% baryonic matter. Since the Universe
is known to be expanding issues of energy conservation arise, given
that $\Lambda$ corresponds to a constant vacuum energy density
\cite{Baryshev06}. \cite{Peacock99} argues that the negative-pressure
equation of state of the vacuum makes it a source of unlimited energy
allowing any region to inflate arbitrarily at a constant energy
density.

The SMoC has convincing aspects in the sense that the augmentary
hypotheses can be incorporated into physics. For example, dark matter
particles can be identified in extensions of the SMoPP (such as
supersymmetry), and dark energy can be identified with Einstein's
$\Lambda$.  These are, however, not at present understood and have no
independent observational foundation apart from the data that were
used to argue for their introduction. Thus, there is neither evidence
for dark matter particles from particle experiments \cite{LUX13}, nor
has experimental evidence for supersymmetry been found.  Given the
present-day null-evidence for physics beyond the SMoPP, despite the
searches at the highest available energies with the LHC, theoretical
approaches appear to be under crisis \cite{Shifman12}. This is
especially relevant for the putative existence of exotic dark matter
particles. Diaferio (2008, \cite{Diaferio08}) reviews this situation
and discusses the possible implications for gravitational physics.
There is also a crisis in the concept of searching for dark matter
particles within the framework of testing the SMoC. The case can be
made that this is an ill-conceived scientific procedure because, as
emphasized by McGaugh \cite{McGaugh14}, the existence of dark matter
particles can not be falsified by direct searches. Their existence
cannot be falsified because in the event of null detections it can
always be argued that the detection cross section lies below the
instrumental threshold (see footnote~\ref{footn:DMexists}).  The model
can therefore be criticized as being largely (to 95~\% in energy
content) based on unknown physics and to not be falsifiable, and it
may therefore be viewed as being unsatisfying, a view also suggested
by \cite{Diaferio08}. Here I argue that the existence of cold or warm
dark matter particles {\it can be falsified} by astronomical data.

One important caveat needs to be remembered: when Einstein formulated
the theory of general relativity (GR, \cite{Einstein16}), the data he
had used indirectly as constraints were provided through Newton's
empirically derived law of gravitation. This law is constrained in its
validity to the scale of the Solar System only. Neither galaxies nor
the dynamics of matter within them had been established in the 1910s
to what we know they are today (see \cite{Kroupa12a} for a discussion
of this discovery).  Indeed, the ``derivation'' of Einstein's field
equation appears rather heuristic and very much driven by Newtonian
properties \cite{HEL06}.  The application of GR to all scales of the
Universe therefore constitutes an extrapolation by more than 5~orders
of magnitude in acceleration and by~6 and more orders of magnitude in
spatial extent. Such an extrapolation of an empirically derived law in
the gradient of the interaction potential or in the range of the
interaction bears the risk of not being correct.

\vspace{2mm} 
\centerline{ \fbox{\parbox{\columnwidth}
{The SMoC  is based on a major extrapolation of the empirical
pre-galactic foundation of the law of gravitation.
 }}}  \vspace{2mm}

Perhaps this is one of the reasons why it had a torturous development
path in the sense that the mathematical description usually led to
failures such that it had to be repeatedly enriched by augmentary
hypotheses beyond the original ones (hypothesis~0i and~0ii). This
could be viewed as the discovery of new physics in some cases, as long
as independent experimental verification is attainable. 

In this context it is known that Einstein's field equation is not
unique (e.g. \cite{HEL06}) and generalizations have been
developed\footnote{\label{footn:general} One generalization of it
  which incorporates the correct dynamics in experimentally tested
  limits, particularly in the classical very weak-field limit avoiding
  the need for cold or warm dark matter, has been discovered and
  published as TeVeS in~2004 by Bekenstein \cite{Bekenstein04,
    Bekenstein11}. Another generalization has been published in~2007
  by Zlosnik \cite{Zlosnik07} as an Aether vector field which couples
  with the space-time metric.  Milgrom discovered a bimetric
  generalization in~2009 \cite{Milgrom09b}.  Nonlocal and pure-metric
  non-Einsteinian gravity models have also been investigated by
  Deffayet et al. (2014, \cite{Deffayet14}).}.  The generalizations
which are already known to account for the observed dynamics on galaxy
scales {\it without cold or warm dark matter} need to pass all
experimental tests.

Today the very major part of the scientific community understands the
SMoC to be an excellent description of the observed Universe
(e.g. \cite{Primack09, Scann12} for an introduction).  But some
authors do acknowledge that significant tensions between the SMoC and
the data exist \cite{Sanders90, McGaugh98, Disney00, Matteucci03,
  Scarpa06, Primack07, Sanders07, Disney08, Kroupa10, PN10, FM12,
  Hernandez14, Trippe14, WL14, Sanders14}.  Perhaps a worrying issue
is that it is unclear whether it is a $\Lambda$CDM or a $\Lambda$WDM
universe \cite{Vega10}. And, there appears to be a significant
under-density of matter in the local Universe \cite{Karachentsev12,
  Keenan13} (see also Fig.~\ref{fig:underdens}).
The SNIa evidence for late-time acceleration, i.e. for dark energy,
may disappear entirely if this under density is real \cite{Keenan13}.
Furthermore, the distribution of galaxies in the Local Volume does not
correspond to what is expected from the SMoC in terms of the
stratification of the type of galaxies in the Local Void and the Local
Filament \cite{PN10}. Also the baryonic to dark-matter-halo masses of
dwarf galaxies in the SMoC do not agree with the observed values
\cite{Ferrero12}, and, given the importance of mergers in the SMoC,
predictions for galaxy populations do not match-up with the observed
populations \cite{Lu12,Scann12,Shankar14}. Interesting is the argument
by Matteucci \cite{Matteucci03} that the observationally deduced rapid
and early formation times of elliptical galaxies and the associated
down-sizing problem contradict the hierarchical merger-driven build-up
of the galaxy population which is, however, the process implied by the
SMoC \cite{LC93,Stewart08,Fakhouri10}.  The extreme correlation
between the density of dark matter and that of baryons as evidenced by
low-surface brightness galaxies has been used as an argument against
the SMoC by McGaugh \cite{McGaugh98}, and that observed disk galaxies
are much simpler than expected in the SMoC has been demonstrated by
Disney et al. \cite{Disney08}.  Despite these problems many authors
warn from discarding the model. A major argument fielded against such
a step is that baryonic processes may influence the distribution of
dark matter significantly \cite{PG14}.  A clear-cut falsification
\cite{Kroupa12a} would have major deep implications for physics.

Being fully aware that this goes against the majority opinion and
against authority \cite{Peebles13}, here I show that the SMoC cannot
be a correct model describing the Universe. A formal falsification of
the SMoC can already today be deduced using two independent arguments:
by applying the dual dwarf galaxy theorem and by the lack of dynamical
friction. Each is discussed in turn in the following.  Consistency
checks are performed. Essentially, the argument excludes the existence
of exotic dark matter particles, the existence of which are however
the carrying pillar of the SMoC.

Reiterating from \cite{Kroupa12a}, ``{\it for a logical construction to
  be a scientific theory it has to be falsifiable. Otherwise
  predictions are not possible and the construction would not allow
  useful calculations.}'' Furthermore,
 
\vspace{2mm} 

\centerline{ \fbox{\parbox{\columnwidth} {{\sc Uniqueness and
        falsification}: If a theory is found to be consistent with some
      data, then this does not prove that there is no other theory
      which has the same success. However, disagreement of a
      prediction with data, if verified, falsifies the theory.  }}}

\vspace{2mm}

A critical researcher may query his or her concept of what a
successful theory ought to or ought not to achieve with the following
Gedankenexperiment\footnote{The word ``predict'' ie here meant to mean
        what it means, i.e. a calculation of a quantity or set of
        quantities {\it before} any observational data on these
        quantities exists.}:

      \vspace{2mm} \centerline{ \fbox{\parbox{\columnwidth} {{\sc
              Challenge for Theory}: Take a single isolated disk
            galaxy for which the distribution of baryonic matter has
            been mapped out with high resolution. {\it Predict} the
            rotation curve. }}}  \vspace{2mm}

\section{The Types of Satellite Galaxies}
\label{sec:types}

Before stating the dual dwarf galaxy theorem a brief review of the
types of dwarf satellite galaxies which ought to exist is given. A
dwarf galaxy is taken to be one with a baryonic mass $M \simless
10^8\,M_\odot$. 

Concerning the definition of what a galaxy is, within the present
context a discrimination criterion is needed which does not rely on
dark matter.  Using the median two-body relaxation time as a decision
criterion is a possibility \cite{Kroupa98,FK11,Kroupa12a}: all systems
commonly referred to as galaxies have two-body relaxation times longer
than a Hubble time, while all systems commonly referred to as star
clusters have had the process of energy equipartition playing an
important role over a Hubble time, and thus have two-body relaxation
times shorter than a Hubble time. This definition of a galaxy implies
the time evolution of the stellar phase-space distribution function of
such a self-gravitating system to be described by the {\it
  collisionless Boltzmann equation} (CBE). Star clusters require
collisional dynamics methods such as the direct $N$-body method
\cite{Aarseth03} to be employed to follow the time-evolving
phase-space distribution of its stellar population.

\vspace{2mm} \centerline{ \fbox{\parbox{\columnwidth}{ {\sc Galaxies
        vs star clusters}: {\it Galaxies} are self-bound
      stellar-dynamical systems, the time-evolution of the stellar
      phase-space distribution functions of which are solutions of the
      CBE.  This is synonymous to galaxies being systems which have
      median two-body relaxation times longer than a Hubble time. {\it
        Star clusters} are self-bound stellar-dynamical systems which
      have median two-body relaxation times shorter than a Hubble
      time. The time-evolution of a cluster's stellar phase-space
      distribution function is not a solution of the CBE, and is
      driven by the energy-equipartition process. }}}
\vspace{2mm}

\subsection{Primordial dwarf galaxies (PDGs)}
\label{sec:PDGs}

A realistic cosmological theory includes a description of the
formation of galaxies. And, in any theory there is likely to be a
lower-mass limit below which dwarf galaxies do not form.  As a
Gedankenexperiment, in a dark-matter free cosmological model one might
envision that the lowest-mass galaxy is given by the local Jeans mass
in the expanding and cooling post-big-bang gas. 

In the SMoC the exotic dark matter particles decouple from the baryons
at a very early stage and form self-bound structures, the first dwarf
dark matter halos. After recombination these accrete cooling baryons
and set the scale of the lowest-mass galaxy evident today subject to
reionisation which may inhibit star formation \cite{Grebel04}. The
predictions are that major galaxies such as the Milky Way (MW) and
Andromeda ought to have thousands of dark-matter sub-halos orbiting
within the dark matter halo which surrounds them with a radius of
about 250~kpc as documented by a number of seminal contributions
\cite{Klypin99, Moore99, Diemand08, Klypin11}.

These satellite galaxies are typically distributed spheroidally about
their host galaxy following the distribution of the host dark matter
\cite{MKJ07, Pawlowski12b, Watson14, Wang14}, because they have
independent and uncorrelated accretion histories onto the growing host
dark matter halo (see also Sec.~\ref{sec:anisotropy}
and~\ref{sec:conflicts}).  In particular \cite{Pawlowski14} re-analyse
all recently made claims in the literature concerning the theoretical
three-dimensional distribution of primordial satellite galaxies within
MW-type host dark matter halos confirming this theoretical spheroidal
and highly non-disk-like distribution (see \cite{Ibata14} for similar
results).

In the SMoC PDGs are dominated by dark matter at all radii and thus
have very large dynamical mass-to-light ratios
($M/L>10\,M_\odot/L_\odot$, taking from hereon $L$ to be the
bolometric luminosity). Indeed, the dwarf spheroidal (dSph) and
ultra-faint dwarf (UFD) galaxies found around the MW and Andromeda do
have $M/L\approx 10-1000\,M_\odot/L_\odot$ by observation when
interpreting the data in terms of Newtonian dynamics \cite{SG07}.

\begin{figure}
\centering{}\includegraphics[width=8cm]{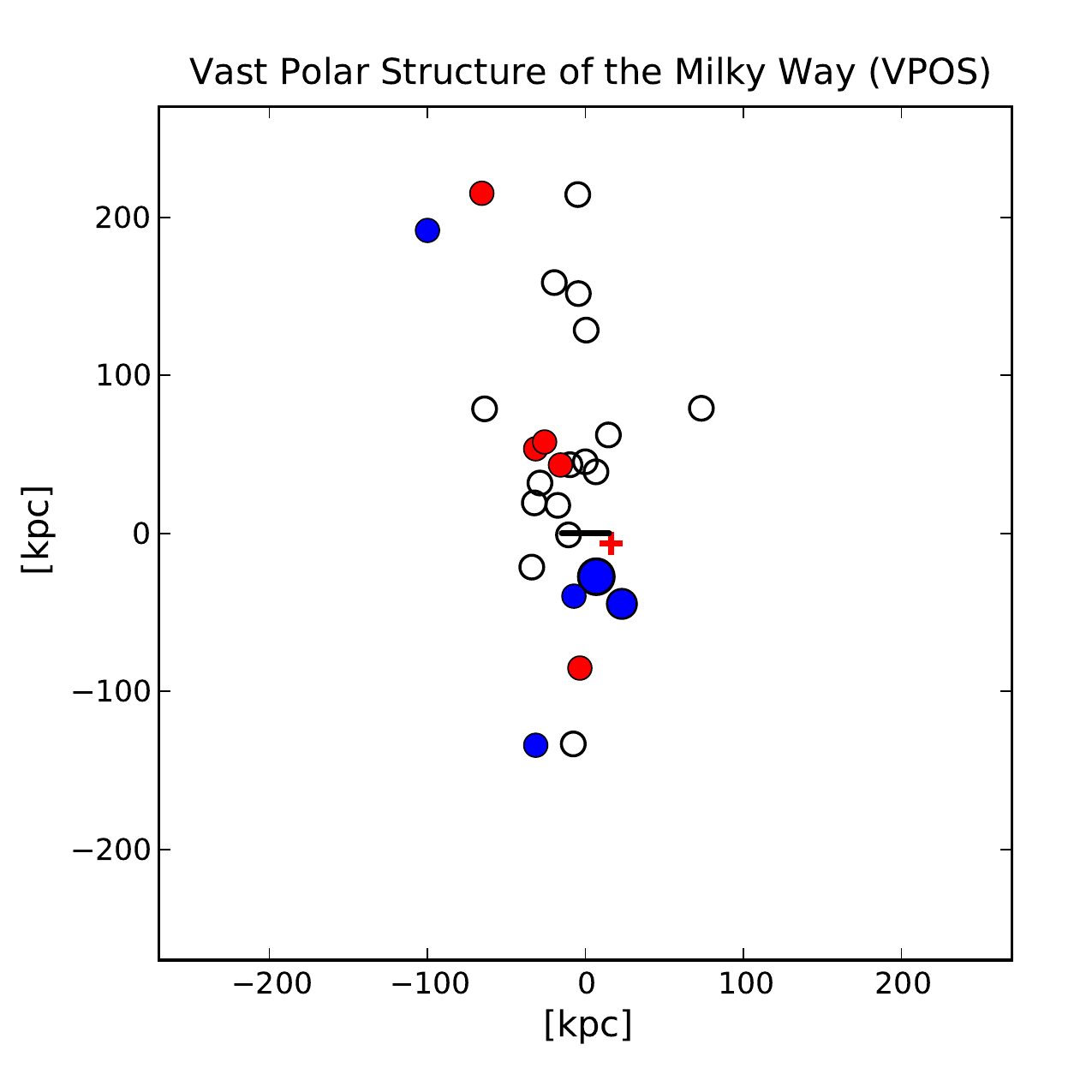}
\caption{\small The vast-polar structure (VPOS) or disk-of-satellites
  (DoS) of the MW seen edge-on (see Sec.~\ref{sec:MW})(``VPOS-3
  orientation'', see \cite{Pawlowski13b}) such that the north pole
  direction of the Local Group and of the MW is along the y-axis
  towards the top (see fig.~6 in \cite{Pawlowski13b}). The face-on
  view is shown in fig.~2 in \cite{Pawlowski13}.  The MW disk is seen
  here edge-on and is shown as a 30~kpc long thick black line at the
  centre of the figure.  The filled circles and the cross are the
  eleven classical (brightest) dSph satellite galaxies. These have
  proper motion measurements such that estimates of their
  three-dimensional motions about the MW are available. Satellites
  colored in red are moving away from the observer who is situated at
  infinity such that Andromeda lies behind and to the left of the MW
  from this viewing direction.  Open circles are ultra-faint dwarf
  (UFD) satellite galaxies for which proper motion measurements do not
  exist yet (see prediction~\ref{sec:pred_pm} in
  Sec.~\ref{sec:predictions}). The Large and Small Magellanic Clouds
  are shown as the largest and intermediate sized blue circles,
  respectively. The red cross depicts Sagittarius, which is close to
  being on a perpendicular orbit to the VPOS and around the MW and
  thus orbits within the here-shown plane apart from a small component
  away from the observer.  The counter-orbiting satellite in the north
  is one of the outermost satellites, Leo~I, which is on a very radial
  orbit (and only has a small velocity component into the plane of the
  figure), while the one in the south may be a case which arises
  naturally if the satellite galaxies are TDGs: the ratio of the two
  counter-orbiting populations yields constraints on the
  MW--other-galaxy interaction which produced the tidal arms within
  which the TDGs formed \cite{Pawlowski11}. The VPOS appears to be
  mass-segregated such that the most massive satellites are near its
  mid-plane (see Sec.~10.1.7 in \cite{Kroupa12a}). The newly
  discovered satellite Crater or PSO~J174.0675-10.8774 with absolute
  visual magnitude $M_V\approx -5.5$ \cite{Belokurov14, Laevens14}
  lies just to the lower right of the fourth open circle from the top
  and enhances the VPOS \cite{Pawlowski14b}.  Andromeda also has a
  great plane of satellites which is discussed in
  Fig.~\ref{fig:GPoA}. This figure was kindly provided by
  M. Pawlowski.}
\label{fig:VPOS}
\end{figure}

\begin{figure}
\centering{}\includegraphics[width=8cm]{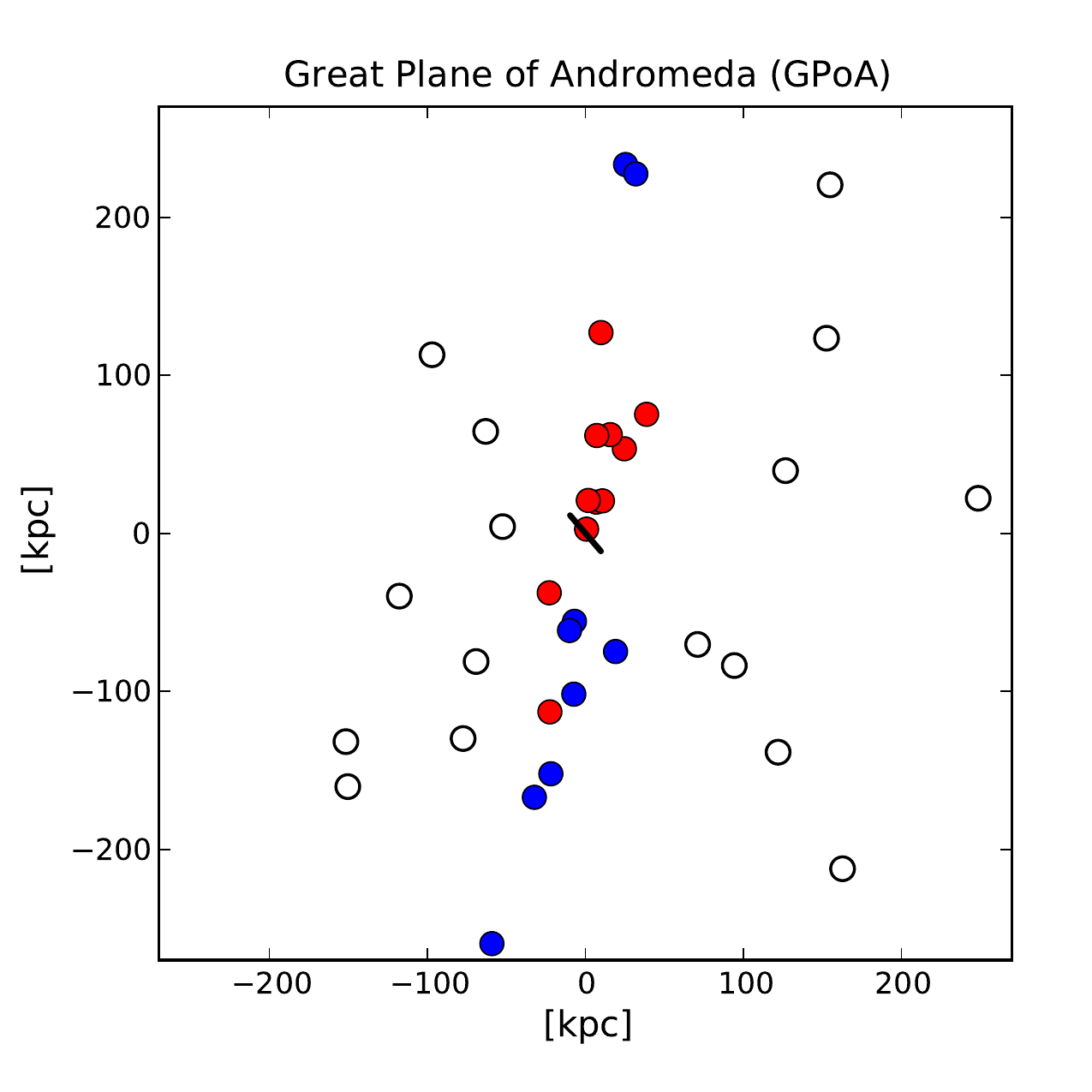}
\caption{\small The great plane of Andromeda (GPoA) or vast thin disk
  of satellites (see Sec.~\ref{sec:GPoA}) confirmed by Ibata et
  al. \cite{Ibata13}. This GPoA contains~19 and thus about half the
  satellite population of Andromeda, has a diameter of about 400~kpc
  and a perpendicular scatter of about 14~kpc and constitutes
  therewith an even more extreme example of a phase-space-correlated
  population of satellite galaxies than the whole VPOS of the MW
  (Fig.~\ref{fig:VPOS}). The GPoA is also rotating mostly in one
  direction: in this rendition the observer is situated near the MW,
  the y-axis points towards MW-north and the satellites moving away in
  Andromeda's rest-frame are red, while the ones moving towards the MW
  are blue. The other satellite galaxies not in this kinematically
  correlated structure are shown as open circles. Note that the GPoA
  and the VPOS (Fig.~\ref{fig:VPOS}) are rotating in the same sense,
  although the two disks are inclined by about~38~deg (fig.~16 in
  \cite{Pawlowski13b}), whereby the GPoA is seen edge-on from the
  MW. Both the GPoA and the VPOS are nearly perpendicular to the MW
  disk. This figure was kindly provided by M. Pawlowski.}
\label{fig:GPoA}
\end{figure}

Because the MW and Andromeda both have merely a few dozen satellite
galaxies each (Figs.~\ref{fig:VPOS} and~\ref{fig:GPoA}), a major
tension between the SMoC and the real Universe is the {\it satellite
  over prediction problem} (popularly known as the {\it missing
  satellite problem}).  
A consensus has emerged in the community that
this problem has been largely solved by various baryonic processes
inhibiting the formation of PDGs in most low-mass dark matter halos,
as is echoed in various research papers (e.g. \cite{Moore06,
  Libeskind07, Tollerud08, Primack09, Belokurov13}).  For example,
\cite{Belokurov13} write about the observed faintest satellite
galaxies of the MW, ``there is now little doubt that these are the
remnants of the galaxies born at high red-shifts in low-mass DM
halos.''

\vspace{2mm} \centerline{ \fbox{\parbox{\columnwidth}{ {\sc Available
        facts}: In the SMoC major galaxies such as the MW and
      Andromeda ought to have a large number (see
      Sec.~\ref{sec:confidencegraph}) of satellite PDGs each with
      their own dark matter sub-halo and each being brighter than a
      typical open cluster, but with a very large dynamical
      mass-to-light ratio.  These satellite galaxies are typically
      distributed spheroidally about their host galaxy reflecting the
      three-dimensional structure of the hosting dark matter halo.
    }}}\vspace{2mm}

\subsection{Tidal dwarf galaxies (TDGs)}
\label{sec:tdgs}

\begin{figure}
\centering{}\includegraphics[width=8cm]{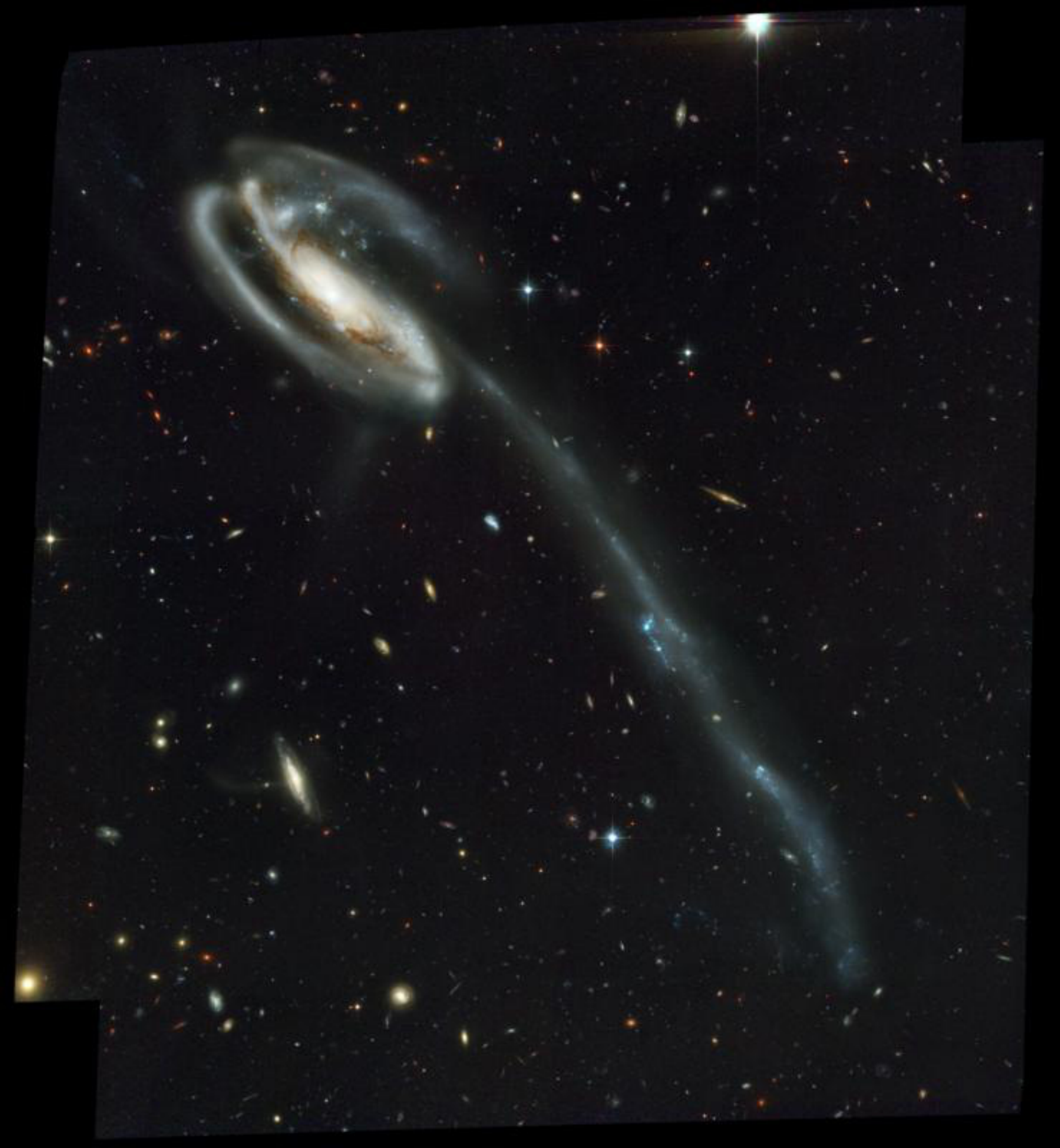}
\caption{\small The tadpole galaxy (UGC 10214) is a MW-class system
  and lies at a distance of about 129~Mpc. Probably a more compact
galaxy crossed in front of the Tadpole Galaxy (from left to right)
  and orbited around the Tadpole to now lie about 90~kpc behind the
  Tadpole. During this process a tidal arm with a length of about
  86~kpc was drawn out. This tidal tail shows regions of star
  formation in star clusters, many of which are clustered within two
  low-mass TDGs. This image \cite{Tadpole04} captures the thinness of
  tidal tails, the star formation within them and the ``lego-principle'',
  according to which stars form in embedded star clusters which are
  the basic building blocks of galaxies (see Fig.~\ref{fig:tadpoleTDG}
  for a close-up and Sec.~\ref{sec:buildingblocks}).  
This image shows a region spanning about $105\,{\rm
    kpc} \times 99\,{\rm kpc}$ (vertical x horizontal).  Image credit:
  NASA, H. Ford (JHU), G. Illingworth (UCSC/LO), M. Clampin (STScI),
  G. Hartig (STScI), the ACS Science Team, and ESA. Publication with
  kind permission by H. Ford.  }
\label{fig:tadpole}
\end{figure}

When two gas-rich galaxies interact angular momentum and energy are
redistributed throughout the interacting system. This involves a
compensation of the temporarily increased binding energy by
acceleration to larger velocities and therewith expulsion of part of
the matter.  Under certain conditions and if one or both interacting
galaxies are disk galaxies the tidal material leaves in a
kinematically cold (velocity dispersion of the pre-encounter disk)
tidal tail.  The tidal arm is then essentially the outward extension
of a spiral pattern which is induced through the galaxy encounter. The
gas in the tidal tail fragments into star clusters, star-cluster
complexes and dwarf-galaxy-scale star-forming clumps, hereafter
referred to as tidal dwarf galaxies (TDGs, \cite{Hunter00,
  Bournaud08}).  A noteworthy features is that the objects born in a
tidal tail constitute highly-correlated structures in phase-space.  A
particularly beautiful example of this is the Hubble Space Telescope
image of the Tadpole galaxy, evident in Fig.~\ref{fig:tadpole}.

Experience shows that high-resolution simulations of galaxy
interactions with gas often lead to the formation of TDGs
(e.g. \cite{Wetzstein07, Bournaud08, Fouquet12,
  Hammer13}).\footnote{Statements do exist in the literature according
  to which very special galaxy--galaxy encounters are needed to
  produce TDGs and that ``a large fraction are certainly
  gravitationally unbound star-forming knots that will quickly
  dissolve''. While the formation and survival of TDGs over~Gyr
  time-spans is a pressing problem to study
  \cite{Ploeckinger14,Duc14}, the rationalisation of these statements
  is not clear, given the already available high-resolution
  computational work, as discussed in this Sec.~\ref{sec:tdgs}.}  A
noteworthy feature of TDG formation is that they and the star clusters
form in tidal arms which constitute highly correlated structures in
phase space. TDGs and star clusters formed in one tidal arm continue
orbiting in the same direction around their host galaxy and form, for
many orbital times, populations of objects that are aligned in a
disk-like structure around the host galaxy.

An important result from all these simulations is that TDGs form by
containing very little exotic dark matter \cite{BH92, Elmegreen93,
  Kroupa97, Hunter00, Bournaud10, Kroupa12a}. Their potential wells
are so shallow (corresponding to maximum circular velocities of about
$v_c \simless$ $30\,$km/s) that the exotic dark matter particles which
are virialised within the much more massive host dark matter halos
($v_c>100\,$km/s) transgress without being captured. Thus, any
rotation curve of matter within a TDG or any dynamical $M/L$ ratio
measurement of a TDG can only be due to a baryonic matter content
assuming the TDG to be in virial equilibrium and the SMoC to be valid
\cite{Hunter00}. For pressure-supported (spheroidal) TDGs (old TDGs
that have lost their gas) repeated tidally induced and thus
non-uniform depletion of their phase space \cite{PP95} over many
orbits leads to quasi-stable solutions that feign domination by dark
matter without it being present \cite{Kroupa97, KK98, Casas12}.

The process of TDG formation is well documented observationally in the
present-day Universe (\cite{Bournaud10}, for a collation of available
observational data see \cite{Dab13}). At large redshift, galaxies are
known to appear perturbed and are most likely interacting at a
significantly larger rate than in the present-day Universe, although a
bias exists for detecting such systems because they are likely to be
brighter with interesting morphologies (see \cite{Disney12} for a
related discussion).  At distant cosmological epochs galaxies would
have been richer in gas. The formation of TDGs would have been much
more frequent then. Indeed, the systematic numerical study Wetzstein
et al.  \cite{Wetzstein07} has demonstrated that the number of formed
TDGs increases with increasing gas fraction.  Today a large fraction
if not all of the dwarf satellite population may be composed of
ancient and in some cases young TDGs even assuming the SMoC to be
valid. This has been calculated assuming a merger tree consistent with
the SMoC by \cite{OT00}, and is also suggested by the correspondence
between structural parameters of dwarf early-type galaxies and
observed TDGs \cite{Dab13}. It is also supported by the observational
finding that dwarf elliptical (dE) galaxies have stellar mass-to-light
ratios within their optical regions such that their DM content is not
significant \cite{Lisker09, Forbes11}.

However, this {\it conjecture that most if not all dwarf satellite
  galaxies are TDGs} needs to be verified with high-resolution
simulations of TDG formation and evolution, as is now being performed
in the Vienna group by Pl\"ockinger, Recchi and Hensler
\cite{Ploeckinger14}. \cite{Duc14} provide significant observational
evidence that TDGs survive for at least a few Gyr.

TDGs would not constitute a major population of satellite galaxies
today if TDGs dissolve after they form.  TDGs may dissolve in three
ways: (a) They may ``fall-back'' on to their host galaxy. (b) They may
become unbound as a result of the star-formation within them driving
out the matter that binds them.  (c) If they survive their formation
they may be destroyed by tidal forces from their host galaxy.

Case (a) is ruled out because tidal material cannot ``fall back'' on
its host galaxy, since orbital angular momentum is conserved and since
TDGs are likely to form and survive only on orbits with large
semi-major axes ($\simgreat 50\,$kpc). A TDG may be tidally shocked
and as a result it may dissolve if it is on a highly radial
orbit. These are rare cases though. Case (b) has been shown to not be
the case by detailed simulations of the first few hundred Myr of TDG
evolution including realistic star-formation and feedback descriptions
by Recchi et al. \cite{Recchi07} and Pl\"ockinger et al.
\cite{Ploeckinger14}.  More simulation work is necessary though to
ascertain the true life-times of TDGs, and the observation of a
few~Gyr old TDGs by \cite{Duc14} suggests survival not to be a
problem.  Case (c) has been shown, by long-term high-resolution
simulations, to not be likely either \cite{Kroupa97, KK98,
  Casas12}. Observational evidence supports these theoretical results
that TDGs do not dissolve upon formation
(e.g. \cite{Bournaud07,Duc11}).

\vspace{2mm} \centerline{ \fbox{\parbox{\columnwidth}{ {\sc Available
        facts}: TDG formation has been common over cosmological
      times. They cannot contain significant amounts of exotic dark
      matter particles and they form in phase-space-correlated
      structures (i.e. in tidal tails).  Available work suggests TDGs
      survive for many Gyr.  It is not known yet which fraction of
      satellite galaxies are ancient TDGs. TDGs and star clusters that
      formed in one tidal tail form highly-correlated structures in
      phase-space.}}}  \vspace{2mm}

\subsection{The dual dwarf galaxy theorem}
\label{sec:dual}

It follows that in any cosmological theory in which galaxies interact,
the following is true:

\vspace{2mm} \centerline{ \fbox{\parbox{\columnwidth}{ {\sc The dual
        dwarf galaxy theorem \cite{Kroupa12a}}: In any realistic
      cosmological model, there exist two types of dwarf galaxies:
      Type~A dwarf galaxies are PDGs that formed above a low-mass
      threshold given by the ambient physical conditions. Type~B dwarf
      galaxies are TDGs.  }}}  \vspace{2mm}

\section{Testing the SMoC}
\label{sec:test}

The SMoC can be tested with astronomical data using two unavoidable
consequences of its underlying hypotheses. The tests are discussed in
Sec.~\ref{sec:testdual} and~\ref{sec:testdynfr}.

\subsection{TEST1: The dual dwarf galaxy theorem and the SMoC}
\label{sec:testdual}

Assuming the SMoC were a correct description of the real Universe,
then dwarfs of Type~A would be dark-matter dominated and would thus
constitute dwarf galaxies which formed in dark matter halos with
masses that are significantly larger than their baryonic masses
\cite{Ferrero12,Behroozi13}. Type~B dwarfs on the other hand would
have stellar/baryonic dynamical mass-to-light ratios ($M/L$ $\simless
5\,M_\odot/L_\odot$) and would have formed through self-regulated
star-formation in a self-regulated potential without a dark matter
halo \cite{Wetzstein07, Recchi07, Ploeckinger14}.

\vspace{2mm} \centerline{ \fbox{\parbox{\columnwidth}{ {\sc Logical
        consequence of the dual dwarf galaxy theorem
        \cite{Kroupa12a}}: If the SMoC is true then type~A and type~B
      dwarfs must have significantly different dynamical $M/L$ values
      and different morphological properties.  }}}  \vspace{2mm}

\vspace{2mm} \centerline{ \fbox{\parbox{\columnwidth}{ {\sc Logical
        inverse of the dual dwarf galaxy theorem \cite{Kroupa12a}}:
      If type~B dwarfs cannot be distinguished dynamically from type~A
      dwarfs (i.e. if type~B dwarfs $=$ type~A dwarfs) then the SMoC
      is falsified.  }}}  \vspace{2mm}

\vspace{2mm}

\noindent
This can be tested with two dual dwarf galaxy (DDG) tests:

\subsubsection{DDG Test~I (late-type dwarf galaxies)}
\label{sec:BTFR}

\begin{figure}
\centering{}\includegraphics[width=8cm]{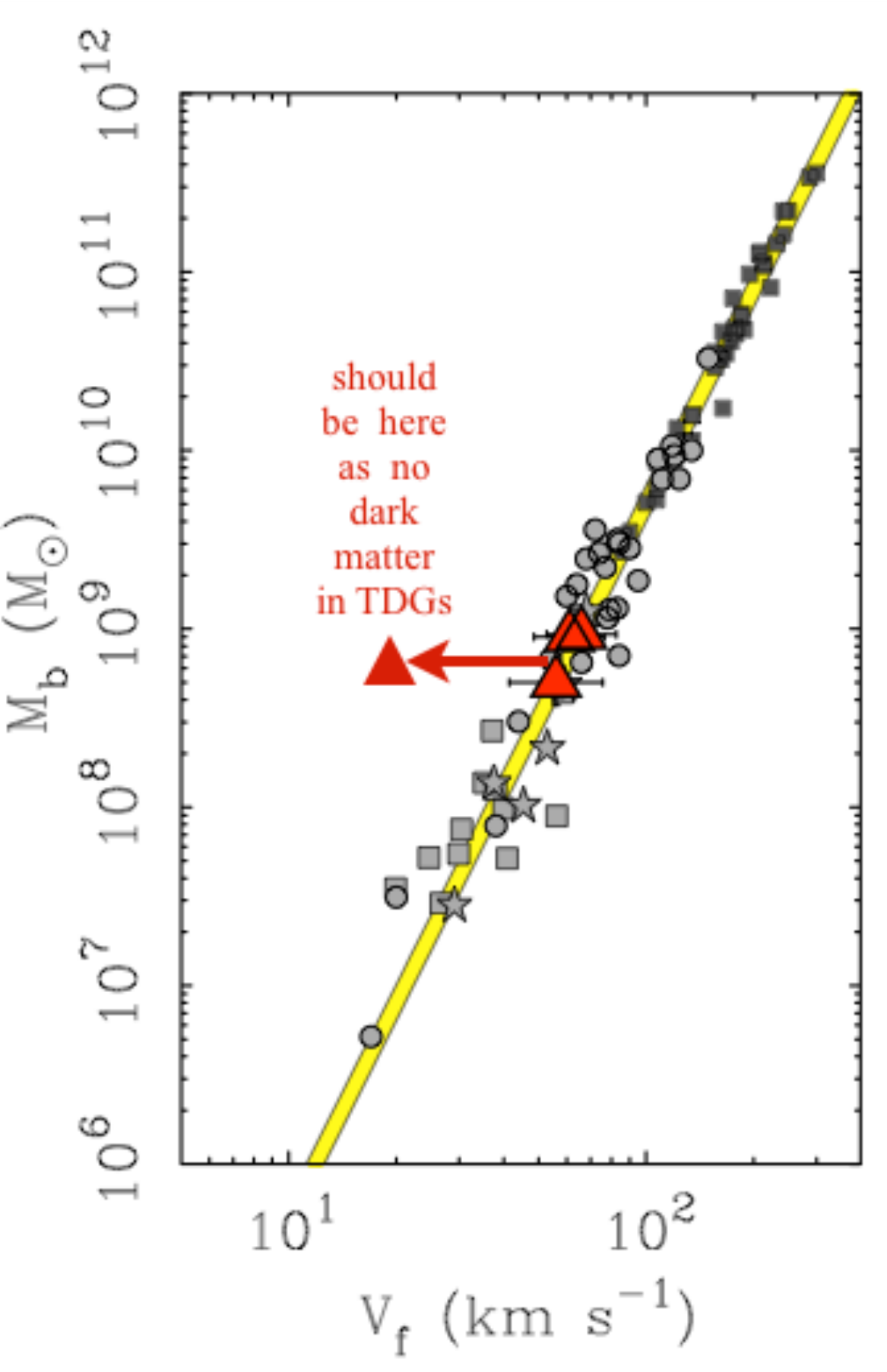}
\caption{\small The baryonic Tully-Fisher relation (BTFR) of late-type
  galaxies. The grey symbols are the total observed baryonic mass
  (stars plus gas) and observed rotation velocity ($v_c\equiv V_f$) as
  collated and explained by McGaugh \cite{McGaugh12}. The three TDGs
  are plotted as red triangles. The red arrow points to the circular
  velocity at the outer radius of about 5~kpc they ought to have in
  this diagramme if they were virialised rotationally supported dwarfs
  of type~B without dark matter and in Newtonian gravity (see
  \cite{Bournaud07} for the rotation curves). The linear (yellow)
  regime is the BTFR (eq.~\ref{eq:TFR}).  The width of the band
  represents the one-sigma uncertainty in the parameter $a_0\approx
  1.21\times 10^{-10}{\rm m}\,{\rm s}^{-2}$ obtained from detailed
  fits to the rotation curves of disk galaxies by \cite{Begeman91}. }
\label{fig:BTFR}
\end{figure}

If the SMoC is true then the baryonic Tully-Fisher relation (BTFR,
\cite{McGaugh12}) is defined by the dark matter halos within which the
luminous galaxies reside, because the circular velocity at the edge of
or outside of the optical galaxy is determined for type~A dwarfs by
the dominating dark matter component. This is well established, given
the observed high dynamical mass-to-light ratios of dwarf galaxies
(deduced by applying Newtonian dynamics in the analysis of the
data). Virialised type~B dwarfs therefore {\it must} lie off the BTFR
by having circular velocities which are a factor $>3$ smaller at the
same galaxy luminosity than a type~A dwarf assuming the conservative
lower value for the dynamical $M/L$ ratio of~$10\,M_\odot/L_\odot$.
Observations have shown three TDGs, for which rotation curves exist,
to lie very close to the BTFR \cite{Gentile07}.  This is shown in
Fig.~\ref{fig:BTFR}.
One TDG could lie close to the BTFR by chance if it is accreting gas
such that the velocity field does not correspond to a virialised
self-gravitating object. But that three TDGs which have been observed
to have well-behaved rotation curves lie about equally close to the
BTFR is unlikely to be chance. Cold molecular gas in the outer regions
of the TDGs \cite{Bournaud07} cannot be the reason for the
super-Keplerian rotation curves because this would require significant
fine-tuning. That is, each TDG would have to have just the correct
amount of cold molecular gas at the right radii in its outer regions
to place it near the BTFR which is defined by the rotation velocity of
the dark matter halo of the normal dwarf galaxies.

\vspace{2mm} \centerline{ \fbox{\parbox{\columnwidth}{
{\it Result}: Since by observation type~B dwarfs $=$ type~A dwarfs in terms
of the BTFR this falsifies the dual dwarf galaxy theorem under the
assumption that the SMoC is true. Thus the SMoC is false. Note that
this deduction is not sensitive to the survival of TDGs, because {\it
  observed} TDGs lie on the BTFR.  Should they dissolve in the future
would not impact this finding.
 }}}  \vspace{2mm}

\subsubsection{DDG Test~II (early-type dwarf galaxies)}
\label{sec:R(M)R}

Consider now pressure-supported dwarf galaxies, that is, dE
galaxies. Type~A dwarfs reside in dark matter halos significantly more
massive by a factor of~10 or more than their baryonic component
\cite{Ferrero12,Behroozi13}. Thus, either type~B dwarfs at the same
baryonic mass as a type~A dwarf must have a velocity dispersion which
is a factor~3 or more smaller than that of the type~A dwarf (for the
same effective radius), or the effective radius of the type~B dwarf
must be smaller by a factor of ten or more than the type~A dwarf for
the same velocity dispersion. However, the comparison of data on known
TDGs with those of dE galaxies shows them to be indistinguishable in
terms of their radius--mass relation, R(M)R \cite{Dab13}.

\vspace{2mm} \centerline{ \fbox{\parbox{\columnwidth}{
{\it Result}: 
 Since by observation the R(M)R of type~B dwarfs $=$ the R(M)R
of type~A dwarfs this falsifies the dual dwarf galaxy theorem if the
SMoC were true. Thus the SMoC is false.  Note that this deduction is
not sensitive to the survival of TDGs, because {\it observed} TDGs lie
on the same radius--mass relation as the putative PDGs.  Should they
dissolve in the future would not impact this finding.
 }}}  \vspace{2mm}

\subsection{TEST2: Dynamical friction}
\label{sec:testdynfr}

If the SMoC is true then a satellite galaxy, be it of type~A or~B,
suffers orbital decay due to Chandrasekhar dynamical friction within
the dark matter halo of its host.  An observational signature of
orbital shrinkage of a satellite galaxy over time at a rate consistent
with the dark halos of the SMoC would thus be a smoking gun for the
existence of exotic dark matter particles. If galaxies are embedded in
massive halos of exotic dark matter particles then they merge whenever
encounters occur with galaxy--galaxy separations smaller than the sum
of the virial radii of the dark halos (about 500~kpc for the MW and
Andromeda, for example) and with relative velocities comparable to the
virial velocity dispersion of the dark halos (a~few hundred~km/s for
the MW and Andromeda, for example) or less. The merging times are
comparable to the crossing times of the dominant dark matter halo (a
few Gyr).

Some observational constraints on this issue are already available:

\subsubsection{The Sagittarius (Sgr) satellite galaxy of the MW}
\label{sec:sgr}

  Attempts at fitting the large number of observational
  constraints on the Sgr streams which are available over the
  past three orbits of the satellite have used models in which
  Sgr has a stellar mass of about $10^8-10^9\,M_\odot$ but no
  or only a minor dark matter halo \cite{Johnston99,Ibata01,
    Helmi01,Fellhauer06,LM10,Correnti10,Vera-Ciro13}. Models without a
  dark matter halo may lead to better agreement with the observations
  \cite{Gomez99}.

  If the SMoC were true, then a type~A dwarf of baryonic mass $M =
  10^8\,M_\odot$ ought to have a dark matter halo mass of about
  $M_{\rm DM}\approx 10^{10}\,M_\odot$, while $M = 10^9\,M_\odot$
  would imply $M_{\rm DM}\approx 10^{11}\,M_\odot$ \cite{Behroozi13,
    Ferrero12}.

  In models in which Sgr is assumed to have a dark matter halo it
  would have had to have fallen-in to the MW dark matter halo about
  2.5~Gyr ago with a first peri-center passage about 1.4 to 2~Gyr ago
  \cite{Purcell11}. But more work needs to be done in order to find
  an improved match to the observational constraints available for the
  Sgr streams. An analytical estimate of the orbital shrinkage time
  until merging of Sgr with the center of the MW leads to merging
  times of a few~Gyr in agreement with the above result \footnote{The
    dynamical friction time scale, eq.~7-26 in \cite{BT87}, is used
    with the following assumptions and results: the maximum impact
    parameter is about the extent of the Sgr dark matter halo taking
    this to be $b_{\rm max}=100\,$kpc \cite{Ishiyama13} and with a
    circular velocity $v_c=220\,$km/s for the MW dark matter halo. If
    the dark matter halo of Sgr has a mass $M_{\rm
      Sgr}=10^{11}\,M_\odot$ then the friction time-scale is $t_{\rm
      fric}=2.5\,$Gyr if Sgr began on a circular orbit at a
    Galactocentric distance $r_i=100\,$kpc.  If the dark matter halo
    of Sgr has a mass $M_{\rm Sgr}=10^{10}\,M_\odot$ then the friction
    time-scale is $t_{\rm fric}=3.3\,$Gyr if Sgr began on a circular
    orbit at a Galactocentric distance $r_i=50\,$kpc.  For $b_{\rm
      max}=25\,$kpc and $M_{\rm Sgr}=10^{9}\,M_\odot$ then $t_{\rm
      fric}=6.5\,$Gyr if Sgr began on a circular orbit at a
    Galactocentric distance $r_i=25\,$kpc.\label{footn:dynfr}}.  An
  alternative model exists in which Sgr was scattered onto its current
  orbit in an encounter near apogalacticon with the Magellanic Clouds
  2--3~Gyr ago \cite{Zhao98}, so that Sgr is not convincing evidence
  for dynamical friction.

\subsubsection{Other MW satellite galaxies}
\label{sec:other}

Assume the satellite galaxies Fornax, Sculptor, Ursa Minor and Carina,
which have measured proper motions, are of type~A such that they have
their own dark matter halos.  By explicitly taking into account
dynamical friction \cite{Angus11} have shown that acceptable orbits,
which are also consistent with their ages and distribution in the
disk-of-satellites or vast polar structure \cite{Pawlowski12a},
appear to be non-existing. The problem lies in the satellites all
needing to walk a tight-rope by not merging with the MW due to
dynamical friction and not being completely dissolved by Galactic
tides but loosing just enough dark matter halo mass to end up on their
observed present long-period orbits.

\subsubsection{The M81 group}
\label{sec:M81}

The M81 group of galaxies consists of a MW-type primary with a few
major star-forming satellite galaxies and strong evidence for past
interactions between the galaxies in the form of pronounced gas
bridges. Attempts at modelling this system with live and
self-gravitating components including dark matter halos appear to have
lead to null results: solutions have not been possible because the
system merges within about a crossing time \cite{Thomson99,
  Yun99}. Solutions are available, but without dynamical friction on
the dark matter halos (\cite{Yun99}, see also \cite{Thomasson93}). No
more-recent computational work is available. The ``hot potato'' was
dropped, it seems.

\subsubsection{Merger-driven evolution of galaxy populations}
\label{sec:mergers}

According to Barnes (1998, \cite{Barnes98}) ``{\it Interacting
  galaxies are well-understood in terms of the effects of gravity on
  stars and dark matter}.'' This is due to our excellent understanding
of Newtonian dynamics and the collisionless nature of stars and of
dark matter particles.  Galaxies of similar masses merge within 0.5~to
3~Gyr \cite{Privon13}. It has indeed become very popular in the
community to refer to mergers or post-mergers when
galaxies-in-interaction or galaxies-past-interaction are discussed,
respectively. But galaxies-in-interaction are merely interacting
galaxies, and ``post-mergers'' may just be galaxies that have had a
strong encounter with another galaxy a few~Gyr in the past whereby
matter may have been transferred from one to the other via the tidal
arms drawn out during the encounter (e.g. \cite{Bournaud03}). If two
galaxies have a relatively close encounter with a relative
pre-encounter velocity of 300~km/s then they will have separated by
300~kpc by one~Gyr or nearly one~Mpc after three~Gyr, unless they
merge. 

Where does the implicit wide-spread understanding that galaxies merge
whenever they interact come from? To the best of my knowledge this
concept can be traced back to Toomre (1977, \cite{Toomre77}) who
interprets a few images of interacting galaxies as galaxies that are
in the process of merging, and he speculates that elliptical galaxies
form from such mergers. Today it is known that this is not the case
(see below in this Section).  Toomre \cite{Toomre77} writes ``These
merger hopes would all be in vain, of course, if the severe kind of
dynamical friction which they seem to require proved simply to be
unattainable.'' In this influential paper, Toomre is worried that in
order to get the observed tidal tail morphologies, off-center
encounters are needed: ``the big worry remains that the strength of
braking may drop off too rapidly with increasing impact parameter or
miss distance, as one seeks circumstances that will also permit the
manufacture of tails of the sort summarized in Table I. . . . , it
seems the center of one galaxy needs to impact the other system no
farther out than about the 1/2 or 3/4-mass radius, lest the rapidity
of their sinking cease to be impressive. . . .  In principle at least,
one can always embed them, prior ro any fateful encounter, within some
appreciably larger and more massive systems like the much-debated
extensive halos.''  This is a remarkable deduction supporting the
evidence for DM, {\it if} galaxies merge. But is it true?

If the SMoC were true then a typical galaxy and its dark halo acquire
their masses through a sequence of mergers driven by dynamical
friction \cite{LC93, Stewart08, Fakhouri10}. The galaxy population at
low-redshift should be a result of this process.  \cite{Shankar14}
confront semi-analytical and semi-empirical models with observational
data on low-redshift galaxies finding significant discrepancies, and
in particular, that the observational data favor long-dynamical
friction time-scales. This is qualitatively consistent with the
conclusion by \cite{Weinzirl09} and \cite{Kormendy10} that too many
($>50$~\%, 94~\% according to \cite{Fernandez14}) of all late-type
galaxies (with baryonic mass $\simgreat 10^{10}\,M_\odot$) do not have
a classical bulge, being at odds with the merging-dominated growth
history of galaxies if the SMoC were true. 

\vspace{2mm} \centerline{ \fbox{\parbox{\columnwidth}{
It is important to
  re-emphasize at this point that the profuse merging behavior of
  galaxies in the SMoC is entirely the result of the assumed existence
  of exotic dark matter particles which form the dark halo potentials
  within which the galaxies exist.}
 }}  \vspace{2mm}

\noindent  The angular-momentum problem of
late-type galaxies is a consequence of this same issue, because
angular momentum is removed in the highly dissipative infall into the
deep potential wells of the dark matter halos of gas streams
(e.g. \cite{Scann12} and references therein). That the haphazard or
stochastic process of galaxy mergers indeed appears not to be an
important process in establishing the present-day galaxy population is
also found by Speagle et al. (2014, \cite{Speagle14}): late-type
galaxies lie on a main sequence with a surprisingly small spread
(Sec.~\ref{sec:mainsequence}). Galaxies appear to evolve in-step over
cosmic time such that, approximately, their star-formation rates
(SFRs) are proportional to their stellar masses.

According to the survey of galaxies with absolute magnitude in the $J$
band brighter than $-20.3$ (stellar mass $\simgreat 1.5\times
10^{10}\,M_\odot$) by \cite{Delgado10}, $\simgreat 90$~\% of all
galaxies are late-type star-forming galaxies, while only 3~\% are
elliptical galaxies (see their fig.5), being completely in accord with
the well-known previous results on galaxy populations (e.g. fig.~4.14
in Binney \& Merrifield 1998, \cite{BM98}).  This population mixture
exists also about 6~Gyr ago \cite{Delgado10}.  A major challenge
remains to understand this population mixture in the SMoC within which
galaxies evolve and grow through a sequence of many mergers of their
dark matter halos. \cite{Puech12} emphasize this tension between
observed galaxies and the SMoC and argue that galactic disks may be
rebuilt after significant mergers implying that half of all disk
galaxies ought to have disks younger than~9~Gyr. This nevertheless
does not solve the problem that more than half of all disk galaxies
with circular velocity $>150\,$km/s have no classical bulge
\cite{Kormendy10}, while \cite{Fernandez14} deduce 94~\% of their disk
galaxy sample to not have a classical bulge.  The evidence of early
formation times at high redshift and short ($<2\,$Gyr) formation
time-scales of elliptical galaxies also contradicts the merger-driven
build-up over time of the galaxy population \cite{Matteucci03}.
\cite{Naab09} confirm that elliptical galaxies cannot have been
forming from disk galaxy mergers.  It is noted here, in advance of the
discussion below, that at any redshift galaxies with peculiar
morphologies may merely be interacting rather than merging, as mergers
are deduced or required only in the framework of the SMoC.

The postulate that elliptical galaxies form rapidly within the SMoC in
an early burst together with dry (i.e. gas-free) mergers at later
times (to be consistent with the merger-driven build-up of dark halos
which continues for the majority of dark matter halos over cosmic
time) may be compromised in view of the majority of the galaxy
population with baryonic masses larger than about $10^{10}\,M_\odot$
not being dry but being late-type disk galaxies containing significant
amounts of gas. In other words, the population of to-merge dry
galaxies does not appear to exist, unless less massive galaxies are
involved mostly. But the not-yet-merged large number of dry dwarf
galaxies with stellar population ages and chemical properties
corresponding to those of the elliptical hosts are not evident
in observations.

The postulate that disk galaxies only occupy dark matter halos which
do not have a major merger after the formation of their thin disks is
incompatible with the majority of dark halos having such mergers and
the great majority of galaxies being disk galaxies, as noted above.
As discussed in \cite{WK15}, SMoC simulations have shown that over the
last 10~Gyr about 95~\% of MW-mass-scale dark matter halos with
a mass of $\approx10^{12}\,M_\odot$ have undergone a minor merger by
accreting a sub-halo with mass $>5\times 10^{10}\,M_\odot$, and 70~\%
of them have accreted a subhalo with mass $>10^{11}\,M_\odot$
\cite{Stewart08}.  According to \cite{Fakhouri10} 69~\% of such halos
have major mergers since z=3 (11--12~Gyr ago) and 31~\% have major
mergers in the past 7-8 Gyr.  Mergers are therefore common for
MW-scale halos in the SMoC.  The simulations have also shown that
major mergers (with equal mass galaxy pairs) disrupt disks completely
and that the remnants of such mergers become early-type galaxies
(elliptical galaxies or bulge-dominated galaxies \cite{CL08}).  Minor
mergers (with a mass ratio 10 : 1) also lead to growth of the bulge
and thickness of the disc (e.g. \cite{Walker96, NB03,
  Kazantzidis09}). Thus, the very large fraction of observed
bulge-less disc galaxies and disk-dominated galaxies (70~\% in edge-on
disk galaxies) is inconsistent with the high incidence ($>70$~\%) of
significant mergers, a point also emphasized by \cite{Kormendy10} (see
also \cite{Fernandez14}).

\begin{figure}
\centering{}\includegraphics[width=8cm]{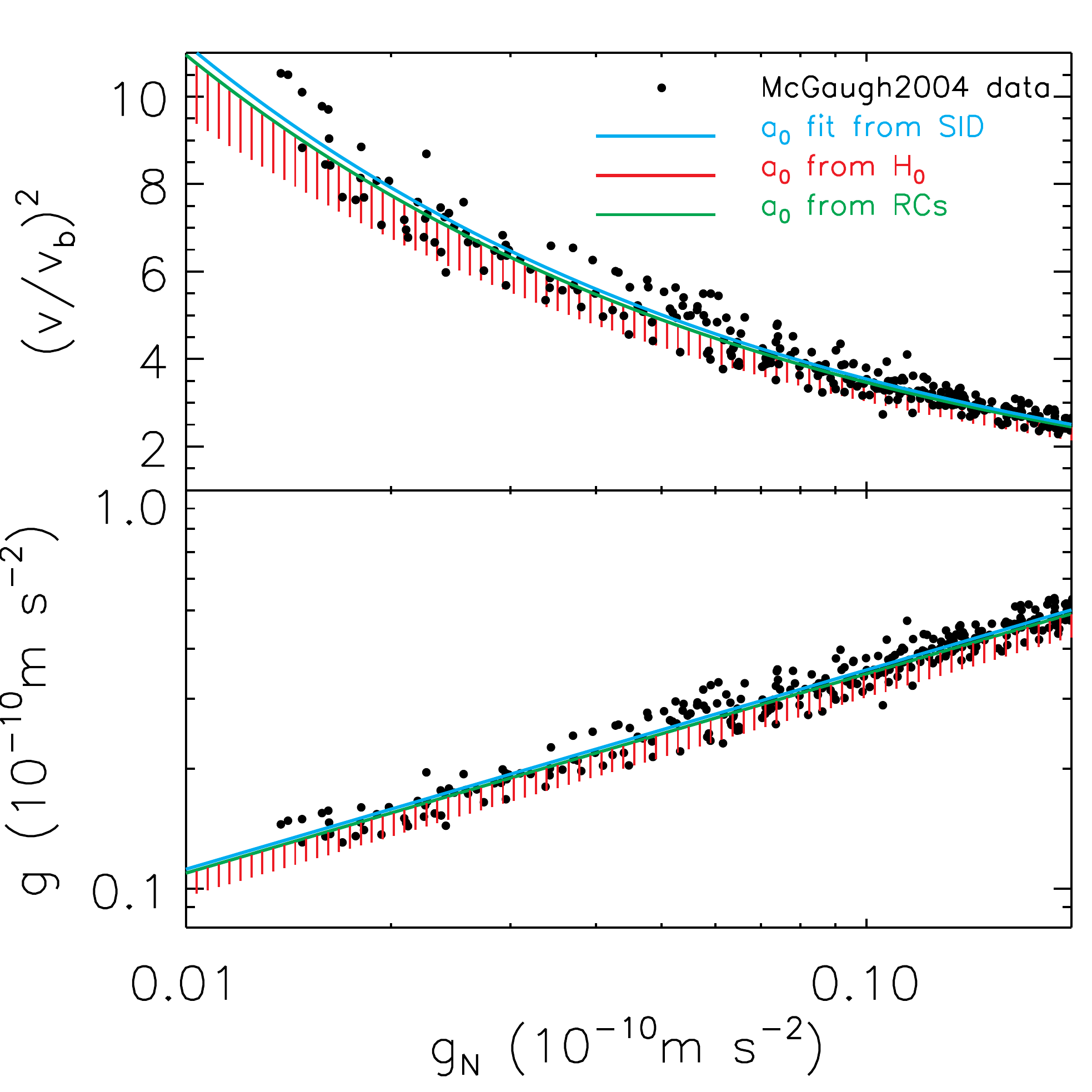}
\caption{\small Observational data (black dots, \cite{McGaugh04})
  confirm that galaxies obey the MDA relation (eq.~\ref{eq:MDA}, in
  this figure $v\equiv v_c, v_{b}\equiv v_{c,b}$).  {\it Upper panel}:
  The MDA relation (eq.~\ref{eq:MDA}) is plotted for different values
  of Milgrom's constant $a_0$ as derived from independent sources
  (i.e. not from the black dots shown here), and as obtained from
  fitting the relation to the data (upper most solid cyan curve:
  $a_0=(1.24 \pm 0.03)\times 10^{-10}\,$m$\,$s$^{-2}$). The red shaded
  region is for $a_0$ according to eq.~\ref{eq:H0} for
  $H_0=76.4\,$km$\,$s$^{−1}\,$Mpc$^{−1}$ (upper limit) and
  $H_0=66.1\,$km$\,$s$^{−1}\,$Mpc$^{−1}$ (lower limit). The green
  curve is for $a_0=1.21\times10^{-10}$m$\,$s$^{-2}$ obtained by
  \cite{Begeman91} by fitting rotation curves of disk galaxies. {\it
    Lower panel}: The true acceleration for a circular orbit, $g$, is
  plotted against the Newtonian acceleration, $g_{\rm N}$, computed
  from the baryonic matter content of the galaxy. The lines show
  eq.~\ref{eq:a0} for the different values of $a_0$ as in the upper
  panel. From \cite{WK15}. }
\label{fig:SID_weakF}
\end{figure}

\begin{figure}
\centering{}\includegraphics[width=8cm]{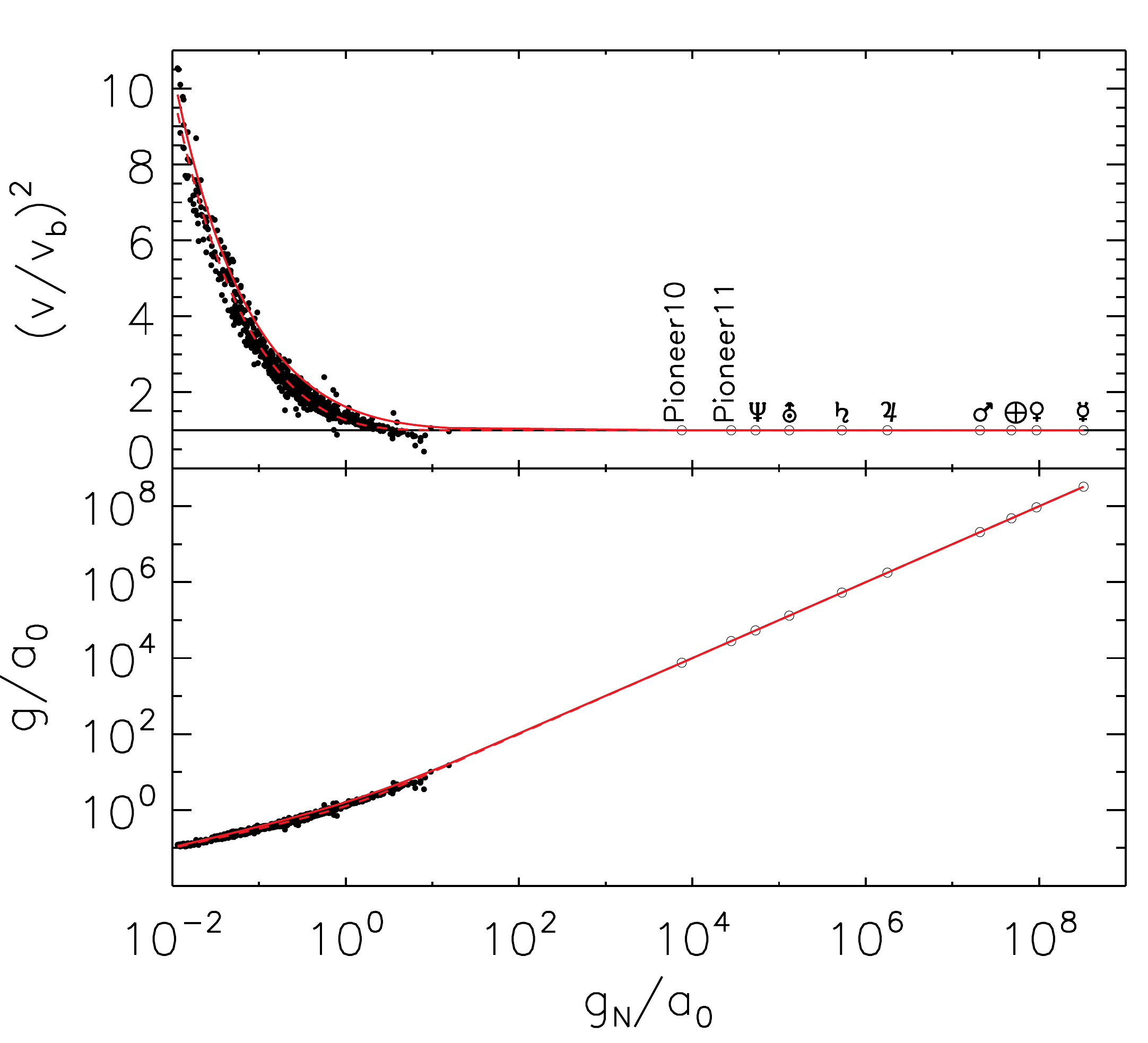}
\caption{\small The role of the function $\mu$
  (eq.~\ref{eq:genPoisson}).  Both panels are as in
  Fig.~\ref{fig:SID_weakF} but for the entire range of classical
  accelerations, including the Solar System objects collated by
  \cite{FM12}.  The solid red line is for the ``simple''
  $\mu$-function and the dashed line for the ``standard''
  $\mu$-function (both lines being very close to another). See
  \cite{FM12} for details. From \cite{WK15}.  }
\label{fig:SID_all}
\end{figure}

More recently, a series of new models using smoothed-particle
hydrodynamics (SPH) and moving-mesh simulations \cite{Agertz11,
  Guedes11, Aumer13} are stated to reproduce realistic disk galaxies
if the host dark matter halo has a quiet merger history, that is, if
it undergoes no major merger after redshift $z > 1$ or~3. For example,
\cite{Marinacci14} apply their new hydrodynamic algorithm AREPO5,
which relies on a moving unstructured Voronoi tessellation together
with a finite volume approach. They explicitly exclude one of their
models at redshift $z=0$ because it is involved in an encounter with
another model galaxy such that the disks are significantly perturbed.
The rotation curves of the gas component of their model disk galaxies
show significant drops at galactocentric radii beyond about 10~kpc in
significant disagreement with the observed HI rotation curves of disk
galaxies which extend as far as a tracer can be detected and often to
50~kpc and beyond (see e.g. the review by \cite{FM12}).  Indeed,
\cite{WK15} show that all disk galaxies hitherto modelled in the SMoC
fail to reproduce the empirical mass-discrepancy--acceleration
correlation (Figs.~\ref{fig:SID_weakF} and~\ref{fig:SID_all}) and thus
do not constitute adequate models of galaxy formation and evolution.

In the above simulations of the formation of disk galaxies, no mergers
with a mass ratio of the sub-structure and host galaxy larger than 1 :
10 are allowed at low redshift. These authors have thus, essentially,
rediscovered the models by Samland \& Gerhard (2003, \cite{SG03}) who
demonstrated that disk galaxies can be grown in growing dark matter
halos without mergers.  But such galaxies are selected from the
unlikely fraction of MW-scale halos with a quiet merging history in
SMoC simulations.

Using the same AREPO5 code, Vogelsberger et al. (2014,
\cite{Vogelsberger14}) report the hitherto largest full-scale
cosmological simulation finding a ``reasonable population of
ellipticals and spirals''.  This work is a milestone in simulating and
visualizing structure formation down to galaxies with a stellar mass
$M_*\approx 10^9\,M_\odot$ in the Universe assuming the SMoC to hold.
The Illustris project \cite{Vogelsberger14b} provides an on-line
galaxy observatory.  The computed population of galaxies results from
mergers of smaller dark matter halos and awaits detailed studies in
terms of their morphological types, their rotation curves and their
location with respect to the empirical baryonic Tully-Fisher relation
(Fig.~\ref{sec:BTFR}).  The authors do not address the major problems
that have arisen in testing the SMoC on real data and their
implementation of the IMF can be challenged
(Sec.~\ref{sec:buildingblocks}).  Their simulated galaxies with low
star-formation rates contain about 52~\% of all the stellar mass which
is in galaxies more massive than $M_*=10^9\,M_\odot$. The authors
state that this agrees well with observations, but according to
\cite{Delgado10} (see also e.g. fig.~4.14 in \cite{BM98}) $\simgreat
90$~\% of all galaxies with $M_*\simgreat 10^{10}\,M_\odot$ today and
6~Gyr ago are late-type star-forming galaxies. This may suggest a
tension between the simulated and the real Universe in conformity with
the previously known angular momentum problem. But the
authors\cite{Vogelsberger14} suggest that the problem of forming
Milky-Way-type galaxies has now been overcome: ``The fact that our
calculation naturally produces a morphological mix of realistic disk
galaxies coexisting with a population of ellipticals resolves this
long-standing issue. It also shows that previous futile attempts to
achieve this were not due to an inherent flaw of the LCDM paradigm,
but rather due to limitations of numerical algorithms and physical
modelling.'' This appears to contradict the inability to form MW-type
disk galaxies within the SMoC by the most recent work of another team
\cite{Roskar14}. This friction between the two SMoC models is not
explained.

Another criticism that may be stated along the lines of
Sec.~\ref{sec:conflicts} is that in \cite{Vogelsberger14b} the
Illustris model galaxies are close to the BTFR, while a previous
recent study \cite{Lu12} cannot reproduce the BTFR within the
SMoC. Vogelsberger et al. \cite{Vogelsberger14b} do not discuss this
discrepancy nor is \cite{Lu12} cited by them.  A discussion as to why
the Illustris models are closer to the BTFR than the Lu et al. models
\cite{Lu12} would have been appropriate in order to clarify why both
SMoC models disagree with each other.  More thorough and independent
comparisons between observed galaxies and the simulated ones await the
publication of the simulated-galaxy catalogue. A perhaps major issue
with the Illustris modelling is that it employs ``a kinetic stellar
feedback scheme, where the stellar wind velocity is scaled with the
local halo environment ($3.7\,\sigma^{\rm 1D}_{\rm DM}$ , where
$\sigma^{\rm 1D}_{\rm DM}$ is the local one-dimensional DM velocity
dispersion'', \cite{Vogelsberger14}, p.~3 in
\cite{Vogelsberger14b}). In effect, the Illustris models assume local
baryonic processes to couple closely to the entire DM halo. But the
exotic DM particles do not interact electromagnetically with baryonic
matter by construction of the SMoC. Neither the SMoPP includes such a
coupling nor is there experimental evidence for such a coupling.  This
may thus be considered as being an unphysical assumption, these models
being internally inconsistent at a fundamental level. Nevertheless,
such models are valuable for gaining a deeper understanding of
hypotheses that may be required to establish the galaxy population.

Collecting the above important work on the formation of elliptical
galaxies and disk galaxies the following statement can be made with
confidence: 

\vspace{2mm} \centerline{ \fbox{\parbox{\columnwidth}{ {\sc The SMoC
        and the Galaxy Population}:
      The SMoC is untenable as a framework for galaxy formation:\\
      \underbar{Elliptical galaxies}: must form quickly early on and
      require dry mergers to grow further and to be consistent with
      the merging history of the SMoC. This postulates a large
      population of dry galaxies which contradicts the observation
      that 6~Gyr ago only about 4~\% of all galaxies more
      massive than $1.5\times 10^{10}\,M_\odot$ are
      ellipticals with this fraction not evolving since. \\
      \underbar{Disk galaxies}: must form in dark matter halos with a
      quiet merging history. Even then the model disks have rotation
      curves not in agreement with observed ones. The required quiet
      merging history is appropriate for a minority of all host halos
      but this contradicts the observation that from~6~Gyr ago the
      very major majority (about 97~\%) of all galaxies more
      massive than about $1.5\times 10^{10}\,M_\odot$ in baryons are
      of late type.  }}}  \vspace{2mm}

\subsection{Conclusion on testing for the existence of dark matter}

\subsubsection{Test1 and Test2}

Given the failed tests~I and~II of the dual dwarf galaxy theorem on observational data
within the framework of the SMoC, and the unfavorable test of the SMoC using
dynamical friction, the following can be concluded:

\vspace{2mm} \centerline{ \fbox{\parbox{\columnwidth}{ {\sc Result}:
      According to both tests~I and~II above: type~A dwarfs $=$ type~B
      dwarfs. This falsifies the dual dwarf galaxy theorem if the SMoC
      were true. Thus the SMoC, and more precisely, the existence of
      cold or warm dark matter particles, appears to be ruled out. The
      lack of evidence for dynamical friction also implies the cold or
      warm dark matter framework, and thus the SMoC, to not be an
      adequate description of galaxies.  }}} \vspace{2mm}

\subsubsection{Implication}

By inference the observational evidence suggests the following implication: 

\vspace{2mm} \centerline{ \fbox{\parbox{\columnwidth}{ {\sc
        Implication}: The ruling-out of the existence of dynamically
      relevant cold or warm dark matter particles implies the dynamics
      of galaxies need to be accounted for by an effective
      non-Newtonian theory.  That the hitherto observed late-type
      (i.e. star-forming) TDGs lie very near to the BTFR defined by
      the PDGs suggests that the dynamics should be given by a
      fundamental principle not related to exotic dark matter
      (Sec.~\ref{sec:sid}).  }}} \vspace{2mm}

\section{Robustness of the falsification}
\label{sec:robust}

The observational evidence thus points towards dynamical friction not
playing a dominant role in galaxy evolution and galaxy--galaxy
interactions. Put differently, if anything, the data are easier to
understand with dynamical friction being suppressed. There is no
convincing observational evidence for dynamical friction at work.
This test is not as clear cut though as the tests based on the dual
dwarf galaxy theorem, since for example the modelling of the M81 group
appears to have stopped once two independent groups reported in
conference proceedings that dark-matter halo solutions have been found
to fail to account for the observed distribution of matter and the
line-of-sight kinematical information.  Similarly, the Sgr satellite
galaxy problem requires much more attention with Sgr being embedded in
its own dark matter halo, because most of the effort has shied away
from this critical approach, although in easy reach of numerical work.

The two-fold test using the dual dwarf galaxy theorem is, however,
robust and significant, but it is subject to the observational
evidence remaining valid.  The observed Universe does not provide any
evidence for two types of dwarf galaxies to exist which can be
differentiated in terms of their dark matter halos. The observed TDGs
appear to have just as much dark matter as PDGs.

\vspace{2mm} \centerline{ \fbox{\parbox{\columnwidth}{ {\sc Overall
        conclusion}: Combining the tests on dynamical friction and on
      the dual dwarf galaxy theorem suggests very strongly that dark
      matter halos are not present. {\it Dynamically significant
        exotic dark matter particles do not appear to exist.}  }}}
\vspace{2mm}

Clearly, while some might be convinced, most astronomers and
physicists will not be. Reluctance to accept this conclusion is
understandable and comes twofold: a well-founded inertia to depart
from GR and a sociological inertia given by a well-established
professional landscape.  \footnote{\label{footn:DMexists} A part of
  the research community appears to have fallen into the trap ``We
  {\it know} there is dark matter therefore {\it it is not possible}
  that there is none''.  As an example of the certitude in parts of
  the community, \cite{Lang14} exemplifies this certainty in the
  community that dark matter exists by writing ``Due to a large number
  of astrophysical observations ... we know today that dark matter
  exists'' (originally: ``Aufgrund einer Vielzahl von
  astrophysikalischen Beobachtungen ... wissen wir heute, dass Dunkle
  Materie existiert'' ) and ``The question is thus not: does dark
  matter exist?  Rather, the issue is to find out: what does it
  consist of?''  (originally: ``Die Frage ist also l\"angst nicht
  mehr: Existiert die Dunkle Materie?  Vielmehr gilt es
  herauszufinden: Woraus besteht sie?'').  As another example, a
  cosmologist who is a life-long director of a North-German research
  institution walked out in protest after about one third of the
  invited colloquium at the director's institute which was being given
  on this topic in February 2014. Although no-one from the packed
  audience followed, and the seat was quickly refilled, this incidence
  does indicate an interesting attempt at exerting peer pressure,
  especially since it may be employed thereafter tactically when
  referring to an ``impossible speaker''. It has also occurred that
  authors of submitted manuscripts were required, on occasion, by the
  referee and editor of major journals to drop references to
  peer-reviewed papers which are SMoC-critical, such that
  e.g. students reading such manuscripts would not be exposed to such
  critical studies.  }

The ``no-dark-matter'' deduction has major implications, both for
fundamental physics and sociologically. The implications for
fundamental physics are highly exciting, since we are, in essence,
facing a possible departure from Einstein's GR. At the least we may
obtain enhanced knowledge of space-time-matter-vacuum physics (see
e.g. \cite{Verlinde11}, Sec.~\ref{sec:MD}
and~\ref{sec:conserv}). Sociologically and politically the implication
is that searches for the exotic dark matter particles would be
expected to be fruitless. But many resources have been invested in
this avenue of research.

Given the gravity of the ``no-dark-matter'' deduction, it is necessary
to make sure that it is held-up by independent
evidence. Self-consistency checks are thus required.

\vspace{2mm} \centerline{ \fbox{\parbox{\columnwidth}{ {\sc
        Self-consistency}: If the above ``no-dark-matter'' deduction
      were to be true, then the SMoC should not be a successful
      description, or in the least, it should not be a unique
      description of the observational data. That is, how well does
      the SMoC truly account for the observed extragalactic
      structures?  }}}  \vspace{2mm}

\noindent While this question has already been answered in
Sec.~\ref{sec:testdynfr} (the SMoC has not successfully accounted for
the galaxy population and its spatial distribution), in the following
a few additional considerations are discussed in order to further
illuminate the ``no-dark-matter'' deduction.

\subsection{Which satellite galaxies are TDGs without dark matter, and
  which might be PDGs with dark matter?} 
\label{sec:selfcons}

That the type~A dwarfs and the type~B dwarfs are, within the
observational uncertainties, apparently indistinguishable in terms of
their dynamical and morphological properties, suggests the SMoC to be
false. Given the current observational situation, we thus have

\begin{equation}
{\rm type~A~dwarf} = {\rm type~B~dwarf}.
\label{eq:equality}
\end{equation}

\noindent This may be due to only type~A (i.e.  dark-matter-dominated
PDGs) existing, with type~B dwarfs (i.e. TDGs) disappearing rapidly
after their formation. However, the bulk of the evidence does not
favor this possibility, because TDGs are observed and it has already
been well established, both observationally and by high-resolution
numerical work, that TDGs survive for many~Gyr
(Sec.~\ref{sec:tdgs}). Furthermore, this would not alleviate the
problem with dynamical friction discovered above.

An alternative is to assume {\it all} dwarf galaxies are TDGs. All
dwarf galaxies cannot, however, be TDGs because isolated late-type
dwarf galaxies and dwarf-galaxy groups \cite{Tully06}, as discussed
e.g. in \cite{Metz09}, cannot have formed as TDGs. 

But it may be posible that the majority and perhaps all {\it
  satellite} dwarf galaxies are TDGs. In this case most satellite
systems ought to show anisotropies such as flattened rotational
populations of dwarf galaxies around their host galaxy because TDGs
from in tidal tails. This will be true if interactions between major
galaxies are typically less frequent than twice per Hubble time,
because one encounter is required to produce the population of TDGs
while a second encounter would disperse the satellit system reducing
its anisotropy (see Sec.~\ref{sec:predictions},
prediction~\ref{sec:pred_frequency}).

An important consistency check on the conclusions reached in
Sec.~\ref{sec:test} is therefore given by the spatial distribution of
dwarf satellite galaxies around major host galaxies, as discussed in
the following.

\subsubsection{Anisotropy of satellite distributions and the origin of
galactic satellites}
\label{sec:anisotropy}

Observational evidence for a population of satellite galaxies in a
correlated distribution would imply these to be most likely TDGs,
because no other mechanism is known to generate significantly
phase-space-correlated distributions of satellite galaxies.  From
Sec.~\ref{sec:tdgs} it follows that populations of TDGs around one
host galaxy are likely to be in phase-space correlated structures, as
long as the host did not experience more than about one encounter per
Hubble time, assuming one major encounter can randomise a previously
formed disk of satellite TDGs.  A small number of encounters per
galaxy is expected to be the case because the vast majority of host
galaxies are late-type rotationally supported disks which preclude
them to have had many encounters (Sec.~\ref{sec:mergers}). Given
eq.~\ref{eq:equality}, we would have a significant consistency
argument supporting the ``no-dark-matter'' deduction if such
anisotropically distributed satellite galaxies were to be found. 

This argument would be very strong if it were to be found that some
satellite population is so anisotropic and correlated in phase space
(e.g. by a large number of satellite dwarf galaxies orbiting their
host galaxy in one and the same sense in a relatively thin disk-like
structure) that the hypothesis that they might be independently
accreted dark-matter-dominated (type~A) dwarfs can be excluded with
very high confidence. This is indeed the case:\\

\subsubsubsection{\it The case with the best data: the Milky Way:}
\label{sec:MW}

Considering the positions of the then-known (bright) satellite
galaxies and of globular clusters of the MW in the 1970s, Lynden-Bell
\cite{LyndenBell76,LyndenBell82} and Kunkel \& Demers \cite{Kunkel76}
realized their arrangement in a Galactic near-polar band on the
sky. The natural interpretation of this structure was it to have been
born from the tear-up of a larger Magellanic Cloud.  However, with the
advent of the SMoC in the 1990s and its prediction that a MW-sized
galaxy should have a large number of dark-matter dominated satellite
galaxies ({\it the satellite over-prediction problem}, \cite{Klypin99,
  Moore99}) and the observation that the real dSph satellite galaxies
have very large dynamical $M/L$ ratios \cite{SG07} this interpretation
was discarded and attention of the research community was drawn to
dealing with the satellite over-prediction problem.  The significant
discrepancy of the SMoC-prediction of a spheroidal dark-matter
dominated satellite distribution with the observed disk-of-satellite
(DoS) distribution, taking into account all then-known satellite
galaxies, was documented for the first time by Kroupa et al. (2005,
\cite{KTB05}). That the high dynamical $M/L$ ratios may be feigned by
the satellites being non-virial-equilibrium and largely unbound
remnants of satellite galaxies gave support to an interpretation of
the dwarfs as ancient TDGs despite their large dynamical $M/L$ ratios
\cite{Kroupa97, KK98, MK07, Casas12, Yang14}

The recent work of Pawlowski et al.  \cite{Pawlowski12a, Pawlowski14,
  Pawlowski14b} has shown that this vast polar structure (VPOS) around
the MW is known today to host virtually all known satellite galaxies
(27 in number, Fig.~\ref{fig:VPOS}, this includes Crater or
PSO~J174.0675-10.8774)
and all the young halo globular clusters (YHGCs) as well as a
significant fraction of all known gas and stellar streams. The highly
significant existence of the very large ($\approx 500$~kpc in
diameter) planar or disk-like (thickness $\approx50$~kpc) distribution
of material around the MW \cite{MKJ07, Pawlowski12a}, which we know to
be mostly rotating in one sense \cite{MKL08, Pawlowski13}, can most
naturally, and probably only, be explained as the MW satellite
galaxies being old TDGs \cite{MK07, Pawlowski11, Hammer13}. Attempts
at accounting for this rotationally supported vast structure around
the MW in terms of the infall of dark-matter, i.e. primordial or
type~A dwarf satellite galaxies, have been excluded with extremely
high confidence \cite{Metz09, Pawlowski12b, Pawlowski14}.  The infall
of a group of or filamentary accretion of dark-matter dominated PDGs
are ruled out as well (Sec.~\ref{sec:filinfall}). \\

\subsubsubsection{\it The Andromeda system (GPoA):}
\label{sec:GPoA}

Recently, Ibata et al. \cite{Ibata13} and Conn et al. \cite{Conn13}
verified and quantified the existence of a great plane of Andromeda
(GPoA, sometimes also referred to as the vast thin disk of satellites,
VTDS) which contains about half of all satellite galaxies of
Andromeda. It consists of old gas free dSph satellite galaxies and is
very thin ($\approx 14$~kpc in thickness) and extended ($\simgreat
400$~kpc in diameter) and the satellites are moving about Andromeda in
bulk. That is, this GPoA structure is rotating (Fig.~\ref{fig:GPoA}).
Given attempts to argue that this structure arises ``naturally'' or
``commonly'' in SMoC simulations, \cite{Ibata14} have studied such
claims. They performed significance tests and they have calculated the
likelihood that such a GPoA structure can be obtained from dark-matter
sub-halos. They write ``We find that 0.04~\% of host galaxies display
satellite alignments that are at least as extreme as the
observations'' considering the ``extent, thickness and number of
members rotating in the same sense''.  The only viable explanation for
the GPoA structure appears to be a remnant of an ancient tidal arm
within which satellite dwarf galaxies formed as TDGs (see
e.g. \cite{Hammer13}).

The GPoA is significant evidence that type~A dwarfs cannot be
distinguished dynamically nor morphologically from type~B dwarfs
(i.e. that eq.~\ref{eq:equality} holds).  This is the case because the
dSph galaxies in the GPoA appear in every respect to be the same as
those apparently dark-matter dominated satellite galaxies that are not
in this GPoA.

That the MW VPOS and Andromeda GPoA may be physically related is
suggested by both systems having the same sense of rotation with
respect to the Local Group coordinate system (Fig.~\ref{fig:VPOS}
and~\ref{fig:GPoA}). This may be understandable to be a consequence of
both satellite systems being born as TDGs during an encounter between
both major galaxies about 11~Gyr ago \cite{Zhao13}. Since then the MW
and Andromeda would have evolved independently. There is evidence that
Andromeda had a more complex past by experiencing further minor
encounters with its own satellites. This evidence is discussed by
\cite{Yin09}, who also summarize our knowledge of the present-day
properties of both galaxies. Evidence for a rich history of encounters
of Andromeda is studied by Hammer et al. \cite{Hammer10} as well. An
innovative solution to the VPOS/GPoA correlation problem is suggested
by \cite{Fouquet12,Yang14}, where it is shown that a merger in
Andromeda about 6~Gyr ago reproducing tidal features around Andromeda
would have expelled a tidal tail within which TDGs formed. The
remarkable aspect of this model is that without additional parameter
adjustments, this tidal tail turns out to have passed the MW such that
the MW satellite galaxies which are today arranged in the DoS/VPOS may
be the TDGs that formed in this tidal tail.  While mergers appear to
be rare in reality, some may occur. It will be useful to investigate
this scenario in a dynamical model without exotic dark matter
particles (Sec.~\ref{sec:sid}).  The highly symmetrical structure of
the dwarf galaxy arrangement in the Local Group (Pawlowski, Kroupa \&
Jerjen 2013, \cite{Pawlowski13b}) may be suggestive of the Local Group
dwarfs having formed as TDGs. Further observational indications of a
TDG-origin of dwarf galaxies nearby the Local Group in terms of the
structure of the Perseus~I and NGC~3109 association has been
discovered \cite{PM14}.

As pointed out by \cite{Pawlowski13b}, there is evidence that
Andromeda has another disk of satellites which is co-planar within the
disk of Andromeda.  Such a DoS would not be observable around the MW
because of the obscuration of it through the
MW. \\

\subsubsubsection{\it The correlated satellite systems of NGC1097, NGC5557,
  NGC4631, NGC4216, NGC5291, M81  and the Tadpole:}

Other galaxies have been found to also have phase-space-correlated
satellite populations. Examples are the dSph satellite galaxies in the
``dog-leg'' stream of NGC1097 \cite{Galianni10}, the linear
arrangement of 2--3~Gyr old dSph satellites around the
post-interaction remnant NGC5557 \cite{Duc11}, the three satellites
in the tidal feature of NGC4631 \cite{Karachentsev14}, and the
arrangement of satellite galaxies around the spiral NGC4216
\cite{Paudel13}. The Tadpole (Fig.~\ref{fig:tadpole}) is another case
of a strongly correlated young population of star clusters,
star-cluster complexes and low-mass TDGs, as is the vast gas ring with
TDGs around NGC5291 \cite{Bournaud07}.  Last not least,
\cite{Chiboucas13} discuss the dwarf galaxy population in the
M81~group of galaxies, which is a sparse group comparable to the Local
Group, and they find evidence that the faint satellite galaxies are
distributed anisotropically. They write ``In review, in the few
instances around nearby major galaxies where we have information, in
every case there is evidence that gas poor companions lie in flattened
distributions.''

{\it This proves} that the two major galaxies of the Local Group are
no exception. The above discussion also suggests what the dominant
origin of satellite galaxies appears to be. All known satellite
galaxies of the MW are members of the VPOS such that at least the
majority of them would be ancient TDGs. The Andromeda satellite system
is richer than that of the MW. About halve of all the satellites of
Andromeda are in the GPoA that rotates in one sense. Thus, at least
about halve of the Andromeda satellites would be ancient TDGs. Since
these do not differ in any discernible way from the other satellites,
it appears reasonable to assume that most of Andromeda's satellite
galaxies are also TDGs.

\vspace{2mm} \centerline{ \fbox{\parbox{\columnwidth}{ {\sc Empirical
        evidence}: At least {\it nine} host galaxies have been
      documented to have satellites which are correlated in phase
      space. Thus, most dwarf satellite galaxies may be TDGs. }}}
\vspace{2mm}

\subsubsection{The mass--metallicity relation of TDGs}
\label{sec:m-Z}

Galaxies are known to follow a mass--metallicity relation which is
largely understood as a result of more massive galaxies having larger
star-formation rates, lesser galactic winds and increasingly top-heavy
galaxy-wide stellar initial mass functions (IMFs; \cite{Koeppen07,
  Recchi09, Spitoni10, Gargiulo14}, see also
Sec.~\ref{sec:buildingblocks}).  An important question is whether
TDGs, which form from pre-enriched gas, also follow such a
relation. This is particularly relevant for the satellite galaxies of
the MW and of Andromeda: if they are TDGs as is suggested by them
populating correlated disk-like distributions around their host
galaxies, then how can they follow a mass--metallicity relation of
PDGs? The essential argument here is that by ruling-out the existence
of exotic dark matter above, it transpires that galaxies all follow
the same gravitational rules (Sec.~\ref{sec:sid}). The evolution in
the mass--metallicity diagramme will thus be indistinguishable, apart
from possible physical processes such as ram-pressure stripping which
contributes to the truncation of star-formation activity in a low-mass
satellite dwarf galaxy, and the tracks will depend on an initial
pre-enrichment of the gas in the tidal arm from which the TDGs form.

\begin{figure}
\centering{}\includegraphics[width=6.5cm,angle=-90]{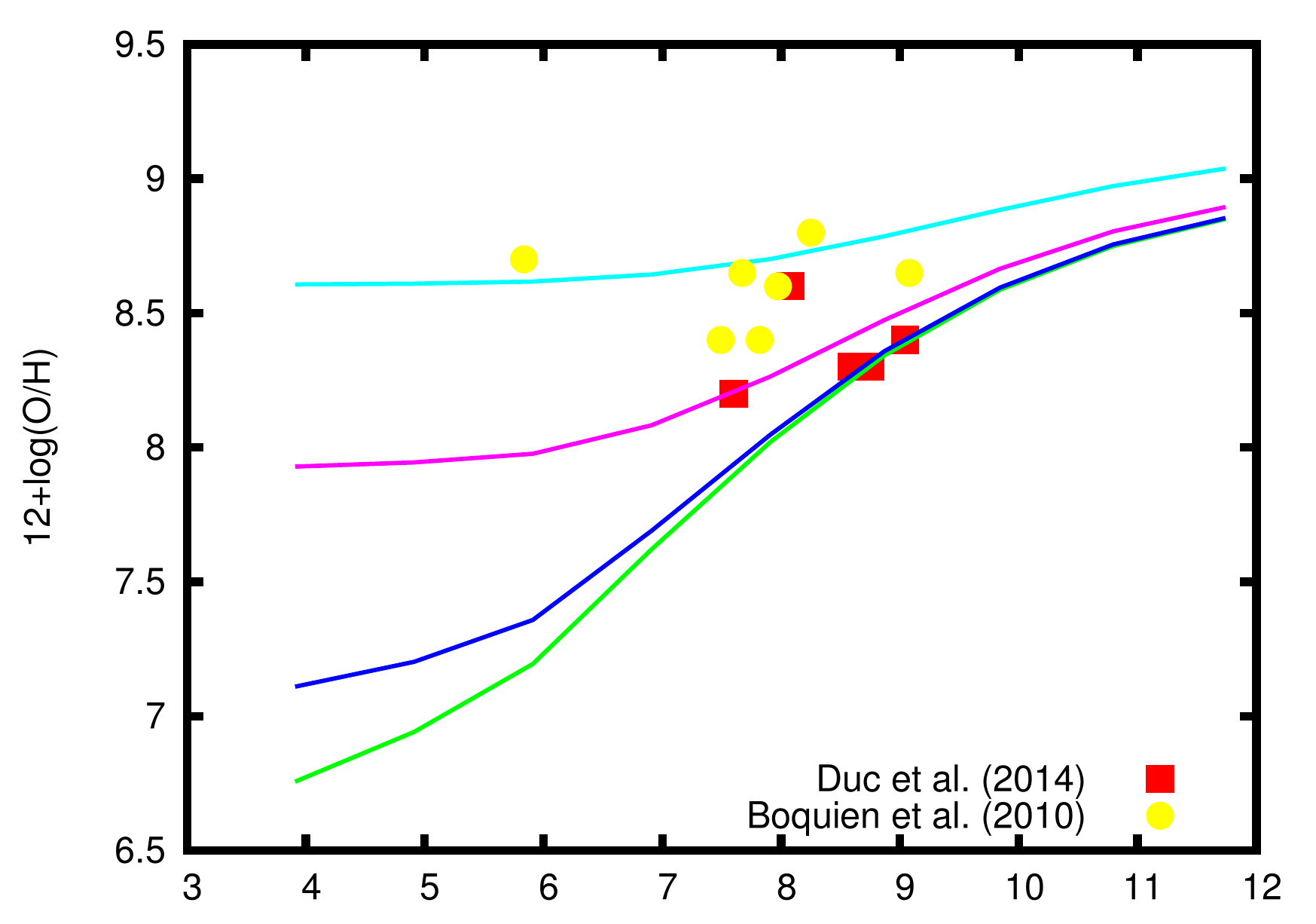}
\centering{}\includegraphics[width=6.5cm,angle=-90]{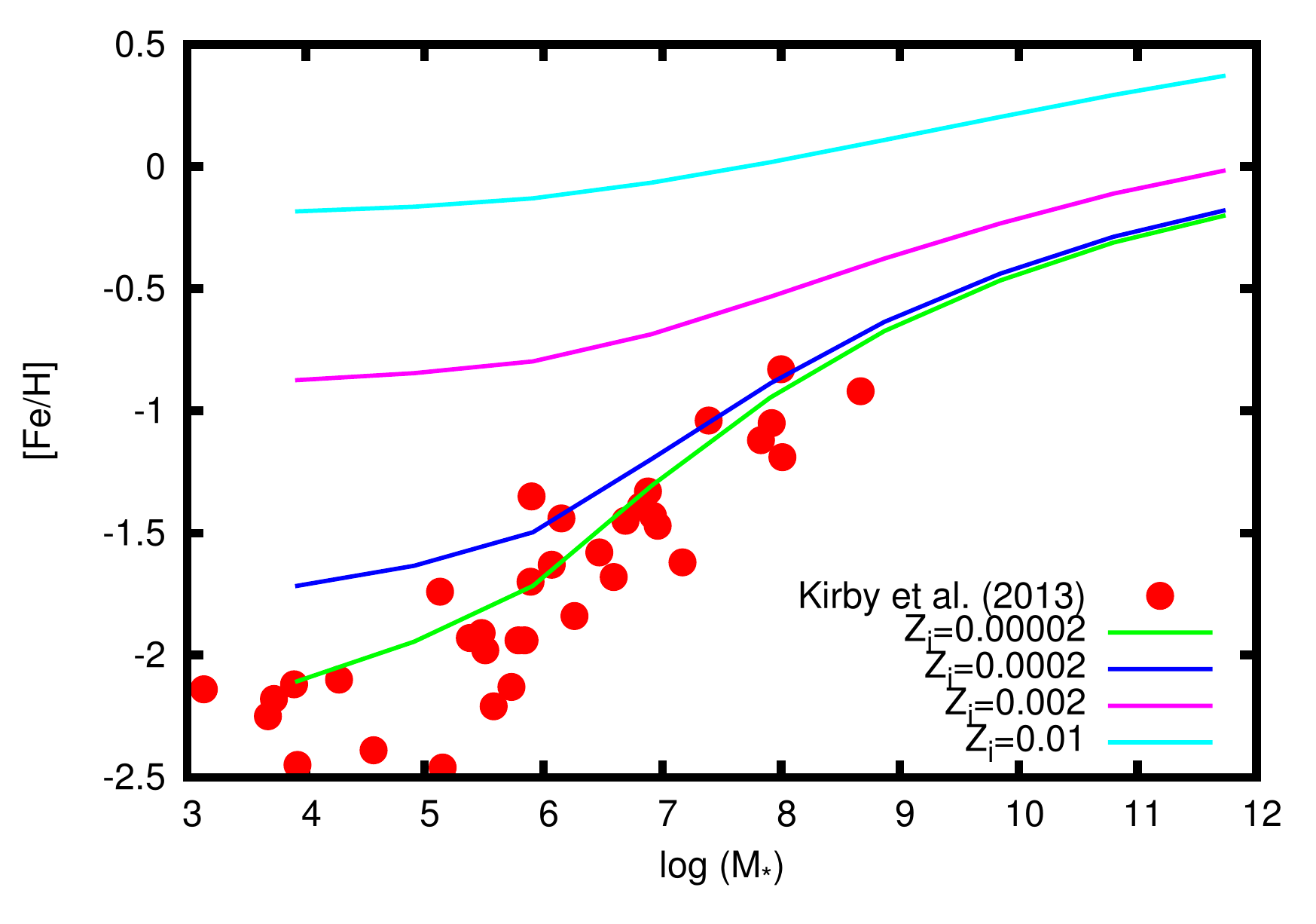}
\caption{\small The mass--metallicity relation of TDGs. The curves are
  one-zone leaky-box chemical evolution models (compare with
  \cite{Recchi08}) starting the chemical evolution at an initial
  metallicity, $Z_i$, as given in the key.  The galaxy-wide stellar
  IMF is assumed to be the star-formation-rate-dependent IGIMF
  (Sec.~\ref{sec:buildingblocks}).  The outflow rate is chosen to be
  larger in low-mass galaxies, smaller in more-massive ones, in
  compliance with the results obtained by \cite{Spitoni10}.  The upper
  panel shows the oxygen abundance in the gas phase, as compared with
  observations of TDGs (from \cite{Boquien10,Duc14}).  The lower panel
  shows the stellar [Fe/H] abundances predicted by the same models in
  comparison with observations of MW and M31 satellites summarized in
  \cite{Kirby13}.  The lower panel shows thus that the MW and
  Andromeda satellite galaxies are consistent with the laws of star
  formation (Sec.~\ref{sec:buildingblocks}), if the pre-enrichment
  $Z_i$ is very low.  The same models able to fit the satellites of
  the MW and M31, are able to fit the oxygen content of younger TDGs
  (upper panel), with different degrees of pre-enrichment (from very
  low to moderately-high).  }
\label{fig:mZ}
\end{figure}

Fig.~\ref{fig:mZ} shows the theoretical mass--metallicity relations of
TDGs in comparison to observed TDGs (upper panel) and to the dSph
satellite galaxies of the MW and Andromeda (lower panel).
The models are computed with the IGIMF theory (the galaxy-wide
star-formation-rate(SFR)-dependent IMF, Theory~C in
Sec.~\ref{sec:sftheories}).  The models are modifications of standard
leaky-box models \cite{Recchi08} in which the hypothesis of an
invariant IMF is relaxed \cite{Recchi15}.  The top panel shows gas abundances in
comparison to observed TDGs. The lower panel shows stellar abundances
derived from the same models in comparison with MW satellites.  The
different curves have different initial metallicities corresponding to
initial pre-enrichment. Thus, the MW and Andromeda dSph satellites
formed as TDGs about 11Gyr ago when the gas in the outer region of the
pre-encounter galaxies was not much pre-enriched, while the
present-day TDGs (upper panel) formed from pre-enriched material, as
is to be expected.

{\it Conclusion}: Using standard chemical evolution models (with the
IGIMF) dSph satellites and TDGs naturally fit the mass-metallicity
relation when one takes into account that they form at different
epochs from pre-enriched gas.

\vspace{2mm} \centerline{ \fbox{\parbox{\columnwidth}{
The observed mass--metallicity relation of dwarf satellite galaxies is
thus consistent with them being TDGs.
  }}} \vspace{2mm}

\subsubsection{SMoC--internal conflicts and inconsistencies}
\label{sec:conflicts}

It is important to assess whether dwarf galaxy satellite structures
similar to the observed VPOS and GPoA can result within the SMoC in
terms of the distribution of dark-matter sub-halos, because the dSph
and UFD satellite galaxies within the VPOS and GPoA appear to be the
most dark-matter dominated galaxies known to exist \cite{SG07}. If
dark-matter-made VPOS or GPoA-like structures cannot arise in the SMoC
then an inconsistency between theory and data arises and the SMoC
would need to be discarded in agreement with the conclusion reached in
Sec.~\ref{sec:test}.  The inference on the dark-matter content of the
satellite galaxies is valid if Einsteinian/Newtonian dynamics is
applied in analyzing the structural and kinematical data of the stars
orbiting within the satellite galaxies.  If the VPOS and GPoA
structures cannot arise in the SMoC, then the observed high dynamical
$M/L$ ratios of the MW and Andromeda satellite galaxies would hint at
effective non-Einsteinian/non-Newtonian dynamics (see also
Sec.~\ref{sec:sid}).

How mutually consistent are the various attempts at accounting for the
VPOS and GPoA structures, that is, are inconsistencies and perhaps
even mutually logically exclusive results evident in applying the SMoC
to the VPOS and the GPoA? Incompatible results, if they exist, point
towards the underlying theory being wrong.

Consulting the available published research it is indeed evident that
consistency issues do arise. In attempting to reproduce VPOS- and
GPoA-like structures with dark-matter sub-halos, various teams may
focus on one or a few aspects thereby possibly neglecting the wider
context or even similar attempts by other teams who may have come up
with solutions that are inconsistent with the new proposal:

\vspace{2mm} \centerline{ \fbox{\parbox{\columnwidth}{How should the
      community judge the scientific situation if a team~T1 proposes a
      SMoC solution~A to some phenomenon, and if team~T2 proposes a
      SMoC solution~B to the same phenomenon, but the solutions~A
      and~B contradict each other, which, however, is neither discussed
      by~T1 nor by~T2?  }}} \vspace{2mm}

Two important contributions illustrate this situation: \cite{Deason11}
find that in SMoC simulations the recent infall (at redshift $z<1$) of
a group of type~A dwarfs may account for an anisotropically
distributed population of satellite galaxies at the present
($z=0$). This confirms \cite{Klimentowski10} who write ``Although the
satellites often form groups, they are loosely bound within them and
do not interact with each other. The infall of a large group could
however explain the observed peculiar distribution of the LG
satellites, but only if it occurred recently.''  But, \cite{NB11}
obtain the result that in order to explain the observed lack of gas in
the MW satellites, they must have fallen in early at $z=3-10$. Both
are mutually exclusive but claim consistency with the data and the
model. This discrepancy lies in \cite{Deason11} not accounting for the
MW satellites being void of gas, while \cite{NB11} do not account for
the VPOS. But the MW satellites are both, gas poor and in the VPOS:

\vspace{2mm} \centerline{ \fbox{\parbox{\columnwidth}{ {\sc Mutually
        inconsistent results}: For the VPOS to be accounted for with
      PDGs the MW satellites must have fallen in at a high redshift
      and at the same time at low redshift. Given the observational
      data, this logically incompatible result invalidates the
      underlying hypothesis (that the SMoC is valid), unless one of
      the studies needs to be revised.  }}} \vspace{2mm}

The Durham team around Carlos Frenk has been addressing the VPOS
challenge for the SMoC with important results.  Of~400 MW-scale dark
matter halos that form in their cosmological simulations, the
occurrence of the DoS/\-VPOS around the MW becomes negligibly likely
if more than 8 satellite galaxies have orbital angular momenta aligned
to within~45~deg (fig.~9 in Libeskind et al. 2009,
\cite{Libeskind09}). \cite{Deason11} confirm this result.  As shown by
\cite{Pawlowski13}, taking into account the additional cuts that are
required to end up with the MW-like satellite system in the SMoC
simulation of the Durham team, the number of MW satellites with
orbital angular momenta as closely clustered as is observed becomes
negligible with a likelihood less than about 0.5~\% (see also a
detailed discussion of this in \cite{Kroupa10}).  Thus, according to
the Durham team neither the VPOS nor the GPoA can be accounted for
with primordial, dark-matter dominated dwarf galaxies. Since the
publication of the Durham analysis, further SMoC teams working on this
issue have not addressed this conclusion, while postulating their own
solutions (e.g. \cite{Wang13, BB14, GB13}).  Libeskind (2011,
\cite{Libeskind11}) write ``{\it While the planarity of MW satellites is no
longer deemed a threat to the standard model, its origin has evaded a
definitive understanding.}'', and study anisotropic infall of PDGs into
the MW halo to better understand its satellite distribution but do not
apply their results for a direct statistical assessment of whether the
VPOS may be explainable this way.

For example, in attempting to explain the GPoA around Andromeda in
terms of PDGs with dark matter, \cite{BB14} write in their abstract
that ``we find that planes with an rms lower than the VTPD are common
in Millennium II'' but in the body of the paper the probability of
this occurring is stated to be 2~\%, which is not common. As pointed
out by \cite{Ibata14} the analysis by \cite{BB14} is flawed by it not
taking into account the appropriate criteria\footnote{Of the three
  criteria that define a disk-like distribution of satellite galaxies,
  \cite{BB14} have adopted two at a time yielding a probability of
  occurrence of such a structure within the SMoC simulation data of
  2~\% which \cite{BB14} describe as being common. \cite{Ibata14}
  confirm the \cite{BB14} likelihoods and show that the three criteria
  that are in truth required to define GPoA-like structures lead to a
  combined likelihood of occurrence of GPoA structures of 0.04~\%,
  consistent with their own independent calculation based on the
  cosmological simulation data and with those of \cite{Kroupa10}. That
  is, obtaining VPOS or GPoA-like structures does not seem to be
  possible in the SMoC if these are to consist of PDGs.
  \cite{Ibata14} give the following example which visualises why using
  the correct selection criteria is essential for inferring the
  workings of nature (cited from their footnote~8): ``To make the
  issue perfectly clear, consider measuring the incidence of animals
  that have stripes and paws and are nocturnal. Clearly, selecting
  only two of these three properties will yield a larger (and
  incorrect) sample of such animals, giving a falsely optimistic
  measurement of how common they are.''}, and the true likelihood of
the occurrence of systems comparable to the GPoA is negligible in the
SMoC (0.04~\%). The proposition by \cite{GB13} that GPoA- and
VPOS-like structures can be understood {\it naturally} within the SMoC
as arising from a few thin filaments connecting to the host galaxy
along which the $\simgreat 20$ dark-matter dominated satellites
accrete has, at the time of writing of this text, not been accepted
for publication. But the manuscript is already being cited in the
community as having shown that GPoA- and VPOS-like structures do arise
naturally in the SMoC (e.g. \cite{ST13, BB14}). Pawlowski et
al. (2014, \cite{Pawlowski14}) analyse and rule-out the proposition by
\cite{GB13} in detail.

Metz et al. \cite{MKJ07} have developed statistics tools from the mathematical
literature to assess the degree of clustering of data distributed on
the sphere in terms of strength and shape parameters. These tools have
been tested carefully and proven to be ideal for robustly determining
the significance of satellite anisotropies given their
three-dimensional spatial information \cite{MKJ07, Pawlowski12a}:

\vspace{2mm} \centerline{ \fbox{\parbox{\columnwidth}{
The scientific community is highly encouraged and invited to compare
their own criteria for quantifying the statistical likelihood of
satellite anisotropy in cosmological models to the robust method of
\cite{MKJ07} (see also the above discussion and \cite{Ibata14,
  Pawlowski14}).
 }}} \vspace{2mm}

Some recent work on the distribution of satellite galaxies continues
to only concentrate on their radial distribution.  For example,
\cite{Watson14} state ``The accurate prediction for the spatial
distribution of satellites is intriguing given the fact that we do not
explicitly model satellite-specific processes after infall, ...''
but do not mention the significant anisotropy of the satellite galaxy
distribution now established for the MW and Andromeda. 

At the same time, \cite{Wang14} stress that ``current semi-analytic
simulations, where satellite profiles always parallel those of the
dark matter''. This recent contribution does not address the question
posed originally by \cite{KTB05} why the satellites of the MW and
Andromeda are in thin planes while the host dark matter halos are
spheroidally distributed. Furthermore, that the model satellite
galaxies violate the observed lack of a correlation between the
satellite masses and their luminosities has not found an explanation
in the SMoC framework, where such a correlation needs to be present
\cite{Kroupa10}.

It has been appreciated since a long time \cite{LyndenBell76,
  LyndenBell82, Kunkel76} that the Magellanic Stream and the orbits of
the Large and Small Magellanic Clouds (LMC and SMC, respectively)
define a plane within which also lie globular clusters and dwarf
spheroidal satellite galaxies. Today this is known as the DoS or VPOS
(Sec.~\ref{sec:anisotropy}).  The most recent work on the orbital
evolution of the Large and Small Magellanic Clouds by Besla et
al. (2012, \cite{Besla12}) studies, for the very first time, the two
galaxies self-consistently within the SMoC. Therewith this is a
landmark study which has advanced knowledge of the Magellanic system
appreciably. The LMC and SMC have, respectively, DM halo masses of
about $2\times10^{11}\,M_\odot$ and $2\times10^{10}\,M_\odot$ with
radii of $R_{200}=117\,$kpc and $R_{200}=57\,$kpc where the enclosed
average density is 200~times the critical density of the universe. For
the MW the authors use a DM halo mass of $1.5\times10^{12}\,M_\odot$
with $R_{200}=220\,$kpc. This halo is treated as a rigid potential,
while the LMC and SMC halos are live. Thus, dynamical friction occurs
between the LMC and the SMC but is neglected for both as they orbit
through the MW DM halo. The results are useful in interpreting whether
the two galaxies are on a first-time passage around the MW, and the
authors find that an encounter between the LMC and the SMC generated
tidal tails which are observed as the Magellanic Stream today. The
authors stress that they are not intending to reproduce the exact
properties of the stream, but the existing models do account for
overall properties of it.  Future work will have to be done to assess
how neglecting dynamical friction of the LMC and the SMC on the MW DM
halo affects the conclusions, and if solutions may be found at
all. Also, a discussion is lacking entirely as to the coincidence of
the plane of the first-time incoming orbit of the LMC and SMC being
oriented within the VPOS such that the resulting Magellanic Stream
also lies within the VPOS although it is created in a pre-infall
encounter between the LMC and the SMC. Thus, the fine-tuning between
the VPOS and the LMC--SMC binary orbit with semi-major axis near
100~kpc, this orbit and the hyperbolic orbit of the LMC/SMC barycentre
around the MW, and the encounter geometry between the LMC and SMC such
that the Magellanic Stream is produced which lies within the VPOS
although it is generated as tidal tails well before the LMC/SMC binary
has crossed the MW $R_{200}$ radius, remains to be discussed.  This is
thus a situation reminiscent of the M81 case of Sec.~\ref{sec:M81}.
As is evident, these models raise important questions and they are
therefore essential and valuable for gaining a deeper understanding of
processes that may be playing a role in establishing the properties of
the Local Group.

Leaving the problem of the VPOS and of the GPoA and other anisotropic
satellite systems, it is not uncommon indeed to perform simulations
within the framework of the SMoC of the motions of satellite galaxies
by ignoring dynamical friction. Thus, \cite{PB13} study the formation
of ultra-compact dwarf (UCD) galaxies from nuclei of stripped dE
galaxies. For computational ease the expansive dark matter halos of
the dE satellites are not treated, but dynamical friction of such
sub-halos on the massive dark matter halo of the host galaxy would
lead to the merging within one or two crossing times thus possibly
eliminating any viable solution to the problem altogether. For
example, a dE galaxy with a baryonic mass of $M=10^8\,M_\odot$ would
have a dark matter halo of $M_{\rm DM}\approx 10^{10.5}\,M_\odot$
while a dE with with $M=10^9\,M_\odot$ would have $M_{\rm DM}\approx
10^{11.2}\,M_\odot$ \cite{Ferrero12, Behroozi13}. By
footnote~\ref{footn:dynfr} a short dynamical friction time scale
obtains assuming the host galaxy to have a dark matter halo $M_{\rm
  DM}\approx10^{12}\,M_\odot$. Since dynamical friction is not taken into
account, it remains unknown if any viable solution to the problem of
UCD formation exists, but the published work states ``we consider
tidal stripping to be a likely origin of UCDs''.

\vspace{2mm} \centerline{ \fbox{\parbox{\columnwidth}{ 
{\it Concluding}, it is thus apparent that while important and useful
work has been done within the framework of the SMoC, typically such
work had to neglect important aspects or has not taken into account
the conclusions reached by others or has even been contradictory, time
and again. The results thus arrived at within this framework are not
convincing, although being important for aiding to assess the possible
validity of the SMoC.
 }}} \vspace{2mm}


\subsubsection{Cold filamentary infall}
\label{sec:filinfall}

In order to account for the gas accretion onto galaxies and their
star-formation histories, \cite{DB08, Dekel09} proposed that gas
accretes into the host dark matter halos in long-lived gaseous filaments that
are thinner than the virial radii of the host dark matter halos. This
fuels the formation of major galaxies. \cite{Gao14} demonstrate with
cosmological simulations that such very thin cold gaseous filaments
arise within a warm dark matter cosmological model and that they feed
MW-type galaxies mostly before a redshift $z=2$ after which the
streams subside. Such filaments reside in the much thicker dark matter
web. 

One attempt at reconciliating the existence of the VPOS and of the
GPoA with dark-matter (type~A) PDGs is to postulate that these
structures are a result of infall from cosmological filaments. The
incompatibility of  late versus early infall noted in
Sec.~\ref{sec:conflicts} persists but is not usually addressed. 

\cite{Pawlowski13} and \cite{Pawlowski13b} analyse in much detail the
claims that the VPOS of the MW and the GPoA of Andromeda may be a
result of filamentary infall and demonstrate that these claims can be
ruled out. Basically, the cosmological simulations show the dark
matter filaments or sheets to be too thick to account for the thin
VPOS (thickness about 50~kpc, diameter about 500~kpc) and the even
thinner GPoA (thickness about 14~kpc, diameter about 400~kpc).

\begin{figure}
\centering{}\includegraphics[width=8cm]{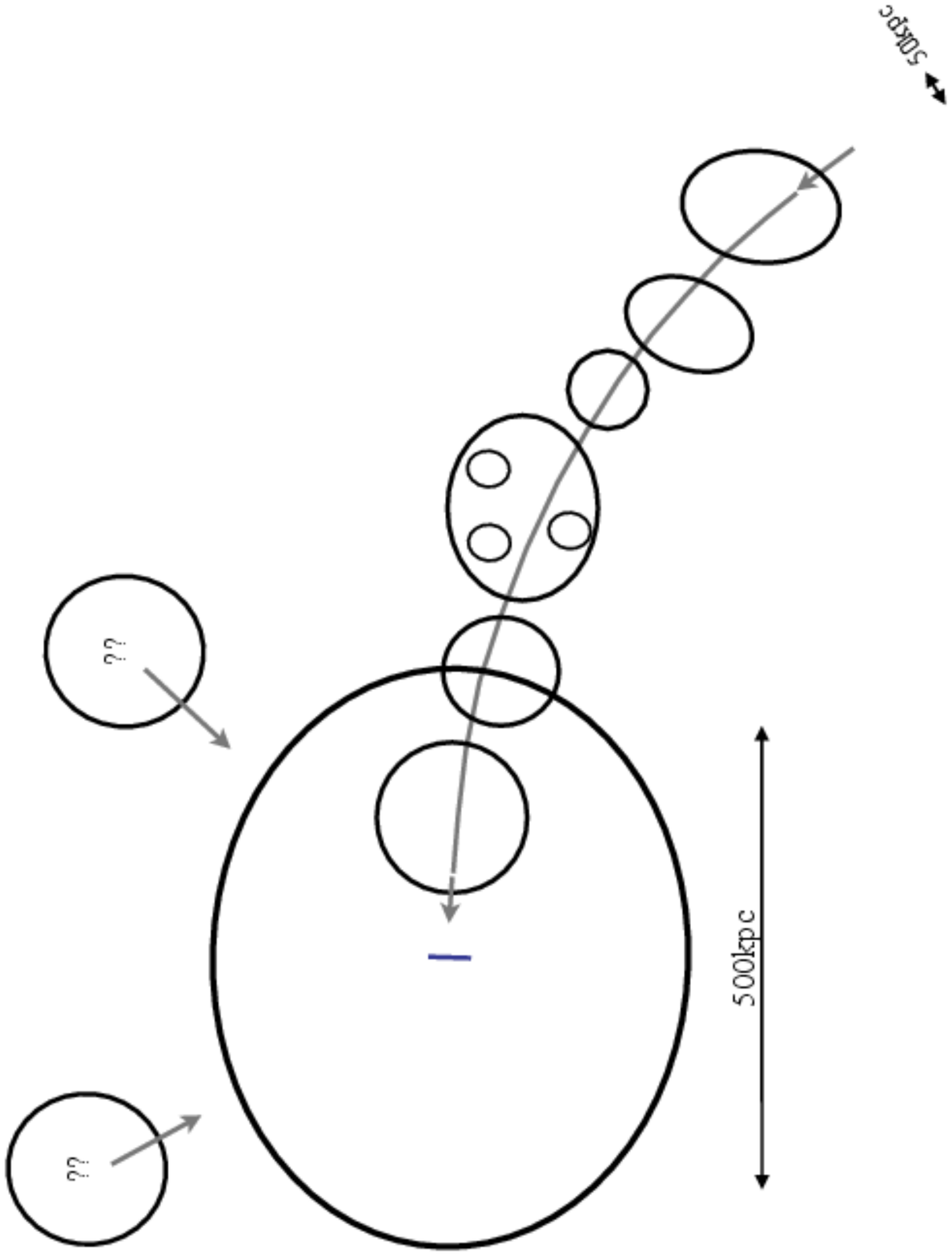}
\caption{\small Cartoon of how the infall of a sequence of about $20$
  or more PDGs would need to have occurred for the resulting
  distribution to become comparable to the VPOS (Fig.~\ref{fig:VPOS})
  or GPoA (Fig.~\ref{fig:GPoA}). The host galaxy is shown as the
  edge-on solid disk and its hosting dark matter halo is the
  near-circular black outline with a diameter of about 500~kpc. The
  grey line indicates a cold filament with thickness no larger than
  50~kpc along which a long sequence of dark-matter sub-halos need to
  fall in. The diameters of these are between 100 and 200~kpc but
  their centres need to follow the cold filament. More than one or two
  dark matter sub haloes cannot accrete from elsewhere onto the host
  halo, in order to preserve the VPOS of the MW. Note that some of the
  sub-halos may have their own sub-sub-halos. }
\label{fig:filinfall}
\end{figure}

Perhaps the last hope would be to assume that the VPOS and GPoA would
stem from accretion of dwarf galaxies along cold filaments
\cite{GB13}. This proposition has at least two problems: Firstly, it
would require the cold filaments of baryons to be populated by dark
matter sub-halos which would then fall into the hosting halo like
beads on a string with thickness less than 50~kpc
(Fig.~\ref{fig:filinfall}). Such beads of dark matter sub-halos, each
with a virial radius of about 100~kpc given that they would have to be
hosting pre-infall dwarf galaxies that are to become MW or Andromeda
dSph and UFD satellites (for their putative dark matter halo masses
see \cite{Ferrero12, Behroozi13} and for their radii see \cite{
  Ishiyama13}), have never been shown to arise in any self-consistent
structure formation computation, and it is difficult to see how such
highly-focussed infall of dark matter halos could arise.
The baryonic-matter stream cannot influence the orbital path of dark
matter sub-halos, given that the baryonic mass component is small
compared to the dark matter mass component. That is, it is difficult
to see a physical mechanism by which the cold gaseous filaments align
at least about~20 PDGs which later, after infall, become all of the
known satellite galaxies in the VPOS and GPoA. The second problem is
that the streams are radial, and infall would have to have occurred at
a redshift larger than $z=2$ \cite{NB11}. It is unclear how disks of
satellites are to arise at a redshift of $z=0$ since a given sub-halo
would fall-into a host halo and would launch on a Rosetta-type orbit
with changing orbital orientation in the typical triaxial host dark
matter halo. Each satellite would end up on a different orbital plane,
subject to dynamical friction.

\vspace{2mm} \centerline{ \fbox{\parbox{\columnwidth}{ {\sc Cold
        filaments}: Infall from cold filaments of dark-matter
      dominated dwarf galaxies can be ruled out as an explanation for
      the existence of the VPOS and GPoA.  }}} \vspace{2mm}

\subsection{Falsification and the theory confidence graph}
\label{sec:confidencegraph}

In principle a single robust observation which significantly
contradicts a predicted phenomenon suffices to falsify the underlying
theory. In practice it is important to understand if the prediction,
in the face of failure, may not be amendable by including previously
not accounted-for physical processes.

Thus, for example, the ``{\it satellite over-prediction problem}''
(i.e. the ``missing satellite problem'', \cite{Moore99, Klypin99})
has not been taken to rule out the SMoC. Instead, many research teams
attempted to reconcile the much larger number of predicted PDG
satellites in MW-type host halos with baryonic-physical processes,
such as suppression of the formation of stars due to re-ionisation, or
suppression of satellite galaxy formation through destructive feedback
processes. It has indeed become possible to account for the small
number of observed MW and Andromeda satellites in this way, but with
the caveat that many hundreds of even fainter satellite galaxies
distributed spheroidally around the MW or Andromeda should be found
once the surveys become more sensitive than the SDSS. Thus,
\cite{Koposov09} provide ``a convincing solution to the "missing
satellite" problem.'', but do not address the DoS/VPOS
problem. Similarly. \cite{Tollerud08} state ``that there are likely
between $\sim$300 and $\sim$600 satellites brighter than Boo II within
the Milky Way virial radius''.  This gives new predictions which can
be falsified with deeper more sensitive surveys.

However, if a theory has a history of failed predictions each of which
need individual attention in searching for possible solutions within
the theory, the scientific community has a right to loose confidence
in the theory, independently of the availability of a single
cataclysmic failure (such as the violation of the dual dwarf galaxy
theorem):

\vspace{2mm} \centerline{ \fbox{\parbox{\columnwidth}{ 
In fact, if a single cataclysmic failure has been shown to
exist, then this failure, if correct, implies the theory to not be
valid. In this case this theory should have a history of failures
since it does not describe reality. A history of failures would be
equivalent to a history of loss of confidence.
 }}} \vspace{2mm}

One possibility to assess the confidence in a theory is through the
{\it confidence theory graph} \cite{Kroupa12a}. To construct this
graph one begins with a confidence of one hundred~\% at the point in
time when the theory is mathematically formulated whereby parameter
values are taken to be part of the definition of the model. Each time
a prediction is found to be inconsistent with observational data a
step downwards is taken in the logarithm of the confidence that the
theory is correct. The step may, in the simplest application, be equal
in every reported failure, or it may depend on the gravity of the
failure. This would, however, require a quantification of the loss of
confidence which is difficult.

One example are the many tests of the SMoPP which have, to this date,
not found significant failures. The SMoPP thus remains near a
confidence of~100~\%. The fact that neutrinos have masses
requires an extension of the SMoPP and can be considered a step
downwards in the confidence graph and the question of the origin of
inertial mass has been clarified though the discovery of the predicted
Higgs boson.

In comparison, the SMoC has been found to fail massively in its
expansion behavior. This is remedied by introducing the novel but not
understood physical property of dark energy.  This ``failure'' is thus
considered to be the discovery of major new physics, in particular
since it is possible to fix the failure in the mathematical
description of the theory through introduction of Einstein's
constant. One could, in this case, allow one step upwards in the
confidence graph, but {\it only if} the new physics can be verified
independently (e.g., if the postulated existence of dark matter were
to be verified by a direct detection).  Nevertheless, if a theory
acquires a long history of failed predictions, each of which need the
addition of additional hypotheses without independent verification or
which cannot be convincingly remedied (e.g. the angular-momentum or
cusp/core problems have no remedies to date), the theory can be seen
as having lost in confidence.

In \cite{Kroupa12a} a list of at least 22~failures (including the
Bullet cluster) of SMoC predictions is provided and the {\it theory
  confidence graph} is introduced to graphically document the long
history of failed predictions (termed ``problems'' in the
literature). If each confidence-loss step would correspond to a loss
of confidence by 50~\%, the history of failures of the SMoC
would amount to a cumulative loss of confidence to a level of
$0.5^{22}=2.4\times 10^{-7}$. 

\vspace{2mm} \centerline{ \fbox{\parbox{\columnwidth}{ 
That is, by the year~2012 the
observational data have reduced the confidence that the SMoC
represents reality to a likelihood of $2.4\times 10^{-5}$ ~\%.
 }}} \vspace{2mm}

Significant failures in the list of \cite{Kroupa12a}, for which
solutions within the SMoC are not evident, are the high fraction of
galaxies with baryonic masses larger than $1.5\times 10^{10}\,M_\odot$
that are late-type star-forming galaxies ($\simgreat 90$~\%)
rather than early-types (\cite{Delgado10}, see also fig.~4.14 in
\cite{BM98}), and the dominant fraction of galaxies without classical
bulges (about 60~\%, \cite{Kormendy10}). \cite{Fernandez14} further
suggest that the very large fraction (94~\%) of disk galaxies
have pseudo bulges which appear to be largely ancient but may be
rejuvenated during galaxy interactions.

The core/cusp, the angular-momentum and the satellite over prediction
problems are still qualitatively conceived to be due to not understood
baryonic processes. Energetic feedback, which blows out gas for it to
cool and condense back to form a disk galaxy, and the expulsion of
which may form the cores of dark matter halos inferred from
observations while dark-matter only simulations yield cusps, have been
suggested as the remedy. \cite{GZ02} studied this problem using
simulations and analytical approaches and conclude ``We find that no
star formation effect can resolve the problems of CDM cusps.''  Apart
from the typically wrong application of a fully sampled stellar
initial mass function (IMF) which would contain unphysical fractions
of massive stars (see Sec.~\ref{sec:buildingblocks}), observations
have shown that gas-outflows that might damage the inner parts of
star-bursting dwarf galaxies are not evident, despite the injection of
$\approx10^{56}\,$erg into the inter-stellar medium (ISM) in the last
$\approx500\,$Myr \cite{Lelli14}. That even repeated outflows from the
central regions of simulated galaxies by supernova-driven winds and
black hole quasar feedback cannot produce dark-matter cores from cusps
is verified by the most recent and novel simulations of galaxy
formation using the AREOP5 algorithm by \cite{Marinacci14}.  Based on
observational and theoretical work it thus follows:

\vspace{2mm}

\centerline{ \fbox{\parbox{\columnwidth} {{\sc Feedback}: These
      problems do not have a solution in baryonic physics, contrary to
      the claims often made in the community even at present times.
    }}}

\vspace{2mm}

The observed {\it Karachentsev--Keenan under-density} of matter around
the Sun on the scale of 50--400~Mpc appears to be completely in
discord with the fundamental hypothesis that the universe is
homogeneous.  The highly significant under-density of baryonic and
dark matter by a factor of three to four within the local volume with
radius of 50~Mpc has been noted in~2012 by Karachentsev
\cite{Karachentsev12}. A highly significant under-density of luminous
and dark matter by a factor of nearly two within 300~Mpc is reported
in~2013 by Keenan et al. \cite{Keenan13} based on measurements of the
photometric K-band luminosity density as a function of redshift (see
also \cite{Keenan12}). Both, the Karachentsev and the Keenan findings
are in significant tension with the SMoC. Either we live in a hole, or
there is an unexplained issue with redshift measurements.  According
to augmentary hypothesis~1 (inflation, Sec.~\ref{sec:SMoC}) the
Universe cannot have fluctuations larger than about 10~\% in the
density of matter on scales $>100\,$Mpc (see also fig.~12 in
\cite{Keenan13} and \cite{Wiltshire09}).  The Karachentsev--Keenan
under density is depicted in Fig.~\ref{fig:underdens}.

An interesting point raised by Keenan et al. \cite{Keenan13} is that
the observed under density is just the right amount of local
inhomogeneity required to annulate the observational evidence for
dark energy by SNIa. How observed cosmological quantities are affected
in an at present inhomogeneous universe over the assumption of a
homogeneous SMoC is discussed in terms of timescape cosmology by
Wiltshire \cite{Wiltshire09}.

In {\it conclusion}, while individual solutions to the problems have
been suggested in the form of additional hypotheses, most are not
generally accepted. The falsification of the SMoC by the tests above
is consistent with the (poor) performance of the SMoC in accounting
for the observed Universe.

\section{Scale-Invariant Dynamics (SID)}
\label{sec:sid}

Stepping back and reconsidering the problem: galactic dynamics, galaxy
evolution and galaxy formation does not seem to be understood within
the context of dark matter halos made of exotic particles. With strong
if not un-refutable observational evidence against the existence of
exotic dark matter, the observed rotation curves of late-type galaxies
and the super-virial velocity dispersion in galaxy clusters need to be
explained. Note that the evidence by weak gravitational lensing for
dark matter in galaxy clusters stems from applying Einstein's GR to
calculate, from the weak lensing signal, the amount of matter within
the lens. But if this theory is challenged we cannot use it to argue
for the presence of dark matter, a trap the carefree researcher can
readily fall into.

\subsection{The dynamics of galaxies}
\label{sec:galdyn}

The rules underlying galactic dynamics turn out to be extraordinarily
simple, as noted by Milgrom at the Weizmann Institute in Rehovot,
Israel (\cite{Milgrom09}, see also \cite{Kroupa12b}):

Scale the space-time vector  by a number $\lambda$,
\begin{equation}
(\vec{r},t) \longrightarrow \lambda(\vec{r},t),
\label{eq:scale}
\end{equation}
where $\vec{r}=(x,y,z)$ are Cartesian coordinates and
$r^2=x^2+y^2+z^2$.  Under this transformation the kinematical
acceleration scales as
\begin{equation}
a \equiv {d^2r \over dt^2} \longrightarrow \lambda^{-1} a.
\label{eq:kinacc}
\end{equation}
The Newtonian gravitational acceleration, $g_{\rm N}=G\,M/r^2$, however,
scales as
\begin{equation}
g_{\rm N}  \longrightarrow \lambda^{-2} g_{\rm N}.
\label{eq:gravacc}
\end{equation}
Here, $G$ is Newton's gravitational constant, $M$ is the relevant
gravitating baryonic mass and $r$ the distance to $M$ where $g_{\rm N}$ is
evaluated. One can force the gravitational acceleration to scale
invariantly to the kinematical acceleration by redefining it as
\begin{equation}
g = \left (a_0\,g_{\rm N} \right)^{1\over2},
\label{eq:a0}
\end{equation}
where $a_0$ is a constant that needs to be introduced on
dimensionality grounds.  Now $g \longrightarrow \lambda^{-1} g$ under
eq.~2. We refer to $a_0$ as Milgrom's constant. Eq.~\ref{eq:a0} thus
arises from scale-invariance of the equations of motions in a
gravitational field, and is referred to as scale-invariant dynamics
(SID)\footnote{This had originally in~1983 been called the ``deep-MOND
  regime'' by Milgrom \cite{Milgrom83}.}.

Because the centrifugal acceleration equals the centripetal
acceleration for a circular orbit with speed, $v_c$, about $M$,
\begin{equation}
g = {\sqrt{G\,M\,a_0} \over r}  = {v_c^2 \over r}.
\end{equation}
Thus
\begin{equation}
v_c = \left( G\,M\, a_0 \right)^{1\over 4},
\label{eq:TFR}
\end{equation}
which is exactly the BTFR (Fig.~\ref{fig:BTFR}).  Thus, if SID were
the correct description of dynamics in the ultra-weak-field limit ($g
< a_0$), then all rotationally-supported late-type galaxies must lie
on the BTFR (exceptions may occur due to the external field effect,
see below). The observed flat rotation-curve data are consistent with
eq.~\ref{eq:TFR} for $g < a_0$, which is the validity of SID. For $g >
a_0$ the system approaches the Newtonian regime.

Since the Newtonian circular velocity around $M$ is the circular
velocity an observer would see in the absence of dark matter,
$v_{c,b}^2 = G\,M/r$, it follows by using eq.~\ref{eq:TFR} that
\begin{equation}
\left( {v_c \over v_{c,b}} \right)^2 = \left(  {a_0 \over g_{\rm N}}
  \right)^{1\over2}.
\label{eq:MDA}
\end{equation}
This is exactly the Sanders-McGaugh mass-discrepancy--acc\-eleration
(MDA) correlation which late-type galaxies are seen to
obey (Fig.~\ref{fig:SID_weakF}). 


Here it is important to re-iterate what has been found above:
\vspace{2mm} \centerline{ \fbox{\parbox{\columnwidth}{ {\sc SID}: By
      adopting a simple classical space-time scaling symmetry and forcing
      the gravitational acceleration to scale as the kinematical one,
      a law emerges which relates the scaled circular velocity to the
      Keplerian circular velocity. While as such this law is merely a
      hypothetical mathematical result, the observational data show it
      to be excellently fulfilled by nature for describing the
      observed gravitational law in galaxies. This is shown in
      Fig.~\ref{fig:SID_weakF} and~\ref{fig:SID_all}.  }}}
\vspace{2mm}

Sanders (1990, \cite{Sanders90}) pointed out for the first time that
disk galaxies obey the MDA relation (eq.~\ref{eq:MDA}) and that exotic
dark matter halos would not be able to account for it. The MDA
relation was verified with many more data by McGaugh (2004,
\cite{McGaugh04}).  To construct this law, the observed circular
velocity, $v_c$, is compared to the circular velocity the star or gas
cloud should have in the absence of dark matter, $v_{c,b}$, and their
ratio is plotted as a function of the Newtonian acceleration, $g_{\rm
  N}$, required for the star or gas cloud to remain on a circular
orbit given the baryonic mass of the galaxy. Scarpa (2006,
\cite{Scarpa06}) studied the correlation between the velocity
dispersion and luminosity of pressure-supported systems (globular
clusters, elliptical galaxies, groups of galaxies and clusters of
galaxies). In his fig.~12 impressive agreement between SID and the
observations over at least 6 orders of magnitude in acceleration is
evident.

Milgrom's constant, $a_0$, which is introduced with eq.~\ref{eq:a0}
has the value of about $a_0=3.8\,$pc/Myr$^2$ (e.g. \cite{FM12,
  Kroupa12b}). It is unknown at present whether $a_0$ is a constant in
space and time, but the measurements of it are consistent with it
being a true constant of nature. This constant has possible deep
associations.  A de~Sitter universe with a cosmological constant
$\Lambda\approx 10^{-19}{\rm pc}^{-2}$ is characterized by a radius
$R_\Lambda = 1/\sqrt{\Lambda/3}\approx5\times10^9\,{\rm pc}$
which is about the radius of the observable Universe.  Thus $a_0
\approx \left(c^2/2\pi\right)/R_\Lambda$, which could point to a
deeper relation between the two (see \cite{Milgrom99}, footnote~9
in \cite{FM12} and Sec.~\ref{sec:MD})." Also, constructing the
value of an acceleration from cosmological and vacuum constants, we
note that 
\begin{equation}
a_0=c\,H_0/(2\,\pi), 
\label{eq:H0}
\end{equation}
where $H_0$ is Hubble's constant such that the observed MDA
correlation data are reproduced by eq.~\ref{eq:MDA}
(Fig.~\ref{fig:SID_weakF}) and $c$ is the speed of light in
vacuum. This also suggests a possible relation of SID and of Milgrom's
constant to cosmology.  Studying the BTFR at high redshifts may
provide constraints on a possible variation of Milgrom's constant
$a_0$ with cosmic time (e.g. \cite{Ziegler02}). A suggestion that
$a_0$ may depend on cosmic time within the framework of pure-metric
non-Einsteinian gravity models \cite{Deffayet14} may provide a
possible test of such field equations.

The physical origin for the departure of gravitation from Newton's
empirical law when $g_{\rm N}<a_0$ is not understood today, but
\cite{Milgrom99} suggests this to be related to quantum-mechanical
processes in the vacuum (see also Sec.~\ref{sec:conserv}), and
\cite{Bekenstein04} has shown a relativistic generalization of
Einstein's field equation to provide this departure (for more
approaches see Footnote.~\ref{footn:general}).

SID, as defined above, implies flat rotation curves (eq.~\ref{eq:TFR})
and is thus mathematically fully equivalent to the baryonic point mass
$M$ sourcing an isothermal dark matter halo about it in
Einsteinian/Newtonian dynamics. Because this halo is not made of
exotic dark matter particles, but is instead a mathematical property
of SID, it is commonly referred to as a {\it phantom dark matter
  halo}.

The phantom dark matter halo has an acceleration at radial distance
$r$ of 
\begin{equation}
g= {v_c^2 \over r} =  {\left( G\,M\, a_0 \right)^{1\over 2} \over r}.
\label{eq:acelSID}
\end{equation} 
The Newtonian observer would interpret this acceleration to be due to
a (phantom) dark mass, $M_{\rm pDM}$, within the radius $r$,
$g=G\,\left( M+M_{\rm pDM}(<r) \right) /r^2$, whereby $M\ll M_{\rm
  pDM}$ for large $r$ where $g < a_0$.  It thus follows \cite{WK15}
\begin{equation}
M_{\rm pDM}(<r) = \sqrt{ {M\, a_0 \over G} } \, r.
\label{eq:pdm}
\end{equation}
The {\it gravitating but not inertial mass} of such a phantom dark
matter halo can be calculated given that the halo is truncated due to
the ambient matter density. The phantom dark matter halo mass as a
function of the baryonic mass is given by eq.~5 in \cite{WK13}.

Eq.~\ref{eq:acelSID} can be written as
\begin{equation}
g= \left( G\,\pi \Sigma_M\, a_0 \right)^{1\over 2},
\end{equation}
where $\Sigma_M=M/(\pi r^2)$ is the baryonic surface density of the
disk galaxy in its disk plane evaluated at large $r$ (see also
sec.~4.3.2 in \cite{FM12}).  \cite{Trippe14} discusses this in more
detail and shows that this prediction of SID is verified by
observational data. It is unclear how the correlation, $g\propto
\Sigma_M^{1/2}$, can emerge in dark-matter models within the framework
of the SMoC.

\vspace{2mm} \centerline{ \fbox{\parbox{\columnwidth}{ {\sc
        Important}: SID implies galaxies to have flat rotation curves,
      to lie on the BTFR, to obey the MDA correlation and $g\propto
      \Sigma_M^{1/2}$.  }}} \vspace{2mm}

If the galaxy of baryonic mass $M$ is immersed in an external
gravitational field with acceleration $a_e<a_0$, which is constant
across the galaxy, then the phantom dark matter halo is effectively
truncated at the galactocentric radial distance $r_t$,
\begin{equation}
r_t = { \sqrt{G\, M\, a_0} \over a_e},
\end{equation}
such that the mass of the phantom dark matter halo is reduced
(eq.~\ref{eq:pdm}). The depth of the equivalent Newtonian (phantom
dark matter halo) potential sourced by this truncated isothermal
phantom dark matter mass is thus reduced. Since SID is a non-linear
dynamics theory\footnote{Note that GR is also non-linear while
  Newtonian dynamics is linear. Newtonian dynamics can thus be viewed
  as a linearized approximation of an otherwise non-linear
  space-time-matter coupling.}, the influence of $a_e$ at all radii
plays a role and the gradient of the potential changes and as a
consequence the density of the phantom dark matter halo is smaller
within $r$ as well. L\"ughausen et al. \cite{Lueghausen14} have
quantified this in detail.  A star orbiting the galaxy of baryonic
mass $M$ beyond $r_t$ will be accelerated in the external field, while
if its orbital radius is within $r_t$ it will be accelerated by the
galaxy.  When $a_e>a_0$ then the galaxy appears as a naked baryonic
system with a Keplerian rotation curve.

A system of baryonic mass $M$, when exposed to an external field, will
be spatially more extended than the same isolated system. This comes
from the virial theorem according to which a system within a potential
must expand if the potential is reduced. A more thorough discussion of
this important SID effect can be found in sec.~3.1 of \cite{Famaey07},
and Sec.~\ref{sec:pred_sizes} contains a prediction due to the
external field effect.

This {\it external-field effect} (EFE) is thus an important {\it
  prediction} from SID. Self-consistent simulations by Wu \& Kroupa
\cite{WK13a} demonstrate the contraction of a system propagating from
the Newtonian regime (strong $a_e$) to the SID regime (weak
$a_e$). Further details on SID and its implication for dwarf galaxies
are available in \cite{Angus14} and in \cite{Lueghausen14,
  WK15}. Given the EFE, local (i.e. terrestrial) tests of MOND become
impossible \cite{Meyer12}. 

Note that the EFE is an unavoidable phenomenon in SID. But is could be
different than the simple discussion above suggests: \cite{Milgrom11}
points out that there are two interpretations of SID (or, more
precisely of the full classical formulation of Sec.~\ref{sec:MD}). The
one interpretation is that SID is classical but a non-Newtonian
gravitation theory. Another interpretation is for SID to be a theory
in which inertial mass does not equal gravitational mass in the SID
regime. This latter interpretation leads to SID being non-local in
time. This implies that the system's state depends on its full
trajectory in phase space. Milgrom writes `` A subsystem, such as a
globular cluster or a dwarf galaxy, moving in the field of a mother
galaxy, or a galaxy in a cluster, may be subject to an EFE that
depends on the accelerations all along its orbit, not only on the
instantaneous value.'' It is possible to construct SID theories which
have ``practically no EFE''.  ``Practically no EFE'' here means that
in theory there is a small unavoidable effect, but in practice it can
be neglected.

Is the EFE evident in observational data?

\subsection{Testing the external field effect (EFE)}
\label{sec:testExternalF}

Detailed and self-consistent calculations are required to test this
important prediction of SID. But it is possible to already
study available data for evidence against or for this prediction.

\subsubsection{Evidence for the EFE}

The satellite galaxies of Andromeda are known to be systematically
larger than those of the MW \cite{Collins11}.  This has been noted
but has found no explanation in the SMoC. That the Andromeda
satellites are systematically larger than the MW satellites appears to
be qualitatively consistent with SID if Andromeda is more massive than
the MW or if the satellites of Andromeda are systematically closer to
Andromeda than those of the MW.

It has also been observed that elliptical galaxies in galaxy clusters
are systematically larger than isolated elliptical galaxies
\cite{Shankar14}. Again, this appears to be consistent with the
EFE.

However, so far no quantifications of the EFE concerning sizes of
galaxies have become available, so that final statements on a
verification of the prediction cannot be
made.\footnote{\label{footn:grants}Here it is perhaps useful to know
  that the scientific community actively studying and testing
  SID/Milgromian dynamics amounts to less than ten persons, in
  contrast to the hundreds working in the SMoC framework. Asking for
  even modest research grants to foster work in SID has typically been
  greeted with rejections.}

\subsubsection{Evidence against the EFE}

Hernandez et al. \cite{Hernandez12a} study the statistical properties
of very wide binary stars in the Galactic field. If the EFE were
active then such binaries, by being immersed in the Galactic external
field, should show the Keplerian relative velocity fall-off with
separation.  But the data suggest the motions of the binary-star
companions to conform to SID in contradiction to the
EFE. \cite{Hernandez12a} thus finds evidence that Kepler's third law
is violated for wide binaries (separation $\simgreat 7000\,$AU) and
for the EFE to be suppressed.

A similar behavior is found by Scarpa et al. \cite{Scarpa11} for the
velocity dispersion profile of seven globular clusters which remains
flat in their outer regions in disagreement with the expected
Keplerian fall-off if the EFE were active. That is, the outskirts of
the globular clusters appear to remain in the SID regime rather than
showing the EFE-induced suppression of SID. \cite{Hernandez12b}
develops self-consistent dynamical models of globular clusters, based
on the gravitational theory of \cite{Mendoza11}, to account for this
SID behavior.

Further observational constraints on the important EFE are needed to
clarify this situation.

\subsubsection{Globular clusters}
\label{sec:tGC}

Related to the above, it has been noted that if a globular cluster
(GC) is of sufficient low density and sufficiently distant then it
ought to be in the SID regime \cite{Baumgardt05}. A number of teams
have since tested this prediction. So far no clear evidence for SID
has been found (\cite{FM12}; see also the Ibata/Sanders disagreement:
\cite{Sanders12}). On the other hand, SID (and the full Milgromian
dynamics theory, see Sec.~\ref{sec:MD}) is valid for continuum
dynamics. A Milgromian-dynamics theory in the collisional regime, in
which two-body encounters and thus the energy-equipartition process,
play a significant role in the dynamical evolution of a system, has
not been formulated yet (see Footnote~\ref{footn:grants}).

\subsection{Milgromian dynamics}
\label{sec:MD}

It is well known that SID breaks down in the Newtonian regime. The
transition from the SID regime into the Newtonian regime is subject to
on-going research. \cite{Milgrom99} studied this transition for the
non-relativistic case. By assuming that inertia is linked to the
physical processes in the vacuum through a balance between the Unruh
radiation and the Gibbons-Hawking background radiation (see also
Appendix~A in \cite{Kroupa10} and Sec.~\ref{sec:conserv} here),
Milgrom suggests a transition function such that the resulting theory
extends from the deep SID regime into the Newtonian regime. It is
commonly known as modified Newtonian dynamics (MOND,
\cite{Milgrom83}).  Reviews of this field are provided by
\cite{Sanders02, Scarpa06, Sanders07, FM12, Hernandez14} and
\cite{Trippe14}. In particular, with their major review, Famaey \&
McGaugh (2012, \cite{FM12}) cover various gravitational theories and
the possible deeper physics of Milgromian dynamics.  The complete
Milgromian description of classical gravity has a Lagrangian
formulation such that this theory is energy and angular momentum
conserving \cite{BM84}.  Minimizing the action leads to the non-linear
generalized Poisson equation,
\begin{equation}
\vec{\bigtriangledown} \cdot 
\left[  \mu
\left(
{ | \vec{\bigtriangledown \phi} | \over a_0} 
\right) \vec{\bigtriangledown \phi} 
\right]
=4\,\pi\,G \, \rho,
\label{eq:genPoisson}
\end{equation}
where $\phi$ is the potential such that
$\vec{g}=-\vec{\bigtriangledown\phi}$ (whereby $\vec{g_{\rm
    N}}=-\vec{\bigtriangledown \phi_{\rm N}}$ is the Newtonian
acceleration derived from the Newtonian potential sourced by the
baryonic matter density $\rho$ and which obeys the Poisson equation
$\bigtriangledown^2\phi_{\rm N} = 4\,\pi\,G \, \rho$). In
eq.~\ref{eq:genPoisson} the function $\mu(x) \rightarrow 1$ for
$x\gg1$ and $\mu(x) \rightarrow x$ for $x\ll 1$ (see \cite{BM84, FM12}
for details).  This formulation (eq.~\ref{eq:genPoisson}) is not
unique, as yet another formulation at the classical level has recently
been presented (see \cite{Milgrom10, FM12} for details) and
implemented in the {\it Phantom of Ramses} (POR) simulation code
(L\"ug\-hausen, Famaey \& Kroupa 2014, \cite{Lueghausen14b}).

Simulations of the evolution of galaxies in Milgromian dynamics have
been performed showing generally better agreement with the observed
galaxies than galaxies modelled in Newtonian dynamics and dark matter
\cite{TC08}. The formation of disk galaxies in growing (phantom) dark
matter halos that have no significant merging histories is successful
\cite{SG03}. The observational result that galaxies are simpler than
expected if the SMoC were true \cite{Disney08} is consistent with
galaxies obeying SID. Such simulations have furthermore demonstrated
that mergers are rare because dynamical friction is significantly
suppressed given the absence of large and massive dark matter halos,
as shown by Tiret \& Combes \cite{Tiret08, TC08}.  Galaxies may thus
interact with each other repeatedly which can be tested with the
longevity of Hickson compact groups (prediction~\ref{sec:pred_Hickson}
in Sec.~\ref{sec:predictions}).  It will remain to be seen if the
peculiar symmetric Local Group structure and the relative orientation
of the VPOS and GPoA \cite{Pawlowski13b} as well as other dwarf galaxy
arrangements nearby the Local Group \cite{PM14} may follow naturally
from a past encounter between the MW and Andromeda in SID
\cite{Zhao13}. Such an encounter may have produced sufficient debris
that follow-up encounters between parts of it and the original major
galaxies may lead to the symmetrical organized distributions of dwarfs
evident in and nearby to the Local Group.

An isolated star is surrounded by its own phantom isothermal dark
matter halo, as would be deduced by an observer who interprets the
data in terms of Newtonian dynamics. By eq.~\ref{eq:acelSID} SID will
break down at a radius, $r_{\rm N}$, at which $g=a_0$, i.e. at
\begin{equation}
r_{\rm N} = \sqrt{G\,M \over a_0}.
\end{equation}
If $M=1\,M_\odot$ and since $G=0.0045\,$pc$^3M_\odot^{-1}$Myr$^{-2}$
and $a_0\approx3.8\,$pc/Myr$^2$, it follows that each star carries a
Newtonian bubble around it with radius $r_{\rm N}\approx
0.03\,$pc. Thus, only in regions where the star--star separations are
$\simless 0.03\,$pc will the system be in the Newtonian regime, while
at lower density it will be in the SID regime and will thus show
evidence for dark matter. Thus, most of the Universe is in the SID
regime.

A note to history: often it is argued that Milgrom's law
\cite{Milgrom83} has been derived from the
mass-discrepancy--acceleration correlation discussed in
Sec.~\ref{sec:sid}. This is not correct. These correlations were used
as tests of Milgrom's original proposal of~1983. Likewise, Milgromian
dynamics has not been derived from the BTFR. In actual fact, that
low-surface brightness galaxies and gas-rich dwarf galaxies such as
DDO210 (baryonic mass $M\approx 5\times 10^6\,M_\odot, v_c\approx
15\,$km/s) lie on the BTFR constitutes a highly significant and
successful verification of Milgrom's law in a regime not known to
exist in~1983.

A theoretical argument sometimes voiced against MOND is that it has a
preferred frame, because it is based on accelerations relative to
$a_0$. In one formulation of Milgromian dynamics, bimond
\cite{Milgrom09b}, the acceleration is relative to another metric in
the form of the subtraction of two Christoffel symbols. Milgrom's
suggestion (1999, \cite{Milgrom99} that SID may result from a
modification of the laws of motion through the quantum mechanical
vacuum probably needs a preferred frame indeed, and TeVeS
\cite{Bekenstein04} has some preferred-frame effects too. But this
does not imply a fundamental obstacle against such theories, since
nature may have this property. That there is a preferred frame is
evident from the CMB, since the motion relative to the CMB is
measurable.  And, Einstein-aether theories (e.g. the review by Eling
et al., 2004 \cite{Eling04}) are quite popular despite having
preferred frames.

As a final note, a citation from Milgrom \cite{Milgrom09} on SID, and
especially on the symmetry eq.~\ref{eq:scale}, is important: ``(1) it
may help put MOND on more sound footings, showing that it need not be
imposed as an ad hoc dictate of phenomenology, (2) identifying an
underlying symmetry may help extend MOND to non-gravitational systems
for which we have no phenomenological guidance, at present, (3)
identifying a partial symmetry valid in the deep-MOND regime may lead
us to identify a larger symmetry in this regime, or to identifying a
symmetry group for the full MOND theory, and this will help constrain
the MOND theory itself, and (4) this description is more readily
amenable to direct predictions.'' The symmetry underlying SID may thus
be pointing to a new or at least improved understanding of
gravitation.

\subsection{Testing Milgromian dynamics}
\label{sec:testMD}

Milgromian dynamics has, so far, been found to be a remarkably
successful despite the many tests that have been tried to disprove it
\cite{FM12}. Perhaps the hardest test has been invented by my research
group in Bonn by using GCs (Baumgardt et al. 2005 \cite{Baumgardt05},
Sec.~\ref{sec:tGC}). A theory-confidence graph has also been
constructed for Milgromian dynamics in Kroupa (2012,
\cite{Kroupa12a}).  It appears to fare significantly better than the
one for the SMoC (Sec.~\ref{sec:confidencegraph}) because virtually
all of the failures of the SMoC on galactic and galaxy-cluster scales
disappear.  The literature contains numerous cases of claims against
Milgromian dynamics which have, time and again, been refuted or even
withdrawn (e.g. on the rotation curve of Holmberg~II: \cite{Gentile12}
as a rebuttal of \cite{Sanchez13} and the retraction of the original
claim \cite{Sanchez14}; on the rotation curve of M33:
\cite{Corbelli07} which is refuted by \cite{FM12}).

A clear-cut falsification of Milgromian dynamics would be available if
a truly isolated disk galaxy is found with a rotation curve which
deviates significantly from the Milgromian one. No such case is
known. At the same time, the galaxy population which arises in a
universe governed by Milgromian dynamics has not been computed yet, so
it is not possible now to perform the same in-depth tests as are
available for the SMoC \cite{Kroupa10, Kroupa12a}. At the present it
is not known how galaxies form in a Milgromian universe, nor is it
known how they evolve over cosmic time.  It is hoped that all of this
may now become possible with the availability of the POR code by
L\"ughausen, Famaey \& Kroupa (2014, \cite{Lueghausen14b},
Sec.~\ref{sec:concs}).

\section{The build-up of stellar populations in galaxies}
\label{sec:buildingblocks}

The accretion of gas onto galaxies is an important cosmological
problem and is inferred largely through the observed galaxy-wide
star-formation rates (SFR) which need to be balanced by the in falling
material for near-constant star-formation histories. For example, the
MW is known to be accreting gas at a rate comparable to the SFR (a few
$M_\odot$yr$^{-1}$ \cite{Yin09}). If dark matter is not present,
then the dark halos cannot aid in focussing the gas flows onto
galaxies. Observationally deduced gas accretion rates, e.g. through
cold filamentary flows (e.g. \cite{DB08, Goerdt12}), thus yield
constraints on the effective gravitational potential of the galaxies.

Inferring SFRs is based on various tracers, most of which rely on
detecting the massive-stellar population.  Assuming a stellar initial
mass function (IMF) which is normalized by the observational estimate
or the quantity of the tracer (e.g. the H$\alpha$ luminosity) then
allows an estimate of the SFR \cite{Rosa02, Calzetti08}. The
time-evolving SFR defines the evolution of a galaxy.

The evolution of galaxies constitutes one of the greatest current
research problems in astrophysics (e.g. \cite{Hensler11, Silk12,
  Recchi14, Speagle14}). The many physical processes involved are
reviewed by \cite{Shlosman13} which suggests that evolution driven by
mergers between dark-matter halos cannot be the correct theory, a
deduction which is consistent with the apparent validity of SID in
galaxies (Sec.~\ref{sec:sid}) and the lack of galaxy--galaxy mergers
in this effective gravitational framework \cite{Tiret08, TC08}.
Galaxies appear to be self-regulated systems \cite{Koeppen95} and are
much simpler than expected if the SMoC were true, by being populations
governed largely by one parameter only \cite{Disney08}.  This is
particularly demonstrated by the too-small number of late type
galaxies with bulges and the need for smooth accretion probably from
filaments to fuel the gas supply in galaxies. Late-type star-forming
galaxies thus constitute the by far dominant ($\simgreat 90$~\%)
population of galaxies with baryonic masses $\simgreat
10^{10}\,M_\odot$ (\cite{Delgado10}, see also fig.~4.14 in
\cite{BM98}) and their formation and evolution is quite well
understood as long as the dynamics of the putative particle dark
matter is suppressed and they are allowed to accrete baryons smoothly
\cite{SG03, Shlosman13}. However, observations do show that in the
past late-type galaxies had more complex morphology with signs of
interactions increasing with increasing redshift
(e.g. \cite{Delgado10,Kutdemir10}) and/or signs of internal
instabilities such as evident in the chain galaxies at redshifts
between 0.5 and~2 \cite{Elmegreen04a, Elmegreen04b}. Noteworthy in
this context is that accounting for the thick disk of the Milky Way
through clustered star-formation (Sec.~\ref{sec:fundbuild}) predicted
the Milky Way to have been a chain galaxy about 11~Gyr ago
\cite{Kroupa02}.

Elliptical or early-type galaxies with baryonic mass $\simgreat
10^{10}\,M_\odot$ have been observationally deduced to form rapidly
early on and not to build-up from mergers over cosmic time of
late-type galaxies \cite{Ziegler97,Matteucci03, Naab09, Thomas05,
  Recchi09}. This remarkable result had been suggested already by the
pioneering work of Matteucci (1994, \cite{Matteucci94}), being later
confirmed by \cite{Vazdekis97, Ziegler97, Thomas05}. Requiring
elliptical galaxies instead to conform to merger-driven buildup
requires the postulate that they have a merger history with early-type
(gas-poor) galaxies. But such a population of galaxies remains
undiscovered, unless the merging sub-units are faint early-type dwarf
galaxies or most of the merging has finished by~6~Gyr ago. Since then
the population of elliptical galaxies does not seem to have evolved
\cite{Delgado10}, which would make the present cosmological epoch
special. \cite{Pipino08} explicitly test this scenario and rule it
out.

The stellar populations which arise in the galaxies are the most
important tracers that allow an observational approach to
galaxies. But how do these build-up?

Most models of star formation and feedback in galaxies have been
assuming the stellar initial mass function (IMF) to be invariant. The
theoretical and observational evidence has however shown this to be
incorrect. Before turning to the observationally confirmed variation of
the IMF in Sec.~\ref{sec:laws}, the hot-topic of a bottom-heavy IMF in
elliptical galaxies is briefly discussed.

\subsection{A bottom-heavy IMF in elliptical galaxies?}
\label{sec:E_IMF}

The cooling-flow problem in galaxy-clusters led to the suggestion that
the X-ray emitting cooling gas may be forming low mass stars in the
central elliptical galaxies as these would not be observable as
star-forming systems. Stellar gravity-sensitive spectral absorption
features in the $\mu$m spectral region were proposed to test the
notion that such elliptical (E) galaxies may have a large population
of low-mass stars (Kroupa \& Gilmore 1994, \cite{KG94}).  Applying
the dwarf-sensitive spectral features Na I, Ca II, and FeH in the
$\mu$m spectral range uncovered the signature of stellar mass
functions heavily dominated by low-mass stars. The work of van Dokkum
\& Conroy \cite{CvD12} (see also \cite{Spinniello14}) unambiguously
showed these spectral indicators to strengthen with increasing E
galaxy mass such that massive E galaxies would have had stellar IMFs
much steeper than the Salpeter index. But, as pointed out by
\cite{Kroupa13}, there has not been a theoretical prediction of such
extreme IMFs; theory ever only predicted the IMF to become top-heavy
with increasing temperature or decreasing metallicity and in
star-bursting systems. The IMF results on E galaxies are thus highly
puzzling but also very interesting. In particular, a bottom-heavy IMF
would pose a challenge for understanding the observationally deduced
rapid chemical evolution of E galaxies such that hybrid models of a
top-heavy galaxy-wide IMF in the initial star burst evolving to a
bottom-heavy galaxy-wide IMF over an extended time of lower-intensity
star formation would be required \cite{Vazdekis97,Weidner13c}.

The notion that the IMF becomes extremely bottom-heavy in massive E
galaxies has been tested by observing the closest massive E galaxy
with a strong-lensing signal \cite{SL13}. This lensing signal allows
the mass of the E galaxy to be constrained, and it is found to yield a
$M/L$ ratio consistent with a canonical IMF (which is botom-light
compared to a Salpeter IMF) at the age of the galaxy. This same galaxy
has, however, the strong spectral absorption bands that appear to
imply a strongly bottom-heavy IMF. Such a strongly bottom-heavy IMF
would be inconsistent with the measured amount of mass within the
strong lens.  Thus, these spectral features must have a different
origin, i.e., they cannot stem from a large population of low-mass
stars. An independent test of the bottom-heavy IMF in E galaxies has
been performed by \cite{Peacock14} by testing whether the number of
low-mass X-ray binaries (LMXBs) varies with E-galaxy mass as it should
if the relative number of massive to about solar-mass-stars in the IMF
changes.  Their results are not consistent with the IMF becoming
strongly botom-heavy in more massive E galaxies.  The origin of the
dwarf-star-sensitive spectral features is suggested by
\cite{Maccarone14} to come from a certain type of evolved binary stars
being disrupted more in lower-mass E galaxies than in massive E
galaxies which have lower stellar number densities in combination with
such stars missing in the spectral libraries used to synthesize the
models that, due to the omission of such stars, imply strong
dwarf-star populations.

{\it In conclusion}, although the spectral signatures do seem to be
convincing evidence, it appears that the independently used
constraints do not support the notion that the IMF becomes
increasingly and significantly bottom-heavy in increasingly massive E
galaxies. The trend of increasing dynamical $M/L$ ratio with
increasing E-galaxy mass \cite{Cappellari12, Sanders14} may be
explained by an increasingly top-heavy galaxy-wide IMF through the
presence of an increasing fraction of stellar remnants in more massive
E galaxies. This is consistent with the IGIMF
(Sec.~\ref{sec:sftheories}) becoming top-heavy in star-bursting
galaxies and with the chemical enrichment histories of massive E
galaxies, as discussed below.  However, this result must also be
tested independently by constraining the mass via strong lensing and
the relative number of remnant and main sequence stars using LMXBs.

\subsection{Is star formation stochastic or self-regulated?}
\label{sec:selfreg}

There are two fundamentally different approaches to account for the
emergence of stellar populations in galaxies. The one is to assume
that star formation is a stochastic process such that the distribution
of stellar masses, i.e. the IMF, is a probability density distribution
function (PDF), which is usually assumed to be invariant.  Complete
logical consistency would follow if the form of the IMF was
stochastically variable.  This is not a well-motivated physical
description but is simplest to model.  The other is to assume that the
IMF is subject to constraints posed by local physical conditions such
that the process of star formation is not stochastic, but is
self-regulated instead in galaxies that are in dynamical
equilibrium. This approach is well-motivated physically but is harder
to formulate since the underlying laws need to be discovered. It opens
the avenue for studies with systematically varying IMFs.

Dynamical population synthesis \cite{Kroupa95} and with the advent of
infrared imaging capabilities and high-resolution ($<1\,$pc)
observations of star-forming molecular clouds have shown that
star-formation occurs in compact embedded clusters (ECs). The review
by \cite{LL03} established this and that the ECs are distributed by
mass according to a power-law embedded star-cluster mass function
(ECMF). The build-up of galaxies therefore occurs from such
structures.

The application of the above notions on stochastic versus constrained
star formation to ECs leads to the two following possibilities: The
emergence of ECs can be modelled either as a stochastic process such
that the ECMF is assumed to be a PDF. Or the ECMF is subject to
constraints posed by local physical conditions such that the process
of clustered star formation is not stochastic.

\subsection{Embedded clusters: the fundamental building blocks of
  galaxies}
\label{sec:fundbuild}

An EC forms from a molecular cloud core on a pc-scale or less and
within a Myr or less, with a star-formation efficiency smaller than
50~\% and probably close to 33~\%. The EC expands significantly
\cite{Brandner08} after self-removal of its residual gas through the
feedback of its own young stars thereby dispersing a large fraction of
them into the galactic field. This has been computed for ECs on a
stellar mass scale of $10<M_{\rm ecl}/M_\odot<100$ \cite{KB03}, for
$M_{\rm ecl}\approx10^3\,M_\odot$ \cite{KAH}, for $M_{\rm
  ecl}\approx10^4\,M_\odot$ \cite{BK13}, and for $M_{\rm
  ecl}\approx10^5\,M_\odot$ \cite{BK14}.  Observations with the
Hershel Space Observatory have shown the groups of stars and clusters
to form in molecular cloud cores along thin filaments with radii or
cross sections near 0.1~pc \cite{Malinen12}, which is consistent with
the independently obtained radius--mass relation of embedded clusters
\cite{MK12}. The masses of embedded clusters range from groups of a
dozen binaries ($\approx 10\,M_\odot$, which appear to an observer as
``isolated'' star formation) to an upper limit which is unknown but
may be as large and perhaps larger than $10^7\,M_\odot$ \cite{Marks12,
  MK12, Weidner13b}.

\begin{figure}
\centering{}\includegraphics[width=8cm]{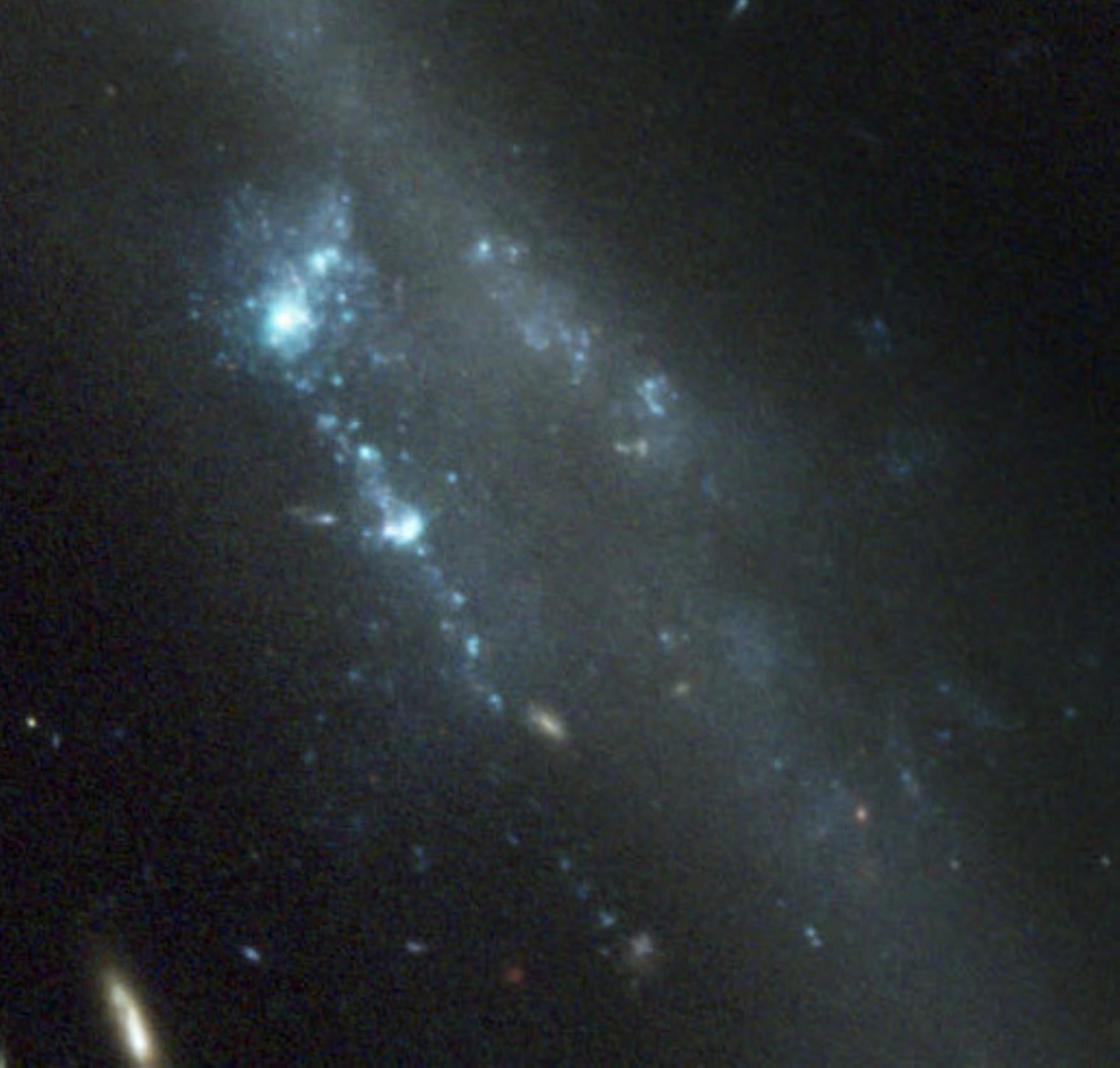}
\caption{\small A close-up of one of the TDGs in the tadpole galaxy
  shown in Fig.~\ref{fig:tadpole}.  This demonstrates the ``lego
  principle'' \cite{Kroupa05} according to which star formation in
  galaxies occurs in embedded star clusters which expel the residual
  gas not used up in the star-formation process and become visible as
  very young stellar populations seen here as the small faint blue
  dots.  \cite{Tran03} estimate the mass of the TDG in the upper left
  part of the image to be about $1.3\times 10^6\,M_\odot$ with a
  radius of about 160~pc and age of about 4.5~Myr. This star-cluster
  complex may be a progenitor of much older objects as it typically
  evolves into spheroidal dwarf galaxies, depending on the details of
  the orbit and the tidal field \cite{Kroupa98, Bruens11}.  Compare
  this to the masses ($<10^7\,M_\odot$) and radii ($\approx 250\,$pc,
  \cite{McConnachie12}) of the MW and Andromeda dSph satellite
  galaxies.  The scale of the figure is approximately $8.6\,{\rm kpc}
  \times 9.1\,{\rm kpc}$ (vertical x horizontal).  }
\label{fig:tadpoleTDG}
\end{figure}

Thus, the formation of the stellar population of galaxies occurs, over
time, through the successive addition of stellar populations
individually born in star-formation events which are correlated in
space and time. In this sense we arrive at the {\it lego principle}
\cite{Kroupa05} according to which stellar populations in galaxies
are obtained by adding together the individual populations formed in
embedded clusters. The ``lego principle'' is visualized in
Fig.~\ref{fig:tadpoleTDG}.

These ``fundamental building blocks'' (i.e. the embedded clusters) are
observed to obey empirically discovered laws which are simple.  When
considered these lead to a {\it computable, predictive description of
  young populations of stars in galaxies}. For example, the thickening
of thin galactic disks over time can be derived from this lego
principle \cite{Kroupa02}, the binary population in different
galaxies can be predicted \cite{Marks11}, and the stellar IMF and its
variation with sar-formation rate of a whole galaxy can be calculated
\cite{Kroupa13,Weidner13b}.

\subsection{The empirical laws of star formation}
\label{sec:laws}

The {\it empirical laws} can be summarized as follows (see
\cite{Weidner13b} for a thorough description): the stellar IMF in each
EC is form-invariant; the mass of the most massive star, $m_{\rm
  max}$, within the EC is a function of the mass of the EC, $m_{\rm
  max}={\rm K_1}(M_{\rm ecl})$ (the $m_{\rm max}-M_{\rm ecl}$
relation); the ECMF is a power-law distribution function with index
$\beta$ ($\beta=2.35$ is the Salpeter index); the most-massive EC
forming in a population of ECs is a function of the galaxy-wide star
formation rate (SFR), $M_{\rm ecl, max}={\rm K_2}(SFR)$ (the
galaxy-wide $M_{\rm ecl,max}-SFR$ relation).

The parameters entering the empirical laws are constrained using
observational data of well-resolved star-formation events. Thus, the
stellar IMF is well-known to be the canonical form (i.e. a
two-part-power law with Salpeter-Massey index $\alpha_2=2.3$ for stars
more massive than $0.5\,M_\odot$ and with $\alpha_1=1.3$ for less
massive stars, \cite{Kroupa02, Bastian10, Kroupa13}). It is
particularly important to be aware of the biases that enter
observational deductions on the shape of the IMF \cite{Kroupa13}:
binary star populations evolve on a cluster dynamical time-scale and
may wrongly suggest IMF variations if not corrected for, massive stars
are efficiently dynamically ejected from clusters such that they are
missing from the star counts, the expulsion of residual gas during
star formation unbinds low-mass stars in mass-segregated clusters
leading to wrong inferences about the IMF shape unless this is taken
into account. The $m_{\rm max}-M_{\rm ecl}$ relation, $m_{\rm
  max}={\rm K_1}(M_{\rm ecl})$, has been established from observed ECs
\cite{Weidner13} and $M_{\rm ecl, max}={\rm K_2}(SFR)$ is constrained
by extragalactic star-forming galaxies \cite{Weidner04}. Furthermore,
the functions $K_1$ and $K_2$ readily follow from mass-constraints on
the respective distribution functions (the IMF and the ECMF).

Accounting for the biases, the observational constraints suggest that the
stellar IMF becomes top-heavy when an EC surpasses the critical mass
of $2.7\times 10^5\,M_\odot$ which obtains from a critical level of
star-formation rate density during the formation of an individual
massive EC. This evidence for top-heavy IMFs in massively
star-bursting clusters appears from four entirely independent data
sets: (i) the elevated dynamical $M/L$ ratios of UCDs \cite{Dab09,
  Dab10}, (ii) the higher incidence of LMXBs in UCDs \cite{Dab12},
(iii) globular clusters in the Milky Way which have ``damaged''
present-day mass functions of low-mass stars \cite{Marks12}, and (iv)
the element anti-correlations observed in massive globular clusters
\cite{PC06}.  This recently discovered tendency of the stellar IMF to
be increasingly top-heavy with increasing star-formation rate density
within (extremely massive) ECs is the key to understanding the
independently obtained evidence from elliptical galaxies and from
large surveys of star-forming galaxies for top-heavy galaxy-wide IMFs
at high SFRs (\cite{Matteucci94, GM97, Thomas05,Ballero07a,
  Ballero07b, HG08, Meurer09, Gun11}, see \cite{Weidner13b} for a
review of the evidence including the cosmological star formation
history).  The top-heavy stellar IMF may result in molecular clouds
that are cosmic-ray heated \cite{Papadopoulos11, Papadopoulos13}.

The only parameters not constrained by observation independently of
the IGIMF framework are $1.5\simless\beta\simless 2.5$ and the minimum
embedded-cluster mass in the population of embedded clusters, $M_{\rm
  ecl, min}\simgreat 10\,M_\odot$, both of which enter the ECMF. These
two parameters may be varied for improved fits, and they are found to
obey systematic variation with the galaxy-wide SFR \cite{Weidner13b}.

\subsection{Star formation theories}
\label{sec:sftheories}

From hereon the following three classes of theories will be discussed:

\begin{description}

\item {\bf A}: {\it Purely stochastic star formation} such that the
  IMF is an invariant PDF, and the ECMF does not play a role. Stars
  may be clustered randomly into ECs without conditions which
  themselves form a random population.

  This is the simplest method to model the forming stellar population
  and is also the most popular. A star-formation event in an
  analytical model of galaxy evolution consists typically of a fully
  populated IMF which may thus have fractions of massive stars and the
  corresponding fractional feedback energy. If variations of the IMF
  are admitted it is either necessary to assume systematic variation
  subject to a rule while the formation of individual stars would be
  stochastic, which leads to an inconsistent approach. Or, the
  variation of the IMF would need to be stochastic as well, which
  leads to consistency issues concerning the shape of the IMF at a
  given drawing of a population.

\item {\bf B}: {\it Stochastic but clustered star formation} such that the IMF is
a PDF but it must fulfill constraints posed by the mass of an
embedded cluster and the ECMF is also a PDF. 

This is the primitive version of the IGIMF theory. According to this
description star formation in a galaxy always occurs in embedded star
clusters, but these may be made up of individual single stars (e.g. an
EC of mass $100\,M_\odot$ may be comprised of a single star of mass
$100\,M_\odot$).

\item {\bf C}: {\it Regulated star formation} such that neither the
  IMF nor the ECMF are PDFs with stellar and cluster masses allowed to
  be sampled over their entire ranges ($\approx 0.1-150\,M_\odot$ for
  stars, $\simgreat 10\,M_\odot$ for ECs). This is the IGIMF theory as
  proposed by Weidner \& Kroupa \cite{KW03, Weidner05} for an
  invariant stellar IMF, while the extension to allow for the
  systematically varying IMF as obtained from data on ultra-compact
  dwarf galaxies and on globular clusters is formulated by
  \cite{Weidner13b, Kroupa13}. The above {\it empirical laws} of star
  formation enter this description.

  According to this description an EC contains a form-invariant IMF up
  until a most massive star, $m_{\rm max}$, such that the mass of the
  stellar population equals the mass of the EC yielding the correct
  function $m_{\rm max}={\rm K_1}(M_{\rm ecl})$. A star-formation
  epoch consists of a full ECMF, the most massive star cluster of which
  scales with the galaxy-wide star formation rate yielding the correct
  function $M_{\rm ecl, max}={\rm K_2}(SFR)$. The {\it integrated
    galactic initial mass function} of stars (the IGIMF), i.e. the
  galaxy-wide IMF, is calculated by integrating over all newly formed
  ECs with their individual stellar IMFs.  An extreme version of the
  IGIMF formulation assumes both the IMF and the ECMF to be optimally
  sampled distribution functions \cite{Kroupa13}.

\end{description}

The three procedures lead to very different results on galactic
scales, because a galaxy model will have a different history of
chemical enrichment and feedback energy release, and the photometric
and spectral emission will differ (e.g.
\cite{Koeppen07,Recchi09,Calura10,Ploeckinger14, Recchi14b}).  Bekki (2013,
\cite{Bekki13}) has incorporated, for the first time, a galaxy-wide
variable IMF based on the empirical laws of star formation
(Sec.~\ref{sec:laws}) showing that chemical enrichment proceeds faster
in massive galaxies which form a smaller mass in stars than if the IMF
were universal. 

The differences between the different approaches become most apparent
for star-forming dwarf galaxies.  With theory~A (purely stochastic
star formation) an ensemble of dwarf galaxies will have, for example,
the same average ratio of H$\alpha$-to-UV emission as massive
galaxies, while in theory~C (regulated star formation) dwarfs will
have a significantly lower ratio. This comes about because dwarf
galaxies have low SFRs \cite{Lee09} such that only low-mass ECs form
which do not contain massive stars. The following result is obtained
\cite{Weidner13b}:

\vspace{2mm} \centerline{ \fbox{\parbox{\columnwidth}{ {\sc
        Galaxy-wide variation of the IMF}: According to the IGIMF
      theory dwarf galaxies have top-light IGIMFs, while massive
      galaxies have top-heavy IGIMFs.  }}} \vspace{2mm}

\subsection{Tests of star formation theories}

Table~\ref{tab:SFTs} tests the three theories sevenfold against
observational data. It follows that no tests are passed by theory~A, a
few tests are passed by theory~B while all are passed by theory~C. The
modest success of theory~B has been used by some groups to erroneously
argue that star formation is stochastic. Such research groups fail to
take into account the other constraints evident in
Table~\ref{tab:SFTs}.  In reality, theory~C accounts for the star
formation observed locally in the MW and at the same time for the
chemical and photometric tracers on galaxy-wide scales. This approach
allows a computable evolution of the SFR with cosmic evolution as it
accounts also for the observed evidence for top-heavy galaxy-wide IMFs
in galaxies with high SFRs.

Scatter in the data, e.g. amongst dwarf galaxies, can be erroneously
interpreted by some researchers to be evidence for stochastic star
formation, especially in combination with a compromised or
oversimplified analysis (see e.g. \cite{Weidner14}).  But scatter in
observational data arises due to measurement uncertainties as well as
from variations in the SFRs due to galaxy-internal processes
(e.g. ``breathing'' or ``gasping'' modes) or from external
perturbations (e.g. \cite{McConnachie06}). For example, the scatter in
observational data around the theoretical $m_{\rm max} = {\rm
  K_1}(M_{\rm ecl})$ relation can be ascribed entirely to the
observational uncertainties such that it cannot be excluded that there
is no intrinsic scatter in the $m_{\rm max}-M_{\rm ecl}$ relation
\cite{Weidner13}.

\begin{table}
\begin{tabular}{ l | c c c c}
Test & Theory  &A  & B & C\\
\hline
{\bf I} & L1641 & N & N & Y \\
{\bf II} & $m_{\rm max}={\rm K_1}(M_{\rm ecl})$  &N &N & Y\\
{\bf III} & very small dispersion of $\alpha_3$ values  & N & N & Y\\
{\bf IV} & maximum cluster mass & N & N & Y \\
{\bf V} & H$\alpha$/UV &N &Y &Y\\
{\bf VI} & mass--metallicity relation &(Y) &(Y)  & Y \\
{\bf VII} & observed variation of IMF &N & N& Y \\
\end{tabular}
\caption{ \small 
  Testing theories~A, B and~C on observational data:
  Test {\bf I}: Molecular cloud L1641: low-density population of $\approx 1600$ stars
  but no O stars \cite{Hsu12}. This cannot be accounted for with
  stochastic star formation.
  Test {\bf II}:
  $m_{\rm max}={\rm K_1}(M_{\rm ecl})$ relation: small observed scatter of data
  rules out stochastic sampling from an invariant IMF
  \cite{Weidner13}. The growth of embedded clusters
  follows such a relation \cite{Kirk14}. 
  Test {\bf III}: for $m>3\,M_\odot$ the observationally derived IMF
  power-law indices show  a very small dispersion about the Salpeter
  value which is in conflict with random sampling of stars from an
  invariant IMF (\cite{Kroupa02b}; especially sec.~9.5 in \cite{Kroupa13}).
  Test {\bf IV}: The maximum masses of young star clusters in the galaxy
  M33 show a significant decrease with increasing radial distance ruling
  out stochastic star formation \cite{Pflamm13}. A similar result is
  found by \cite{Vaisanen14} in the variation of the most-massive
  young cluster with galaxy SFR.
  Test {\bf V}: H$\alpha$/UV flux ratio for dwarf galaxies decreases systematically
  with decreasing SFR \cite{Lee09} -- this can be explained by theory~B
  (Pflamm-Altenburg, priv. comm.) but has been predicted by theory~C
  \cite{Pflamm09b}.
  Test {\bf VI}: Galaxies have an increasing metallicity with increasing
  mass which can only be explained with fine-tuned metallicity-dependent
  winds for theories~A and~B
  but follows naturally in the IGIMF theory \cite{Koeppen07, Recchi09}.
  Test {\bf VII}: Multi-photometric-band surveys of thousands of
  late-type galaxies by independently working teams have uncovered strong
  evidence of a systematic change of the galaxy-wide IMF with increasing
  galaxy mass and increasing SFR \cite{HG08, Lee09, Meurer09,
    Gun11}. Pioneering deductions of this had already
  brilliantly been achieved
  by Matteucci (1994, \cite{Matteucci94}) being confirmed by \cite{Thomas05}. 
  This is incompatible with stochastic/random sampling of an invariant
  IMF but has been predicted qualitatively by the IGIMF theory
  \cite{Weidner05} and has been quantitatively accounted for by the
  most modern version which includes top-heavy IMFs in massive star-bust clusters
  \cite{Weidner13b}.
}
\label{tab:SFTs}
\end{table}

\subsection{The main sequence of galaxies}
\label{sec:mainsequence}

That theory~C passes all tests leads to a simple and computable
understanding of star-formation on galaxy scales, being consistent
with the observational result that galaxies appear to be simple
objects as already concluded by Disney et al. from analyzing a major
observational sample \cite{Disney08}. A significant implication of the
above is that dwarf galaxies have a deficit of massive stars such that
the H$\alpha$ tracer leads to a wrong normalisation of the SFR if it
is assumed (wrongly) that the IMF remains fully sampled.  In truth,
dwarf galaxies may have SFRs up to two orders of magnitude larger than
the standard calibrations imply. Late-type dwarf galaxies may thus be
as efficient in forming stars as are major disk galaxies:

\vspace{2mm} \centerline{ \fbox{\parbox{\columnwidth}{ {\sc
        Important}: Theory~C (the IGIMF theory) thus leads to the
      result that the gas-depletion time-scale is $\tau_{\rm
        gas}\approx 3$~Gyr for all star-forming galaxies
      \cite{Pflamm09}.  }}} \vspace{2mm}

Since $\tau_{\rm gas}=M_{\rm gas}/SFR$ and for a gas fraction per
stellar mass, $\eta=M_{\rm gas}/M_*$, this result implies for the
specific SFR, sSFR,
\begin{equation}
{\rm sSFR}\; \equiv {SFR \over M_*} = {\eta \over \tau_{\rm gas}},
\end{equation}
such that sSFR$\approx 10^{-10.18}\,$yr$^{-1}$ for $\eta=0.20$. Thus,
by analyzing galaxies within the IGIMF theory it follows that the
galaxies are on a main sequence, as is observed
\cite{Guo13,Speagle14}.  Note that the assumption $\eta=0.2$ (20~\%
of the stellar mass) may not hold for late-type dwarf
galaxies. Dwarf galaxies are observed to have increasing $\eta$ with
decreasing stellar mass such that sSFR would be decreasing with
increasing galaxy stellar mass,
\begin{equation}
{\rm sSFR} = {\eta(M_*) \over \tau_{\rm gas}}.
\end{equation}
Applying the IGIMF theory to the observed fluxes from the dwarf
galaxies, Pflamm-Altenburg \& Kroupa (2009, \cite{Pflamm09}) show that
such dwarfs would have acquired up to 10 or more times the stellar
mass in late-type stars than an invariant standard IMF would imply,
such that the baryonic gas fractions, $f_g=M_{\rm gas}/(M_{\rm
  gas}+M_*)=\eta/(1+\eta)$, reported in \cite{McGaugh97} may need some
revision.

The IGIMF-analysis by \cite{Pflamm09} implies that late-type galaxies
may be replenishing their gas supply on this same time-scale,
$\tau_{\rm gas}$, and that the observed problem with stellar-mass
build-up time are solved within the IGIMF theory. This in turn has
implications for understanding the cosmological matter cycle, but is
not evident in the standard, stochastic descriptions based on
dark-matter merging trees.  

\vspace{2mm} \centerline{
  \fbox{\parbox{\columnwidth}{ {\sc In conclusion}: galaxies follow a
      simple gravitational law as well as a simple star-formation
      law. They appear to be largely self-regulated systems through
      the SFR-dependent IGIMF.  }}} \vspace{2mm} 

Self-regulated star formation in galaxies has been introduced by
Koeppen, Theis \& Hensler (1995, \cite{Koeppen95}), and emerges in
simulations of the smooth formation of Milky Way type galaxies
(Samland \& Gerhard, \cite{SG03}).  How self-regulation emerges within
the IGIMF theory, which leads to a stronger coupling between the SFR
and feedback energy, is studied with the pioneering work by
Pl\"ockinger et al. (\cite{Ploeckinger14, Ploeckinger15}), who
incorporated a SFR-dependent IGIMF formulation into gas-dynamical
simulations of galaxy formation with feedback. This work demonstrates
that dwarf galaxies can build-up a larger stellar mass than is
possible with a universal IMF. Previous models on TDGs applying the
IGIMF theory have been computed by Recchi et al. (2007,
\cite{Recchi07}).  Recchi et al. (2009, \cite{Recchi09} has shown that
the alpha-element abundances in early-type galaxies are better and
more naturally matched within the IGIMF theory, and that the need for
downsizing is relaxed somewhat. That the mass--metallicity relation of
galaxies emerges naturally within the IGIMF theory has been shown by
K\"oppen et al. (2007, \cite{Koeppen07}, see also
Fig.~\ref{fig:mZ}). The observational data thus appear to indicate
that self-regulated, SFR-dependent IGIMF description of galaxy
evolution may be more realistic than an invariant IMF. This is nicely
consistent with the independent evidence from galaxy surveys which
suggest a systematic change of the galaxy-wide IMF with SFR
(Sec.~\ref{sec:laws}, Tab.~\ref{tab:SFTs}).

Excursions from self-regulated evolution, e.g. through galaxy--galaxy
encounters, and a possible initial spread of galaxy-forma\-tion times
\cite{Speagle14} lead to a dispersion of properties about the main
sequence, which are easily accommodated with the SFR-dependent IGIMF
theory rather than being evidence for stochastic star formation. The
dispersion is indeed observed to be remarkably small \cite{Speagle14},
in agreement with the conclusions reached previously \cite{Disney08}.

\section{A conservative cosmological model?}
\label{sec:conserv}

If cold or warm dark matter particles do not exist, then the SMoC is
ruled out as a consequence. Does an alternative and realistic cosmological model
exist? 

A Milgromian-dynamics-based cosmological model has indeed been
computed. The CMB can be accounted for well, as shown in 2009 by Angus
(\cite{Angus09}, see also \cite{AD11}).  Structure formation remains a
computational challenge for this model which adopts a hot big bang and
Einstein's GR to describe the cosmological dynamics of the expansion
at all times as well as the growth of perturbations until
recombination. The model also requires dark energy and a hot dark
matter component in the form of about 11~eV sterile neutrinos
\cite{Angus13}, which can be linked to the SMoPP through neutrino
oscillations though. It has a flat geometry and is thus very similar
to the SMoC, except that the problems with cold or warm dark matter
discussed in the previous sections do not arise.  Computations of
structure formation in Milgromian dynamics are compromised not only by
the computational challenge posed by non-linear gravitational dynamics
but also by the matter being baryonic rather than non-interacting, as
is the case in the much easier SMoC computations of structure
formation dealing only with exotic dark matter and linear equations of
motions. Available results based on assuming hypothesis~0ii suggest
structures to grow more rapidly than in the SMoC \cite{Llinares08,
  Angus13}. Issues that arise with the CMB that have been pointed out
by \cite{Dodelson11} have thus partly or wholly been adressed (see
also the major review \cite{FM12}). The dynamics in galaxy clusters,
including the Bullet cluster, are accounted for well with the sterile
neutrinos making up the small amount of missing mass in Milgromian
dynamics \cite{Angus10}.

The following may be stated: the ``Angus cosmological model'' is much
more conservative than the SMoC because:
\begin{enumerate}

\item It does not change GR.

\item It does not invoke exotic particles.

\item The sterile neutrinos can be linked to the standard model via the neutrino oscillations.

\item Milgromian dynamics for the formation of structures (and
  dynamics thereof) can be viewed as simply arising as a vacuum effect
  (Sec.~\ref{sec:MD}), rather than changing GR. The modification would
  only be effective at low accelerations due to the presence of the
  vacuum acting as a lower inertial resistance at very low
  accelerations.

\end{enumerate}

The last point is the most contentious one, as it implies a particular
interpretation of Milgromian dynamics and SID as being due to
quantum-mechanical properties of the vacuum. This idea was proposed by
Milgrom (1999, \cite{Milgrom99}), but does not constitute a proper
theory with general covariance. It provides a direction and an
intuition, something that has been missing from theoretical physics
for quite some time. 

If Milgromian dynamics arises due to the vacuum, then how can one
understand the motion of a star in the SID regime while its
constituent matter, its atoms, are embedded in a strong gravitational
field?  If the underlying fundamental theory can be cast in terms of a
modified Poisson equation of Bekenstein \& Milgrom (1984, \cite{BM84},
see their section~IV for a discussion of the present issue) then the
centre of mass motion of a star can be shown to follow Milgromian
dynamics despite its interior being in a self-generated strong field.
This works because the force on a body (a star) made of smaller bodies
(atoms) can be written as an integral over any surface surrounding the
body (excluding all other bodies). One can take the integral on a
surface surrounding the star far from the star itself, and show that
it is proportional to the total mass of the star, as long as this mass
is small compared to that of the galaxy.  Note that exactly the same
already happens with GR in the strong field and in Newtonian
dynamics. Consider a binary BH orbiting a Newtonian galaxy.
Each BH is in a very strong field and calculating the
exact orbit of each black hole in the three-body system
galaxy---BH--BH is a complicated general relativistic calculation. But
the center of mass of the binary BH obeys Newtonian dynamics in the
field of the galaxy. This analogy is made clearer by replacing GR by
Newtonian dynamics and Newtonian dynamics by Milgromian dynamics
resulting in the same situation for a star and its atoms.  Returning
to the vacuum notion, each atom can be in a strong field and move
according to Newtonian dynamics (or GR) in the frame of the star, but
if the centre-of-mass of the star is in a weak gravitational field,
then the star orbits according to Milgromian dynamics. As this
discussion demonstrates, Milgromian dynamics provides a rich research
field for theoretical and mathematical physicists.

As a final note, it is to be pointed out that the existence of ``dark
energy'' could have been {\it predicted} by Milgrom by virtue of his
argument of 1999 \cite{Milgrom99}: By proposing that the transition
from the Newtonian regime to the SID regime in the weak gravitational
regime comes from the quantum mechanical properties of the vacuum, it
would have been possible to compute the energy density of the vacuum
given observational constraints on $a_0$ from the baryonic
Tully-Fisher relation or from rotation curves of galaxies. With this
vacuum energy in hand, it would have been possible to predict the
accelerated expansion of the Universe. Remember that the underlying model is
the very conservative Milgrom-based model by Angus \cite{AD11},
i.e. GR is not changed but SID arises as a vacuum effect
\cite{Milgrom99}. 

But how would the significant local matter under density
(Fig.~\ref{fig:underdens}) fit-in here? This would need to be
calculated along the principles given by e.g. Wiltshire
\cite{Wiltshire09}. But in the conservative model the under density
would only be associated with baryonic matter and not with a lack of
local dark matter, such that the effect on SNIa redshifts may be
reduced. That is, the vacuum would nevertheless possibly lead to
late-time acceleration.  As discussed above, the conservative model
appears to lead to more structure on large scales, and this may,
qualitatively at least, be consistent with the existence of
fluctuations in the matter density as is observed
locally. Nevertheless, the assumptions on homogeneity and isotropy
inherent in the Angus model do need to be analysed in future work. 

As a final comment, the inquisitive reader may be interested to
consider the discussion of the CMB in terms of its possible physical
properties and origin provided by \cite{Fahr09}.

\section{Predictions} 
\label{sec:predictions}

The contents of the text above suggest that the astrophysical
evolution of galaxies may require major revision. While simulations
need to be performed allowing for self-regulated galaxy evolution
within the SID and the IGIMF theoretical frameworks, a few predictions
can already be stated which ought to be open to experimental scrutiny.

Before turning to new predictions, it is worth remembering that once
exotic DM is abandoned, successful predictions seem to emerge. A
particularly noteworthy prediction of a Hercules-type dSph without DM
was obtained by modelling the long-term dynamical evolution of TDGs
(Kroupa 1997, \cite{Kroupa97}). This turned out to match observed dSph
and UFD satellit galaxies quite well \cite{MK07}.  The models (see
also \cite{KK98, Casas12, Yang14}) of ancient TDGs have shown that the
TDGs loose stars at every periastron passage until only about 1-10~\%
of the original population remains in a quasi-stable remnant with a
highly non-isotropic phase space distribution function. An observer
applying the usual models for virial systems then incorrectly deduces
large dynamical M/L ratios. The remnants are not satellite galaxies in
the process of disruption.

The Hercules satellite galaxy was discovered a decade later
\cite{Belokurov07} and has the near-exact morphological, luminous and
kinematical properties as the model of~1997 (fig.~6 in
\cite{Kroupa10}). Another case may be Bootes~III which also resembles
the TDG-remnant model. Grillmair (2009, \cite{Grillmair09}) discovered
this stellar overdensity to have a complex sub-structured morphology
(fig.~10 in \cite{Grillmair09}) which resembles the DM-free models of
ancient remnant TDGs. This is an important point valid for the other
satellite galaxies as well: Kroupa (2012, \cite{Kroupa12a}) stresses
that given the large velocity dispersion observed within the satellite
galaxies and their large distances from the MW, ``any internal
sub-structure would phase-mix away within 100~Myr. Given their ages of
about 10~Gyr, it follows that the satellites ought to appear as smooth
and symmetric as GCs.'' The measured velocity dispersions therefore
{\it cannot be due to the satellites being immersed within their own
  DM halos}.  Grillmair and \cite{Carlin09} suggest that Bootes~III
may be a remnant satellite with a long extension along the line of
sight, in agreement with the models of Kroupa (1997, fig.~12 in
\cite{Kroupa97} and figs.~16 and~18 in \cite{KK98}).

Furthermore, without exotic DM an entirely new dynamical deceleration
process for satellite galaxies emerges: discovered for the first time
by Yang et al. (2014, \cite{Yang14}), a gas-rich dwarf galaxy which
orbits past a major galaxy with coronal gas may have its ISM stripped
through ram-pressure. If the ISM is sufficiently heavy it's systematic
displacement behind the stellar component leads to ``{\it
  ram-friction}'' on the stellar component; the dwarf galaxy
decelerates and may be captured by the major galaxy despite the
absence of dynamical friction on a DM halo.  

These predictions and theoretical discoveries were obtained in
Newtonian dynamics, which is not consistent with abandoning
DM. However, they demonstrate above all that the large observationally
deduced dynamical M/L ratios of dSph and UFD satellite galaxies do not
unequivocally imply them to be DM dominated. For consistency with
abandoning DM, the above simulations need to be re-done in Milgromian
dynamics. This has not been possible until recently due to the lack of
appropriate Milgromian dynamics simulation codes.  Early results are
available suggesting improved fits of the Milgromian satellites with
the observed ones \cite{MW10}, although the details of the binary
populations within the satellite galaxies which affect the observed
velocity dispersions need to be taken into account \cite{Angus14b,
  Lueghausen14}. The POR code \cite{Lueghausen14b} will become an
important simulation tool for this purpose, but it is already apparent
that the EFE (Sec.~\ref{sec:galdyn}) leads to the satellite loosing
its phantom dark matter halo as it approaches periastron such that it
is likely to be disrupted more easily than a Newtonian system,
although the phantom dark matter halo reappears as the satellite
orbits past periastron and towards apoastron.  The EFE may thus lead
to a critical distance from the MW within which dwarf satellite
galaxies are disrupted. It is observed that the satellite population
appears to largely extend beyond about 50~kpc suggesting this to be
the critical distance for the MW, but quantitative work needs to be
done to asses the impact of the richness of Milgromian dynamics on the
satellite population.

Other remarkable predictions concern galactic rotation curves:
returning to the Gedankenexperiment ``Challenge for Theory'' in
Sec.~\ref{sec:SMoC}, it is true that the rotation curve of every
late-type galaxy for which the spatial baryonic matter content is
known, Milgromian dynamics allows the rotation curve to be predicted
without a free parameter.

\subsection{Prediction: the number of MW satellite galaxies}
\label{sec:pred_NrSat}

There is today no doubt any longer that the VPOS is established to be
a physical property of the MW galaxy (Sec.~\ref{sec:anisotropy}). It
will nevertheless be necessary to search for more satellite galaxies
even though finding fainter members will not affect the fact that the
at present brighter ones are distributed within a mass-segregated VPOS
\cite{Kroupa12a}. If the VPOS is an ancient structure such that it is
phase-mixed then the number of satellite galaxies ought to be
symmetrically distributed about the MW.  The north Galactic hemisphere
is known today to have in total~19 dSph and UFD satellite galaxies
(Fig.~\ref{fig:VPOS}, this includes Crater or PSO~J174.0675-10.8774).
Until the present day no additional satellites have been found
significantly off the VPOS.  If the population of satellites
northwards of the Galactic disk is complete, then the southern
Galactic hemisphere should have in total about 19 satellites as
well. Nine are already known (Fig.~\ref{fig:VPOS}). This leads to the
prediction made here for new satellites with luminosities in the range
of the existing satellites \cite{McConnachie12}:

\begin{description}
\item Assuming the VPOS to be a phase-mixed and thus ancient
  structure, about~10 additional satellite galaxies should be found in
  the southern Galactic hemisphere. 
\end{description}

\noindent {\it Alternatively} hundreds of dark-matter ultra-faint
satellite galaxies have been predicted within the framework of the
SMoC to surround the MW spheroidally (\cite{Koposov09, Tollerud08},
Sec.~\ref{sec:PDGs}).

However, even in the unlikely event that this were to be the case, the
observed fact is that the VPOS is made of the brightest satellite
galaxies such that the models based on dark matter would have to
account for why the most luminous satellites would be in a disk of
satellites (Sec.~\ref{sec:anisotropy}). None of the existing research
papers working within the SMoC framework address this issue (e.g. the
most recent land-mark simulation of the galaxy population by
Vogelsberger et al. (2014, \cite{Vogelsberger14}).

\subsection{Prediction: proper motions of MW satellite galaxies}
\label{sec:pred_pm}

Of the known 28~satellite galaxies that are found within about 270~kpc
of the centre of the MW,~11 have proper motion measurements. Of
these,~9 are found to be orbiting within the VPOS within the
uncertainties. Leo~I is on a very radial orbit, and Sgr may have been
diverted onto an energetic orbit \cite{Zhao98} and it is now on a
near-perpendicular orbit to both the MW disk and to the VPOS
(Fig.~\ref{fig:VPOS}). This state-of-affairs allows the proper motions
of the remaining known satellite galaxies to be predicted:

\begin{description}
\item Assuming the satellite galaxies to orbit within the VPOS, the
  proper motions of almost all satellite galaxies of the MW can be
  predicted. These predictions are available in \cite{Pawlowski13}.
\end{description}

\noindent {\it Alternatively} the satellite galaxies that do not have
measured proper motions would have random motions.  

However, even in the unlikely event that this were to be the case, the
observed fact is that, with the exception of Sgr and Leo~I, the
brightest satellites have velocity vectors within the VPOS. The models
based on dark matter would have to account for why the fainter
satellites are in the VPOS today while they would have motions that
take them out of the VPOS on a time scale of $\Delta
\tau\approx\,$thickness/velocity dispersion $\approx50\,{\rm
  kpc}/$ $300{\rm pc/Myr} = 167\,$Myr, which is significantly shorter
than the age of the system and than the orbital time scale.

\subsection{Prediction: frequency of anisotropic dwarf satellite systems}
\label{sec:pred_frequency}
In Sec.~\ref{sec:anisotropy} it has been found that the incidence of
anisotropic satellite systems around host galaxies that have been well
studied appears to be typical. In fact, nine such systems are already
known, and in the Local Group, where the best 3D data with kinematical
information is available, disk-of-satellites are the rule.  This
suggests that most satellite galaxies may be TDGs. This leads to the
following prediction:

\begin{description}
\item If most dwarf satellite galaxies are TDGs then anisotropic and
  rotating flattened satellite populations should be common, as long
  as encounters between major galaxies are typically less frequent
  than twice per Hubble time (Sec.~\ref{sec:selfcons}).
\end{description}

\noindent {\it Alternatively} the known host galaxies with correlated
satellite populations are exceptions. 

But this is unlikely because, as emphasized by Chiboucas et al. (2013,
\cite{Chiboucas13}), in every case for which data exist so far,
anisotropic and flattened satellite populations have been found (see
Sec.~\ref{sec:anisotropy}).

\subsection{Prediction: satellite galaxies and galactic bulges}
\label{sec:pred_NsatBulge}

\begin{figure}
\centering{}\includegraphics[width=8cm]{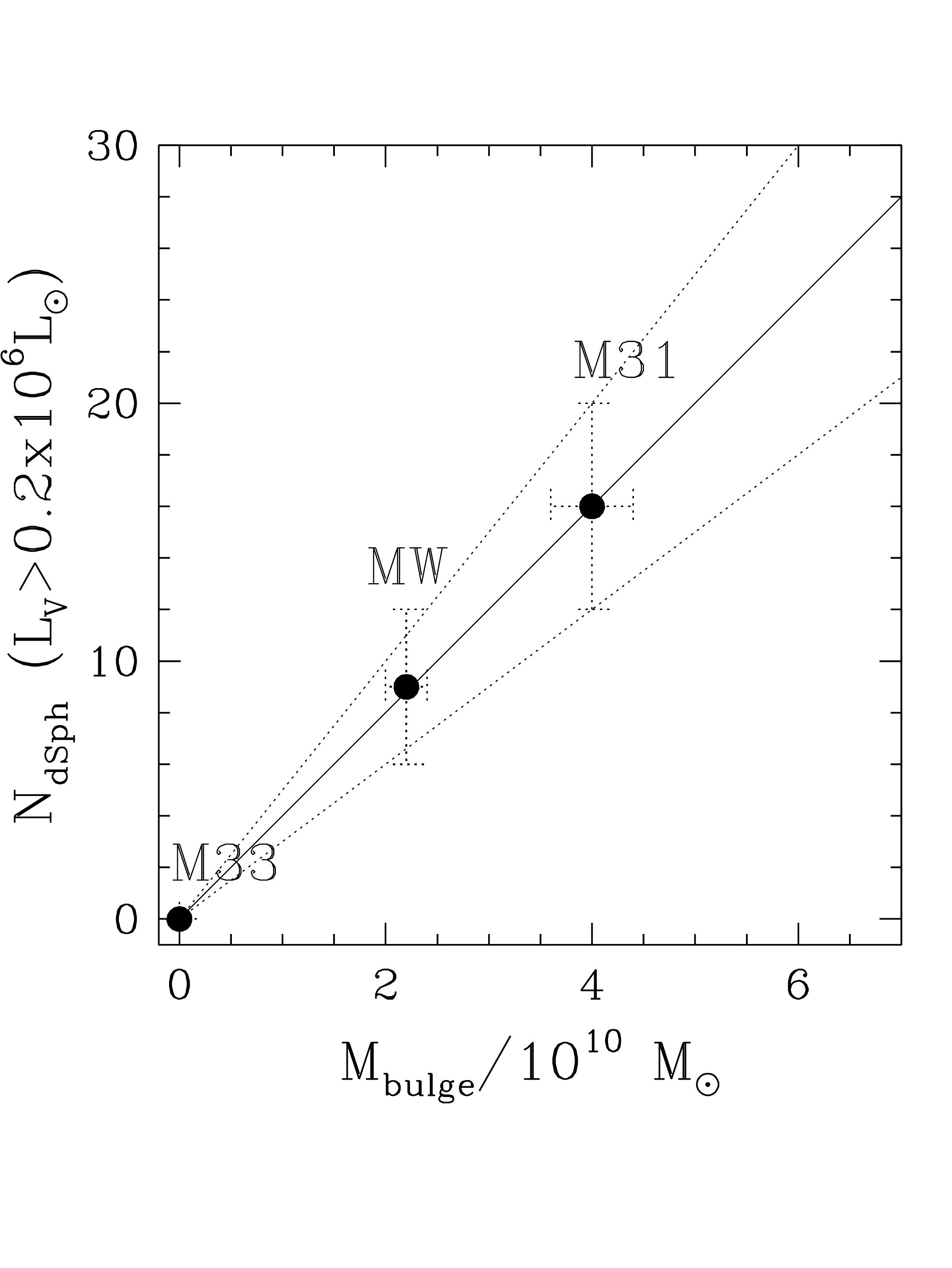}
\vspace{-15mm}
\caption{\small The number of dSph and dE satellite galaxies, $N_{\rm
    sat}$, each more luminous than $0.2 \times 10^6\, L_\odot$, is
  plotted versus the bulge mass, $M_{\rm bulge}$ of the host
  galaxy. Only satellites within a distance of 270 kpc of the MW and M
  31 are used. The solid line is $N_{\rm sat} =(4.03\pm 0.04) \times
  M_{\rm bulge} /(10^{10}\,M_\odot)$ (eq.~17 in \cite{Kroupa10}). See
  \cite{Kroupa10} for more details. Figure reprinted from
  \cite{Kroupa10}.  Reproduced with permission from Astronomy \&
  Astrophysics, \copyright ESO.  }
\label{fig:bulgeNrSat}
\end{figure}

If the majority of dwarf satellite galaxies are TDGs, then the
formation of satellite systems is associated with galaxy--galaxy
encounters. The formation of galactic bulges is also associated with
galaxy--galaxy encounters because the torques in such encounters
induce radial gas motions which can build-up a central spheroidal
component on a time scale of 100s of Myr. Thus, a correlation between
the mass of the bulge and the number of satellite galaxies is expected
in this scenario. In the Local Group such a correlation exists
(Fig.~\ref{fig:bulgeNrSat}). 
It is noteworthy that the MW has a bulge which appears to have formed
within less than a~Gyr \cite{Ballero07a, Ballero07b} about 10~Gyr ago
and also before the thin disk of the MW. Together with the existence
of an older thick disk this may be the smoking gun of a major
encounter the young MW may have had about~9 to 11~Gyr ago. This is a
viable solution to the origin of the Local Group \cite{Zhao13}. Note
that the bulge of the MW may be a pseudo bulge formed from the early
pre-thick disk structure as a result of a bar instability induced by
the encounter. 

The following prediction emerges \cite{Kroupa10}:
\begin{description}
\item If most dwarf satellite galaxies are TDGs then their number
  around a host galaxy should correlate positively with the mass of
  the bulge of the host galaxy.
\end{description}

\noindent {\it Alternatively}, in the SMoC not the bulge mass but the
circular velocity should correlate with the number of satellite
galaxies because heavier dark matter halos have a larger number of
sub-halos within them \cite{Ishiyama13}.

This is unlikely to be the case because heavier dark matter halos are
more likely to have had major mergers and are thus less likely to have
major disk galaxies without classical bulges. But the majority of disk
galaxies do not have a classical bulge \cite{Weinzirl09, Kormendy10},
while 94~\% of isolated disk galaxies which typically must also have
grown via mergers if the SMoC is to be valid do not appear to have
classical bulges \cite{Fernandez14}.

\subsection{Prediction: Hickson compact groups}
\label{sec:pred_Hickson}

Given that the existence of exotic dark matter particles appears to be
inconsistent with the data, it follows that dynamical friction would
not be a significant process for the evolution of galaxy pairs or
groups of galaxies (Sec.~\ref{sec:testdynfr}). This allows to be
tested by the following experiment suggested by
\cite{Pompei03,Kroupa10, Kroupa12a}:

\begin{description}
\item Hickson compact groups of galaxies should not merge within a
  crossing time (see also the M81 case in Sec.~\ref{sec:M81}). Thus, a
  survey of such groups about 1~Gyr in the past, i.e. at a redshift of
  about~0.1, ought to show about the same abundance of such groups if
  there is no dark matter.
\end{description}

\noindent {\it Alternatively}, Hickson compact groups of galaxies
should merge within a crossing time (see also the M81 case in
Sec.~\ref{sec:M81}).  A survey of such groups about 1~Gyr in the past
ought to show many more such groups if the SMoC were true. The
remnants of merged Hickson compact groups should be evident as
elliptical galaxies with significant young stellar populations.  This
would require Hickson compact groups to constantly reform from the
field population of galaxies.

However, no prediction to this effect exists, and this seems already
to be ruled out by the observation that the population of early-type
galaxies does not evolve since the last 6~Gyr \cite{Delgado10}. The
observational constraints provided by \cite{Pompei10} do not appear to
support the notion that Hickson compact groups merge. Indeed,
\cite{Pompei03} write ``we can conclude that SCGs\footnote{SCG,
  southern compact group.} are mostly gravitationally bound structures
with a longer lifetime than that predicted by current numerical
simulations.''

As a cautionary note: there is evidence for young and inter\-mediate-age
stellar populations in some elliptical galaxies (e.g. \cite{Duc11}),
and this is often interpreted to be due to mergers. While some mergers
are likely to occur, due to Chandrasekhar dynamical friction on the
extended stellar components and the ejection of orbital angular
momentum and binding energy with massive tidal tails, mergers are much
less frequent in Milgromian dynamics due to the absence of dynamical
friction on the expansive dark matter halos. This have been studied by
Tiret \& Combes (2008, \cite{Tiret08}).  The observed populations of
young or intermediate-age stars in some E galaxies may also be formed
when a pre-existing E galaxy interacts with a gas-rich disk galaxy
without merging. Encounters involving a wet galaxy can lead to one
of the gaseous tidal arms being partially accreted where it can form a
new generation of stars in the other galaxy. That tidal accretion is
likely the dominant physical mechanism for making polar ring galaxies
has been shown by Bournaud \& Combes (2003, \cite{Bournaud03}; see
also \cite{Spavone11, Pawlowski11}).

\subsection{Prediction: TDGs and the BTFR}
\label{sec:TDGsBTFR}

If SID/Milgromian dynamics is the correct effective description of
galaxies, then a trivial prediction follows:

\begin{description}
\item Virialised TDGs must lie on the BTFR (Fig.~\ref{fig:BTFR}),
  subject to the EFE.
\end{description}

\noindent{\it Alternatively}, if the arguments presented in this
contribution are incorrect and if the SMoC were valid, then virialised
isolated TDGs cannot lie on the BTFR but must be off set from it
towards significantly smaller velocities.

\subsection{Prediction: sizes of satellite galaxies and galaxies in clusters}
\label{sec:pred_sizes}
The exposition of SID in Sec.~\ref{sec:sid} led to the possible
existence of the ``{\it external field effect}''. If this effect is
true then the following prediction ensues:

\begin{description}
\item Elliptical galaxies in galaxy clusters and spheroidal satellite
  dwarf galaxies should be physically larger than their isolated
  counterparts due to the external field effect
  (Sec.~\ref{sec:testExternalF}).
\end{description}

\noindent{\it Alternatively}, if dwarf galaxies are in their own dark
matter halos then the size of their baryonic component is typically
unaffected through galactic tides and through the stripping of their
dark matter halos even after the tidal stripping of over 99~\%
of the initial dark halo mass \cite{Kazantzidis04}.

\subsection{Prediction: number of massive stars in dwarf galaxies}
\label{sec:pred_NOst}

The rate with which galaxies accrete baryons can be inferred from the
rate with which they are forming stars. The SFR is obtained from
observations but depends on assuming whether the galaxy-wide IMF is an invariant
stochastically sampled probability density distribution function, or
whether it is a computable systematically varying IGIMF
(Sec.~\ref{sec:buildingblocks}). The various possibilities can be
distinguished with the following star-count test based on the IGIMF prediction:

\begin{description}
\item If the IGIMF theory is correct then late-type dwarf galaxies
  with SFR$<10^{-2}\,M_\odot$yr$^{-1}$ have on average a significantly
  smaller number of~O stars than expected for an invariant
  stochastically sampled galaxy-wide IMF \cite{Weidner13b}.
\end{description}

\section{Conclusions}
\label{sec:concs}

\begin{figure}
\hspace{-6mm}
\includegraphics[width=7cm, angle=90]{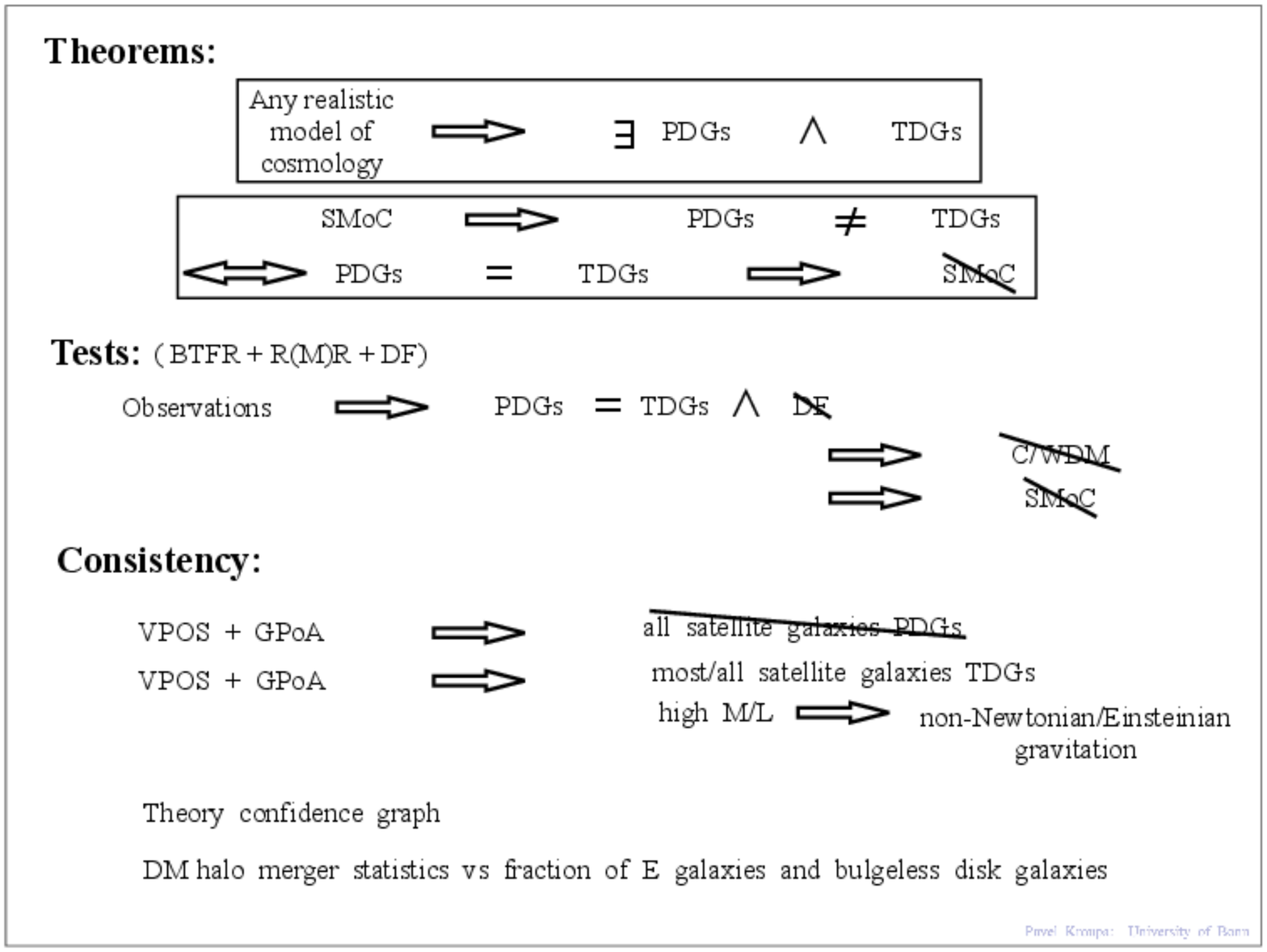}
\caption{\small Structure of the falsification of dynamically relevant
  cold or warm dark matter. The theorems are discussed in
  Sec.~\ref{sec:dual} and~\ref{sec:testdual}. The tests using the
  baryonic Tully Fisher relation (BTFR), the radius-mass relation
  (R(M)R) of pressure supported dwarf galaxies and using dynamical
  friction (DF) are performed in Secs.~\ref{sec:BTFR},~\ref{sec:R(M)R}
  and~\ref{sec:testdynfr}, respectively. The consistency checks based
  on the vast polar structure or the Milky Way (VPOS) and on the great
  plane of Andromeda (GPoA) are in Sec.~\ref{sec:anisotropy}. The
  theory confidence graph is discussed in
  Sec.~\ref{sec:confidencegraph} and the evidence for a lack of mergers
  in the observed galaxy population is covered in
  Sec.~\ref{sec:mergers}.  Other acronyms: PDGs=primordial dwarf
  galaxies (Sec.~\ref{sec:PDGs}) TDGs=tidal dwarf galaxies
  (Sec.~\ref{sec:tdgs}), C/WDM=cold or warm dark matter.  }
\label{fig:flogics}
\end{figure}

Based on assuming Einstein's field equations are valid in an
extrapolation by many orders of magnitude in acceleration and spatial
scale to galaxies and beyond, significant departures of observed
motions from the expectations are found. The additional hypothesis
that this indicates the presence of exotic dark matter particles is
currently very popular and forms the basis of the standard model of
cosmology. This SMoC is held by many to be an excellent account of the
Universe, although many tensions between data and model have been and
are being reported. Applying the dual dwarf galaxy theorem to the SMoC
suggests it to be falsified: the available observational data do not
allow a distinction to be found between the dynamical properties of
primordial dwarf galaxies and tidal dwarf galaxies, although they must
differ if the SMoC is true.  At the same time the arrangements of
satellite dwarf galaxies in the VPOS around the MW and in the
GPoA/VTDS around Andromeda and other anisotropic dwarf galaxy
distributions are strongly suggestive of them being TDGs; the
predicted large numbers of independently or group accreted, DM
dominated spheroidally distributed dwarfs are most inconspicuously
absent.  Confident evidence for dynamical friction is found neither in
interacting galaxies nor in the motions of satellite galaxies. The
overall galaxy population is best explained with dynamical friction
being suppressed. The most notorios case is the galaxy group of M81
(Sec.~\ref{sec:M81}). Two independent groups have been unable to find
solutions for the matter distribution within the M81 system as a
result of the interactions between group members, assuming the SMoC to
be valid. But solutions do exist for the case without dynamical
friction.  That the SMoC may not be the correct description of the
Universe finds consistency in the documented history of many failures
of the SMoC in accounting for observational data, as can be visualized
with the theory-confidence graph
(Sec.~\ref{sec:confidencegraph}). None of these failures has been
solved to-date.  The structure of the exotic-dark-matter-falsification
argument is depicted in compact form in Fig.~\ref{fig:flogics}.

The conclusion from these arguments based only on the available
astronomical data, namely that exotic dark matter particles that are
dynamically relevant on galaxy scales do not exist, would have deep
consequences for current physics and uncomfortable sociological
implications, but it is the reality that needs to be faced. This
astronomical falsification of the cosmological concept of dark matter
is echoed in the present-day null-evidence for physics beyond the
SMoPP, despite searches at the highest available energies. The
community had hoped to explain a number of issues in theoretical
physics with physics beyond the SMoPP, but the null-results on
evidence for such physics such as for supersymmetry is now putting
theoretical approaches under crisis \cite{Shifman12}.  

\vspace{2mm} \centerline{ \fbox{\parbox{\columnwidth}{ Thus, neither
      is there direct evidence for the existence of exotic dark matter
      particles, nor has experimental evidence for extensions of the
      SMoPP, which accommodate possible candidates for dark matter
      particles, been found.  }}} \vspace{2mm}

As pointed out by Milgrom \cite{Milgrom09}, gravitation on the scale
of galaxies and beyond follows a very simple law which is referred to
as scale-invariant dynamics. This SID leads to a number of interesting
consequences, such as the baryonic Tully Fisher relation and possibly
physically larger galaxies when they are subject to an external field
from a nearby host. The tight correlation of observational data on the
BTFR, which cannot be understood in terms of dark matter in the SMoC
because the circular velocity of the dark matter halo is not well
correlated with the baryonic galaxy within it \cite{Lu12}, is a
necessary consequence of this law of nature which all galaxies must
abide by.  The originally voiced need for a dark matter component to
ensure disk stability by Ostriker \& Peebles (1973, \cite{OP73}) and
the requirement that this component be fine tuned with the baryonic
one (Bahcall \& Casertano 1985, \cite{Bahcall85}) are both immediately
and naturally solved by SID/Milgromian dynamics.  That TDGs have been
found to lie near the BTFR finds a trivial explanation in terms of
SID.  In SID, and for the same environmental conditions, primordial
dwarf galaxies and virialised TDGs {\it must} look alike dynamically
and morphologically, because they are required to follow this law of
nature.  Also, in SID pressure supported primordial dwarf galaxies and
old TDGs must look alike, since there is no physical distinction
between them apart from their epoch of formation. In SID there is
therefore no tension between the dual dwarf galaxy theorem and
observational data.  In SID galaxies do not merge readily when they
interact, because dynamical friction on dark matter halos does not
occur. The present-day understanding of the dynamical behavior and of
the evolution of galaxies needs to be revised because the dynamics of
exotic dark matter particles do not play a role.  Returning to the
Gedankenexperiment ``Challenge for Theory'' in Sec.~\ref{sec:SMoC}, it
is by now clear that the rotation curve of the galaxy cannot be
predicted within the SMoC, but it has a unique prediction in SID.

\begin{figure}
\centering{}\includegraphics[width=8.5cm, angle=0]{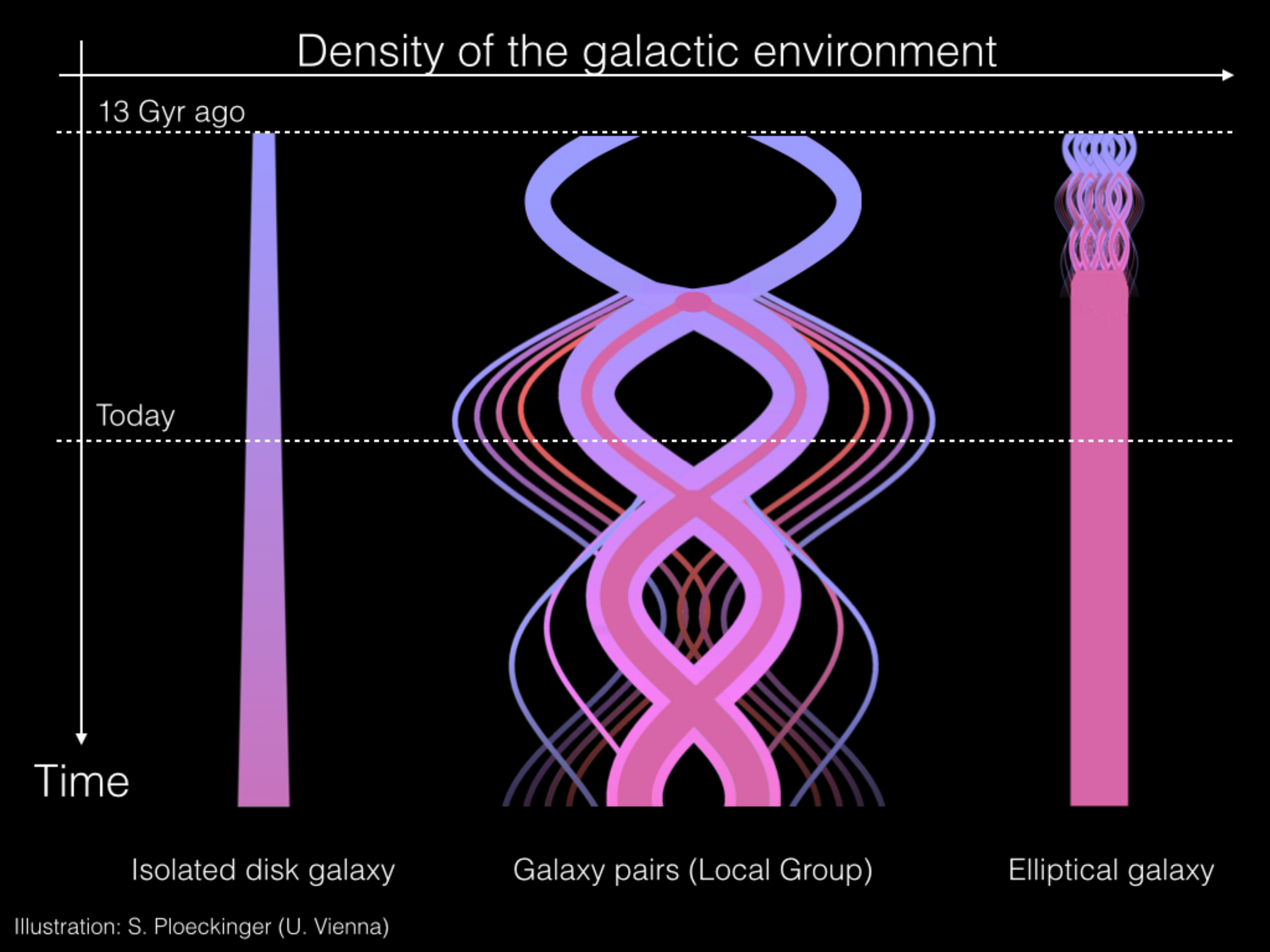}
\caption{\small Cartoon of galaxy evolution in SID in a universe
  without cold or warm dark matter (not to scale). The vast majority
  of galaxies with baryonic mass $M>10^{10}\,M_\odot$ are late-type
  since about 6~Gyr ago (about~97~\%, \cite{Delgado10}, fig.~4.14 in
  \cite{BM98}) and mergers do not play a significant role due to the
  absence of dynamical friction \cite{Tiret08, TC08}. The left
  vertical blue track represents a growing largely non-interacting
  ``island-universe'' galaxy \cite{SG03,Disney08,Speagle14}. It
  reddens with time as the old stellar population builds up. It does
  not attain a classical bulge, but may appear as a chain galaxy
  \cite{Elmegreen04a, Elmegreen04b} if it experiences a disk-wide
  star-burst about 11~Gyr ago, therewith building up a thick disk
  \cite{Kroupa02}.  The central two tracks depict a Local Group with
  the motion of the MW and Andromeda around each other once they
  decouple from the cosmic expansion \cite{Zhao13}. They spawn a new
  generation of TDGs at each encounter. These TDGs typically evolve to
  the present-day ($t=\,$today) faint red dSph and UFD satellite
  galaxies with bolometric luminosities $<10^8\,L_\odot$ but remain
  dIrr galaxies at larger distances from their hosts \cite{Hunter00}.
  After each encounter the two galaxies obtain an increased
  bulge. A bulge-mass--satellite-number correlation arises
  (Fig.~\ref{fig:bulgeNrSat}).  The two major galaxies redden with
  time as the isolated galaxy on the left.  This model of the Local
  Group would have looked as a peculiar galaxy about 11~Gyr ago when
  the early MW and Andromeda interacted.  A minority of galaxies with
  $M> 10^{10}\,M_\odot$ (about~3~\%, \cite{Delgado10}, fig.~4.14 in
  \cite{BM98}) form in dense environments very early on through rapid
  accretion and gas-dominated mergers to become elliptical galaxies
  which evolve passively without significant later evolution. This had
  been deduced already by Matteucci (1994, \cite{Matteucci94}) and is
  depicted in the right-most track. These may have satellite galaxies
  which formed as TDGs during the early gaseous mergers. The observed
  size-evolution of these galaxies may be related to the external
  field effect in SID, but this has not been studied to
  date. Downsizing is accounted for in this cartoon. This cartoon
  supersedes the cosmological merger tree and fig.~9 in
  \cite{Kroupa10}. This figure was kindly provided by S. Pl\"ockinger}
\label{fig:galevol}
\end{figure}

This revision is associated with a more thorough understanding of
stellar populations in galaxies and the star-formation-rate dependent
IGIMF, which is leading to the insight that the chemo-dynamical
evolution of galaxies may be more self-\-regulated and thus simpler
than appreciated hitherto.  In a major study of the evolution of the
main-sequence of galaxies over cosmic time, Speagle et
al. (2014. \cite{Speagle14}) re-emphasize the major tension between
the theoretical hierarchical-buildup of stellar mass in the SMoC in
comparison with the observed rather simple self-similar evolution of
galaxies, confirming the conclusions reached by Disney et al. (2008,
\cite{Disney08}).  Galaxy evolution appears to be fundamentally
self-regulated and not a stochastic processes, such that
dynamical-friction-driven mergers play a small if not an insignificant
role.  Cold or warm dark matter halos therefore do not appear to
exist. Star-formation appears to be self-regulated, even on the
smallest sub-pc scales, and follows well defined laws such that
composite stellar populations, such as what galaxies are, become
computable and predictable over cosmic time via the IGIMF theory and
related concepts (e.g. dynamical population synthesis of binary star
populations).  Gauging the few parameters that enter the IGIMF theory
using local observationally well resolved star formation, it follows
that the galaxy-wide IMF (the IGIMF) changes systematically from
top-light to top-heavy with increasing galaxy-wide star-formation
rate. Late-type dwarf galaxies thus have a deficit of massive stars
when compared to an assumed invariant fully sampled IMF while massive
late-type galaxies have a top-heavy IMF. It emerges that there is no
mystery in how the IMF is to know what type of galaxy it is forming
in. Rather, the galaxy-wide IMF, i.e. the IGIMF, follows naturally
from the simple addition of all star-forming events that are occurring
in a galaxy which are distributed according to distribution functions
that are subject to physical constraints, the most important one being
the amount of mass being converted into the stellar masses of embedded
clusters over about ten Myr timescales.  The star-formation behavior
of late type galaxies (the vastly dominant type of galaxy) yields the
galaxy main sequence. It is consistent with the IGIMF theory,
confirming the self-regulated evolution of galaxies rather than the
haphazard processes associated with the SMoC.  This advanced concept
of galaxy evolution, which does not have an underlying merger tree
\cite{LC93}, is shown schematically in Fig.~\ref{fig:galevol}.

The transition of SID to the Newtonian regime around gravitating
masses may be derivable from the physical processes of the vacuum, as
suggested by Milgrom (1999, \cite{Milgrom99}, see Sec.~\ref{sec:MD}
and~\ref{sec:conserv}). The complete description in the classical
non-relativistic re\-gime, SID plus Newtonian dynamics, is referred to
here as Milgromian dynamics. This framework accounts well for the
observed internal properties of the satellite galaxies of Andromeda
and the MW (e.g. \cite{MW10,MM13}).  It naturally leads to the
kinematical properties of polar ring galaxies \cite{Lueghausen13}, and
the anisotropic distribution of satellite galaxies can also be
accounted for well in Milgromian dynamics
\cite{Kroupa12b}. Furthermore, the hitherto not expected highly
symmetrical structure of the Local Group \cite{Pawlowski13b} may be a
consequence of an early encounter between the MW and Andromeda
\cite{Zhao13}, which is not possible in the SMoC because the two
galaxies would have merged due to dynamical friction on their dark
matter halos.  Interpreting Milgromian dynamics as Einsteinian
dynamics in the classical regime plus quantum-mechanical effects from
the vacuum which become effective in the weak gravitational regime, a
conservative cosmological model emerges which appears to be broadly
consistent with available observational data on large and small scales
(Sec.~\ref{sec:conserv}). Generalizations of Einstein's field equation
(Sec.~\ref{sec:SMoC}) may explicitly incorporate the physics of the
vacuum, or they may have Milgromian gravitation as the correct
classical limit.

The conclusion that exotic dark matter does not exist seems to be
inescapable, if the data used for the tests remain valid and/or are
verified, and thus new avenues for understanding gravitation in the
weak field (acceleration $\,< a_0\approx3.8\,$ pc/\-Myr$^2$) regime
are encouraged. SID appears to be the correct effective description of
the observed dynamics in this regime. An excellent in-depth review of
various approaches is provided by \cite{FM12}. Some ideas can be found
for example in \cite{Blanchet09, Blanchet14} on an approach for
understanding exotic dark matter and dark energy in terms of
gravitational polarisation leading to SMoC-properties on large scales
while recovering the correct SID-behavior on galactic scales, in
\cite{Chua13} on mirror dark matter, in \cite{Moffat06} on modified
gravity, in \cite{Trippe13, Trippe13b} on an approach with massive
gravitons, in \cite{Verlinde11} on interpreting gravitation as an
emergent property through a holographic scenario, and in
\cite{Hehl09a, Hehl09b, RM14} on a nonlocal generalization of
Einstein's theory of gravitation in which a Yukawa-type repulsion
appears and which reproduces galactic rotation curves and galaxy
clusters without exotic dark matter.  \cite{Trippe14} gives an
independent review and account of the issues covered here with a brief
discussion of various approaches to non-standard models such as $f(R)$
gravity. Other independent reviews of related issues can be found in
\cite{Scarpa06, Sanders07, Hernandez14}.

As a final note, the scientific community has put an incredible amount
of effort to support and develop technologies to search for exotic
dark matter and to perform computer simulations within the SMoC. There
are nearly countless major research groups and centers world wide
performing such simulations of the Universe. Attempts at establishing
funding for even a small {\it independent} group in which numerical
tools would be developed to perform similar simulations within the SID
context have been rejected, on multiple occasions. In Bonn the POR
code is being developed (L\"ughausen, Famaey \& Kroupa 2014,
\cite{Lueghausen14b}), on a small budget and thanks to local support
from the university's rectorate. This code will allow simulations of
galaxy formation and evolution within a Milgromian universe and is to
become publicly available, but the future of this and similar projects
is unclear.  Cold or warm dark matter has not been found to this date,
and if it does not exist, as this contribution argues based on
astronomical data, then none of the simulations within the SMoC would
represent reality. In a more realistic cosmological model the CMB may
require a new interpretation \cite{Fahr09} and, independently of this,
distance--redshift--age relations may be different to those in the
SMoC:

\vspace{2mm} \centerline{ \fbox{\parbox{\columnwidth}{ {\sc The future
        of cosmology}: It follows that, without exotic cold or warm
      dark matter and with a 
      systematically varying IGIMF with SFR, {\it all} observational
      quantities that are derived at present, such as star-formation
      rate densities, distances and ages from redshifts, and galaxy
      masses, are likely to require possibly major revision.  }}}
\vspace{2mm}

\section*{Acknowledgements}

I thank the staff at the Institute for Astrophysics at the University
of Vienna, and especially G. Hensler, for the hospitality during my
guest professorship in March and April~2014.  Much of this text was
written in Vienna. M.~Pawlowski, S.~Pl\"ockinger, S.~Recchi, X.~Wu and
I.~Stewart are thanked for helpful comments which improved this
contribution. Benoit Famaey is than\-ked especially for providing me
with important insights into the fundamental aspects of Milgromian
dynamics.  I also thank the staff at the Institute for Advanced Study
(IAS) in Princeton for hosting me as a Visitor in April 2014, the
place where Milgrom discovered his dynamics in the
1980s. Sec.~\ref{sec:conserv} is a result of the discussions at the
IAS. At the University of Bonn I would like to extend my gratitude to
the Rektor, J\"urgen Fohrmann, the Dean, Ulf Meissner, and J\"urgen
Kerp and Cristiano Porciani for playing helpful and decisive roles
during the past 12 months.

\bibliographystyle{unsrt}
\bibliography{/Users/pavel/PAPERS/BIBL_REFERENCES/kroupa_ref}

\appendix
\section{Acronyms}
\label{sec:app}

This appendix lists the used acronyms, their meaning and first-time
occurrence in the text.

\begin{center}
\begin{tabular}{l p{5.5cm} l}
Acronym   & Meaning    &Sec.\\
\hline

BH &black hole & \ref{sec:SMoC}\\

BTFR &baryonic Tully–Fisher relation &\ref{sec:BTFR}\\

CDM & cold dark matter &\ref{sec:introd}\\

CMB &cosmic microwave background  & \ref{sec:SMoC}\\

DDG &dual dwarf galaxy theorem & \ref{sec:dual}\\

dE &dwarf elliptical galaxy & \ref{sec:tdgs}\\

DoS   &disk of satellites of the Galaxy; it contains all known satellite
galaxies of the Galaxy  &  \ref{sec:other} \\

DM & exotic (cold or warm) dark matter particles; not contained in the
SMoPP &\ref{sec:introd}\\

dSph &dwarf spheroidal galaxy; baryonic mass $10^5\simless M/M_\odot
\simless 10^7$; known as common satellites around nearby larger
galaxies; MW dSph satellites were discovered with photographic plates
and are the ``classical'' satellite galaxies& \ref{sec:PDGs}\\

E & elliptical  &\ref{sec:SMoC}\\

EC      &embedded cluster  &\ref{sec:selfreg} \\

ECMF   &embedded cluster mass function  & \ref{sec:selfreg}\\

EFE &external field effect & \ref{sec:galdyn} \\

GC & globular star cluster & \ref{sec:MW} \\

GPoA   &great plane of Andromeda, sometimes also called the VTDS  & \ref{sec:GPoA}\\

GR &Einstein's theory of general relativity & \ref{sec:SMoC} \\

IGIMF  &integrated galactic IMF; the IMF of a whole galaxy & \ref{sec:m-Z}\\

IMF  &initial mass function & \ref{sec:m-Z}\\

ISM &inter stellar medium & \ref{sec:confidencegraph}\\

$\Lambda$CDM &Lambda ($\Lambda$) cold dark matter cosmological model; the SMoC
& \ref{sec:SMoC}\\

LMC  & Large Magellanic Cloud & \ref{sec:conflicts}\\

LMXB &low-mass X-ray binary  & \ref{sec:E_IMF}\\

$\Lambda$WDM &Lambda ($\Lambda$) warm dark matter cosmological model; here
also referred to as the SMoC as it is virtually identical to the LCDM
model apart from the exotic DM particles having a longer streaming
length initially &~\ref{sec:SMoC}\\

MDA &mass-discrepancy-acceleration &\ref{sec:sid}\\

MOND & modified Newtonian dynamics; Milgrom's effective 
theory of classical gravitational dynamics & \ref{sec:MD}\\
 
MW   &Milky Way, the Galaxy  & \ref{sec:PDGs}\\

M31  &Andromeda galaxy & \ref{sec:m-Z} \\

M33  &Triangulum & \ref{sec:testMD}\\

PDF & probability density distribution function &\ref{sec:selfreg} \\

PDG &primordial dwarf galaxy; dominated by DM if the SMoC is true & \ref{sec:PDGs}\\

POR &Phantom of Ramses simulation code\cite{Lueghausen14b} & \ref{sec:MD}\\

R(M)R &radius-mass relation &\ref{sec:R(M)R}\\

SFR  &star formation rate  &\ref{sec:mergers} \\

Sgr &Sagittarius &\ref{sec:sgr}\\

SID &scale-invariant dynamics & \ref{sec:sid}\\

SMC  &Small Magellanic Cloud  & \ref{sec:conflicts}\\

SMoC &The standard model of cosmology; LCDM or LWDM models &\ref{sec:introd}\\

SMoPP &the standard model of particle physics &\ref{sec:introd} \\


TDG &tidal dwarf galaxy; cannot have a dynamically significant DM
content if the SMoC is true  & \ref{sec:tdgs}\\

\end{tabular}
\end{center}

\begin{center}
\begin{tabular}{l p{5.5cm} l}
Acronym   & Meaning    &Sec.\\
\hline

UCD  &ultra compact dwarf galaxy  & \ref{sec:conflicts}\\

UFD &ultra faint dwarf galaxy; baryonic mass $10^3\simless M/M_\odot
\simless 10^5$; MW UFDs are discovered with deep automatic/robotic sky surveys & \ref{sec:PDGs}\\

VPOS  &vast polar structure of the MW; it contains all known satellite
galaxies, young halo globular clusters and a large fraction of the
known stellar and gas streams  &  \ref{sec:other}\\



WDM  &warm dark matter &\ref{sec:introd}\\

\hline

\end{tabular}
\end{center}

\end{document}